\documentclass[aps,showpacs,nofootinbib,eqsecnum]{revtex4-2}

\usepackage{amsmath,amssymb}
\usepackage{bm}
\usepackage{cancel}
\usepackage{cases}
\usepackage{color}
\usepackage{comment}
\usepackage{dcolumn}
\usepackage{graphicx}
\usepackage{here}
\usepackage{multirow}
\usepackage{slashed}
\usepackage{subfigure}
\usepackage{textcase}
\usepackage{tikz-feynhand}
\usepackage{ytableau}
\usepackage[mathlines]{lineno}
\usepackage[dvipsnames]{xcolor}
\usepackage[colorlinks=true,pdfstartview=FitV,linkcolor=blue,citecolor=blue,urlcolor=blue]{hyperref}
\usepackage{CJKutf8}

\begin{document}

\begin{flushright}
    KOBE-COSMO-25-20, KUNS-3086
\end{flushright} 

\title{
Axion-photon conversion in stochastic magnetic fields
}

\author{Wataru Chiba$^{1}$}
\thanks{{\color{blue}245s114s@stu.kobe-u.ac.jp}}

\author{Ryusuke Jinno$^{1}$}
\thanks{{\color{blue}jinno@phys.sci.kobe-u.ac.jp}}

\author{Kimihiro Nomura$^{2}$}
\thanks{{\color{blue}k.nomura@tap.scphys.kyoto-u.ac.jp}}

\affiliation{$^{1}$ Department of Physics, Graduate School of Science, Kobe University, 1-1 Rokkodai, Kobe, Hyogo 657-8501, Japan}

\affiliation{$^{2}$ Department of Physics, Kyoto University, Kyoto 606-8502, Japan}

\date{\today}

\begin{abstract}
We investigate axion-photon conversion in stochastic magnetic fields, focusing on the evolution of the photon intensity and polarizations induced by conversion into axions. 
Assuming Gaussian magnetic fields characterized by the power spectra of their helical/non-helical components, we express the expectation values and variances of the photon intensity and linear/circular polarizations after conversion in terms of these spectra.
We find nontrivial dependencies of these statistical quantities on the characteristic magnetic field correlation length, the propagation distance, and the axion mass.
Moreover, we find that nontrivial polarizations emerge even if the photons are initially unpolarized, that the variances of these observables become suppressed in specific frequency regions, and that a peak structure arises in the expectation value of the circular polarization in the presence of statistically helical magnetic fields.
We also point out consistency relations among these statistical quantities that hold independently of the specific forms of the magnetic field power spectra.
\end{abstract}

\maketitle

\section{Introduction}
Axions are hypothetical pseudoscalar particles originally proposed to resolve the strong CP problem in quantum chromodynamics (QCD)~\cite{PhysRevLett.38.1440,PhysRevLett.40.223,PhysRevLett.40.279,PhysRevLett.43.103,SHIFMAN1980493,DINE1981199,Zhitnitsky:1980tq}.
Remarkably, many theories beyond the Standard Model, including string theories, ubiquitously predict large families of light pseudoscalar particles~\cite{Svrcek:2006yi}.
These are commonly referred to as axion-like particles (ALPs), and in this work we use the term ``axions'' to collectively denote both the QCD axion and ALPs.
Beyond their high-energy particle-physics origin, axions have profound implications in cosmology~\cite{Marsh:2015xka} and astrophysics~\cite{Carenza:2024ehj}.
Indeed, light axions can constitute dark matter~\cite{Preskill:1982cy,Abbott:1982af,Dine:1982ah}, thereby playing an important role in cosmic structure formation. 
Heavy axions can realize slow-roll inflation in the very early universe naturally owing to their approximate shift symmetry~\cite{PhysRevLett.65.3233,Kim:2004rp,Dimopoulos:2005ac}. 
On the other hand, axions also have the potential to affect stellar processes such as white dwarf cooling~\cite{isern2013axioncoolingwhitedwarfs} and supernova energy emission~\cite{Carenza:2023lci}. 
Axion searches are therefore highly motivated not only by particle physics considerations but also by their potential to address open questions in cosmology and astrophysics.

A key feature underlying the multifaceted role played by the axion is its coupling to photons described by the interaction Lagrangian
\begin{align}
    \mathcal{L}_{\mathrm{int}}=-\frac{1}{4}g_{a\gamma\gamma}aF_{\mu\nu}\tilde{F}^{\mu\nu}~,
\end{align}
where $g_{a\gamma\gamma}$ is the axion-photon coupling constant, $a$ is the axion field, $F_{\mu\nu}$ is the electromagnetic field strength tensor, and $\tilde{F}_{\mu\nu}$ is its dual. This interaction leads to conversion between axions and photons in the presence of an external magnetic field~\cite{MAIANI1986359,PhysRevD.37.1237}, with the conversion probability typically given by
\begin{align}
    P_{\gamma\rightarrow a}\propto (g_{a\gamma\gamma}Bd)^2~,
\end{align}
where $B$ is the strength of the background magnetic field and $d$ is the propagation distance.
This coupling provides the theoretical foundation for many experimental searches~\cite{PhysRevLett.51.1415,Irastorza:2018dyq}, such as CAST (searches for solar axions)~\cite{CAST:2017uph,Armengaud:2014gea} and ADMX (searches for dark-matter axions)~\cite{Asztalos_2010}, and so on (see e.g. Ref.~\cite{ParticleDataGroup:2024cfk} and references therein).

Given the above relationship, environments with strong magnetic fields or long photon propagation distances are advantageous for axion searches based on the conversion mechanism.
Such conditions are realized, for example, in astrophysical regions surrounding pulsars~\cite{PhysRevD.37.1237}, black holes \cite{Nomura:2022zyy}, and active galactic nuclei \cite{Hooper:2007bq,Hochmuth:2007hk}, which naturally provide strong magnetic fields. 
In addition, photon propagation over cosmological distances allows the effects induced by axion-photon conversion to accumulate, thereby enhancing their observational signatures.
Moreover, magnetic fields are also suggested to exist on cosmological scales, including in the intergalactic medium, and cosmic voids~\cite{Neronov:2010gir, Taylor:2011bn}.
Such environments compensate for the relatively limited magnetic field strengths and path lengths achievable in laboratory experiments, making the universe itself a natural laboratory for probing axion-photon conversion.

In recent years, observational techniques for detecting high-energy photons from distant astrophysical sources have advanced rapidly.
For instance, the space-based gamma-ray telescope Fermi-LAT~\cite{ballet2024fermilargeareatelescope} has enabled precise spectral measurements over cosmological distances, while ground-based Cherenkov telescope arrays such as MAGIC~\cite{MAGIC:2020xry} have extended observations into the TeV regime.
Moreover, next-generation ultra-high-energy gamma-ray observatories including LHAASO~\cite{LHAASO:2023rpg} and Carpet-3~\cite{Dzhappuev_2025} now provide access to the sub-PeV to PeV range, opening a new observational window on photon propagation over cosmic distances.

These observational developments make it possible to trace how photons are affected during their cosmological propagation through magnetized environments.
If cosmological magnetic fields are present, high-energy photons are exposed to them throughout their propagation history, allowing photon-to-axion conversion effects to accumulate over enormous distances.
This accumulation enhances the observability of signatures imprinted in photon intensity and polarizations, and therefore axion searches that utilize cosmic magnetic fields are expected to play an increasingly important role in future research.
Building on the formulation by Raffelt and Stodolsky~\cite{PhysRevD.37.1237}, several approaches have been developed to model axion-photon conversion in cosmic magnetic fields.
In particular, ``domain-like'' models~\cite{Mirizzi:2009aj,PhysRevD.96.043519,PhysRevD.98.043018,seong2024axionmagneticresonancenovel}, which treat the cosmic magnetic field as a sequence of coherent domains, and statistical approaches~\cite{Li:2022mcf,setabuddin2025axionphotonconversionflrwprimordial,Mirizzi_2007}, which characterize the magnetic field through its power spectrum, have been actively explored. 
However, a statistical formulation of photon polarizations over an ensemble of cosmic magnetic fields remains unexplored.

In this work, we develop a systematic framework to quantify how axion-photon conversion alters the polarization statistics of photons propagating through randomly realized magnetic fields characterized by a power spectrum.
We are primarily interested in the process in which a fraction of the photons initially present is converted into axions.\footnote{ Throughout this paper, we use the term ``axion-photon conversion'' to refer to this process.}
In particular, we analytically derive not only the expectation values of the Stokes parameters but also their variances, which arise from the statistical nature of the magnetic fields.
A notable feature of our framework is that it allows for a statistical treatment of magnetic fields characterized by their power spectra, including helical components, and provides an analytical description of higher-order statistics of the Stokes parameters in terms of these quantities, in contrast to previous studies that have neglected helicity~\cite{PhysRevD.37.1237,setabuddin2025axionphotonconversionflrwprimordial,Mirizzi_2007}, or those that have not analytically treated polarization and higher-order statistics, even though helicity is included~\cite{kushwaha2025probingaxionphotonconversioncircular,seong2024axionmagneticresonancenovel,PhysRevD.96.043519,Li:2022mcf}.
Within this formulation, we clarify how axion parameters are imprinted on the polarizations of photons through their conversion in randomly realized magnetic fields.
We further derive consistency relations linking the expectation values and variances of the Stokes parameters, which hold independently of the detailed shape of the magnetic-field power spectrum.
Our results demonstrate that nontrivial polarization structures can emerge, with characteristics that depend sensitively on model parameters such as the axion mass, the axion-photon coupling, and the correlation length of the magnetic field.
By employing our formalism, electromagnetic signals propagating over cosmological distances---through statistically generated magnetic fields such as cosmological magnetic fields---can be used to place meaningful constraints on the axion parameter space.

The present paper is organized as follows.
In Sec.~\ref{sec:photonreview}, we review the propagation of photons in the presence of plasma, background magnetic fields, and the cosmic microwave background (CMB).
In Sec.~\ref{sec:axionreview}, we review axion-photon conversion in background magnetic fields. We introduce the Stokes parameters to describe the polarization state of propagating photons and calculate how these parameters change due to photon-to-axion conversion.
In Sec.~\ref{sec:random}, we introduce stochastic magnetic fields and study axion-photon conversion in these backgrounds. 
Assuming that the magnetic field is Gaussian and characterized by its power spectra, we derive the formulae for the statistical quantities (expectation values and variances) of the Stokes parameters.
We also derive consistency relations among these statistical quantities, which hold independently of the specific forms of the power spectra.
In Sec.~\ref{sec:numerical}, we numerically investigate the behavior of these statistical quantities.
Sec.~\ref{sec:discussions} is devoted to discussion and conclusions.

\section{Photon propagation in the presence of background}
\label{sec:photonreview}

In this section, we review the propagation of photons in a cosmic environment before discussing their conversion into axions. 
In particular, we consider the effects of plasma, a background magnetic field, and the CMB, and show how these components modify the photon propagation equation.
Throughout this paper, we assume photon sources at low redshift in the present universe and neglect cosmological expansion effects. 
In particular, neither the photon frequency nor the background media (plasma, a magnetic field and the CMB) are subject to redshift evolution. Therefore, all quantities are evaluated at the present epoch. 

We start with the Maxwell Lagrangian in vacuum,
\begin{align}
    \mathcal{L}_{0}=-\frac{1}{4}F_{\mu\nu}F^{\mu\nu}~,
\end{align}
where $F_{\mu\nu}=\partial_\mu A_\nu-\partial_\nu A_\mu$ is the electromagnetic field strength tensor and $A_\mu$ is the photon vector field.
To describe photon propagation, we adopt the radiation gauge condition,
\begin{align}
    \partial_i A^i=0~,\quad A^0=0~,
\end{align}
and introduce orthonormal basis vectors $e^{(n)}_i$ $(n=x,y,z)$ as 
\begin{align}
    e_i^{(x)}=
    \begin{pmatrix}
        1\\
        0\\
        0\\
    \end{pmatrix}
    ~,\quad
     e_i^{(y)}=
    \begin{pmatrix}
        0\\
        1\\
        0\\
    \end{pmatrix}
    ~,\quad
    e_i^{(z)}=
    \begin{pmatrix}
        0\\
        0\\
        1
    \end{pmatrix}
    ~.
\end{align}
Throughout this paper, we take the propagation direction of the photon to be the $z$-direction.
In this case, the radiation gauge condition implies $A_z = 0$, and the photon field can be expanded as 
\begin{align}
    A_i(t, \bm{x})=\sum_{n=x,y}A_n(t,z) \, e_i^{(n)}~.
\end{align}
With this notation, the Maxwell Lagrangian reduces to
\begin{align}
    \mathcal{L}_0=\frac{1}{2}\sum_{n=x,y}(\dot{A}_n^2-A_n^{\prime 2})~,
\end{align}
where $\dot{A}\equiv\partial_0 A$ and $A^\prime\equiv\partial_z A$.

\subsection{Effect of spatially homogeneous plasma}

When plasma with electron number density $n_e$ is present in the propagation environment, it modifies the dispersion relation of the photons.
This effect can be incorporated into the Maxwell Lagrangian by introducing an effective mass as~\cite{PhysRevD.37.1237}
\begin{align}
    \mathcal{L}_{\mathrm{pl}}\equiv
\frac{1}{2}\sum_{n=x,y}(\dot{A}_n^2-{A_n^{\prime 2}}-m^2_\mathrm{pl}A_n^2)~,
\end{align}
where the plasma-induced effective photon mass $m_{\mathrm{pl}}$ is given by
\begin{align}
    m_\mathrm{pl}=\sqrt{\frac{4\pi \alpha n_e}{m_e}} ~,
\end{align}
with $\alpha\approx1/137$ the fine-structure constant and $m_e$ the electron mass.
Throughout this paper, we assume that the plasma is spatially homogeneous and isotropic (see, e.g, Ref.~\cite{brahma2024photonconversionaxionsdark} for the case of anisotropic plasma).

\subsection{Effect of background magnetic fields}

It is known that virtual electron loops induce the so-called Euler-Heisenberg term in the effective action of photons, and its effect cannot be neglected when background electromagnetic fields are present~\cite{PhysRevD.37.1237}.
This term is given by
\begin{align}
\mathcal{L}_{0+\mathrm{EH}}=\mathcal{L}_{0}+\kappa\Big[(F_{\mu\nu}F^{\mu\nu})^2+\frac{7}{4}(F_{\mu\nu}\tilde{F}^{\mu\nu})^2\Big]~,
\label{eqEHLag}
\end{align}
where $\kappa= \alpha^2 / (90m_e^4)$, and $\tilde{F}^{\mu\nu}=(1/2)\epsilon^{\mu\nu\rho\sigma}F_{\rho\sigma}$ is the dual of the electromagnetic field strength tensor. 
In the presence of background electromagnetic fields, we decompose the field strength into the background and perturbations as
\begin{align}
    F_{i0}=\bar{E}_i+\delta E_i~,\quad F_{ij}=\epsilon_{ijk}(\bar{B}_k+\delta B_k)~,
\end{align}
where $\bar{E}_i$ and $\bar{B}_i$ denote the background electric and magnetic fields, respectively, and $\delta E_i$ and $\delta B_i$ represent their perturbations.
Expanding the Lagrangian \eqref{eqEHLag} up to second order in perturbations and neglecting higher order terms, we obtain
\begin{align}
  \mathcal{L}^{(2)}_{0+\mathrm{EH}}&=\frac{1}{2}(\delta E^2-\delta B^2)+8\kappa(\bar{B}^2-\bar{E}^2)(\delta B^2-\delta E^2)+16\kappa(\bar{B}\cdot\delta B-\bar{E}\cdot \delta E)^2\notag\\
  &+28\kappa(\bar{E}\cdot\delta B+\bar{B}\cdot \delta E)^2+56\kappa(\bar{E}\cdot\bar{B})(\delta E\cdot\delta B)~.
  \label{eq:EH}
\end{align}

Let us consider photons propagating in the $z$-direction through a background magnetic field. In this case, we set $\bar{E}_i = 0$ in Eq.~\eqref{eq:EH}.
It is also convenient to introduce orthonormal basis vectors $e^{(\parallel)}_i$ and $e^{(\perp)}_i$ in the $x$-$y$ plane such that the background magnetic field $\bar{B}_i$ has no component along $e^{(\perp)}_i$ and can therefore be decomposed as 
\begin{align}
    \bar{B}_i(z)=B_\parallel (z) \, e_i^{(\parallel)}+B_z (z) \, e_i^{(z)}~. 
    \label{eqbgBexp}
\end{align}
Let $\phi$ be the angle between $e^{(\parallel)}_i$ and the $x$-axis.
Then, the basis vectors $e^{(\parallel)}_i$, $e^{(\perp)}_i$, and $e^{(z)}_i$ are written in the $(x,y,z)$ coordinate system as
\begin{align}
e_i^{(\parallel)}=
     \begin{pmatrix}
        \cos\phi\\
        \sin\phi\\
        0\\
    \end{pmatrix}
    ~,\quad
    e_i^{(\perp)}=
    \begin{pmatrix}
        -\sin\phi\\
        \cos\phi\\
        0\\
    \end{pmatrix}
    ~,\quad
    e_i^{(z)}=
    \begin{pmatrix}
        0\\
        0\\
        1
    \end{pmatrix}
    ~.
    \label{eqparaperp}
\end{align}
Using this basis, the propagating photon field in the radiation gauge can be expanded as 
\begin{align}
A_i(t, \bm{x})=\sum_{m=\parallel,\perp}A_m(t,z) \, e_i^{(m)}~.
\label{eqphotonAexp}
\end{align}
In terms of the coefficients $A_m(t,z)$, the perturbations of the electromagnetic field are given by $\delta E_{\parallel,\perp}= - \dot A_{\parallel,\perp},\,\delta B_\parallel=- A^\prime_\perp,\,\delta B_\perp= A^\prime_\parallel$.
Substituting these expressions and $\bar{E}_i = 0$ into Eq.~\eqref{eq:EH}, we obtain
\begin{align}
  \mathcal{L}_B\equiv\mathcal{L}^{(2)}_{0+\mathrm{EH}}|_{\bar{E}=0}
  &=\frac{1}{2}\sum_{m=\parallel,\perp}(\varepsilon_{Bm} \, 
  \dot{A}_m^2-\mu_{Bm}^{-1} \,{A_m^{\prime 2}})~,
  \label{eq:basispp}
\end{align}
where the magnetic field-induced permittivities $\varepsilon_{B\parallel,\perp}$ and permeabilities $\mu_{B\parallel,\perp}$ are
\begin{align}
\varepsilon_{B\parallel}&=1-16\kappa {B}^2+56\kappa B_\parallel^2~,\\
\varepsilon_{B\perp}&=1-16\kappa {B}^2~,\\
\mu_{B\parallel}^{-1}&=1-16\kappa {B}^2~,
\\
\mu_{B\perp}^{-1}&=1-16\kappa {B}^2 -32\kappa B_\parallel^2~,
\end{align}
with $B^2 \equiv {B}_\parallel^2 + {B}_z^2$.
In Eq.~\eqref{eq:basispp}, we omit the spatial dependence of $\phi$, assuming that the length scale of the magnetic field variation is much longer than the wavelength of the propagating photons.

\subsection{Effect of the CMB}

The space through which photons propagate over cosmological distances contains not only cosmic magnetic fields but also the CMB, and hence the latter should also be treated as part of the background electromagnetic field in Eq.~\eqref{eq:EH}.
Note that the characteristic frequency scales associated with the CMB photons and the cosmic magnetic fields are hierarchically different and thus they can be treated as independent.
In fact, the typical frequency of the CMB photons is of order $10^{11}\text{--}10^{12}~\mathrm{Hz}$ for a temperature $T \simeq 2.726\,\mathrm{K}$, whereas the characteristic frequency corresponding to the coherence scale of cosmic magnetic fields is as small as $\sim 1~\mathrm{Mpc}^{-1}\sim  10^{-14}~\mathrm{Hz}$.
Let us model the CMB as a statistically homogeneous and isotropic background characterized by the following statistical properties:
\begin{align}
  \left<\bar{E}_i\bar{E}_j\right>_\mathrm{CMB}=\frac{1}{3}\left<\bar{E}^2\right>_\mathrm{CMB}\delta_{ij}~,\quad\left<\bar{B}_i\bar{B}_j\right>_\mathrm{CMB}=\frac{1}{3}\left<\bar{B}^2\right>_\mathrm{CMB}\delta_{ij}~,\quad\left<\bar{E}_i\bar{B}_j\right>_\mathrm{CMB}=0~,
\end{align}
where the bracket $\left< \cdots \right>_\mathrm{CMB}$ denotes an ensemble average over photons in the CMB frequency range.
The effect of the CMB on the propagating photons can then be incorporated by taking the angular average of Eq.~\eqref{eq:EH} as
\begin{align}
\mathcal{L}_\mathrm{{CMB}}\equiv\left<\mathcal{L}^{(2)}_{0+\mathrm{EH}}\right>_\mathrm{CMB}=\frac{1}{2}\sum_{n=x,y}(\varepsilon_{\mathrm{eff}}\dot A^2_n-\mu_{\mathrm{eff}}^{-1}{A^\prime}^2_n)~.
\end{align}
Here the effective permittivity $\varepsilon_{\mathrm{eff}}$ and 
permeability $\mu_{\mathrm{eff}}$ induced by the CMB are given by
\begin{align}
  \varepsilon_{\mathrm{eff}}&\equiv1+\kappa
  \bigg(\frac{80}{3}\left<\bar{E}^2\right>_\mathrm{CMB}+\frac{8}{3}\left<\bar{B}^2\right>_\mathrm{CMB} \bigg) =1+\kappa
  \frac{88}{3}\rho_{\mathrm{CMB}}~,\\
  \mu^{-1}_{\mathrm{eff}}&\equiv1-\kappa \bigg(\frac{80}{3}\left<\bar{B}^2\right>_\mathrm{CMB}+\frac{8}{3}\left<\bar{E}^2\right>_\mathrm{CMB}\bigg)=1 - \kappa \frac{88}{3}\rho_{\mathrm{CMB}}~,
\end{align}
where we have introduced the CMB energy density $\rho_{\mathrm{CMB}}\equiv\left<\bar{E}^2\right>_\mathrm{CMB}=\left<\bar{B}^2\right>_\mathrm{CMB}$.
Using the CMB temperature $T=2.726 \, \mathrm{K}$~\cite{Dobrynina:2014qba} at the present epoch, this reads $\rho_{\mathrm{CMB}}=(\pi^2/15)T^4=0.26 \, \mathrm{eV~cm}^{-3}$.

\subsection{Total effects}

So far, we have investigated the effects of the plasma, background magnetic fields, and the CMB on photon propagation separately.
Combining all the effects, the Lagrangian for the propagating photons becomes
\begin{align}
    \mathcal{L}&=\frac{1}{2}\sum_{m=\parallel,\perp}(\varepsilon_m \dot{A}^2_m-\mu_m^{-1}{A_m^{\prime 2}}-m_\mathrm{pl}^2A^2_m)~.
    \label{eqtotallag}
\end{align}
where the total permittivities $\varepsilon_{\parallel, \perp}$ and permeabilities $\mu_{\parallel,\perp}$ are given by
\begin{align}
\varepsilon_{\parallel}&=1+\kappa \bigg(\frac{88}{3}\rho_{\mathrm{CMB}}-16 {B}^2+56 B_\parallel^2 \bigg)~,\\
\varepsilon_{\perp}&=1+\kappa \bigg(\frac{88}{3}\rho_{\mathrm{CMB}}-16 {B}^2 \bigg)~,\\
\mu_{\parallel}^{-1}&=1-\kappa \bigg(\frac{88}{3}\rho_{\mathrm{CMB}}+16 {B}^2 \bigg)~,
\\
\mu_{\perp}^{-1}&=1-\kappa \bigg(\frac{88}{3}\rho_{\mathrm{CMB}}+16 {B}^2+32 B_\parallel^2 \bigg)~.
\end{align}
The equation of motion for the photon field $A_m$ $(m=\parallel , \perp)$ derived from the Lagrangian \eqref{eqtotallag} is 
\begin{align}
    -\ddot{A}_m+n_m^{-2}A_m^{\prime\prime}-\frac{m_\mathrm{pl}^2}{\varepsilon_m}A_m=0~, 
    \label{eqeomAm}
\end{align}
where we have used $\partial_z(\bar{B}A_m)\approx \bar{B}\partial_zA_m$ assuming that the magnetic field varies only on scales much longer than the photon wavelength.
We have also defined the refractive indices $n_m\equiv\sqrt{\varepsilon_m\mu_m}$, which are approximately given by
\begin{align}
    n_\parallel-1&\approx
    \frac{88}{3} \kappa \, 
    \rho_{\mathrm{CMB}}+28 \kappa B_\parallel^2 \equiv\chi_\mathrm{CMB}+\chi_\parallel~,\\
    n_\perp-1&\approx \frac{88}{3} \kappa \, \rho_{\mathrm{CMB}}+16 \kappa B_\parallel^2 \equiv\chi_\mathrm{CMB}+\chi_\perp~,
    \label{eqAmeffmass}
\end{align}
provided that $|\chi_\mathrm{CMB}+\chi_m|\ll1$.
In fact, since a typical case with $\rho_{\mathrm{CMB}} = 0.26 \,\mathrm{eV \, cm^{-3}}$ and $B = 1 \,\mu\mathrm{G}$, which is also estimated at the present epoch, gives $\chi_{\mathrm{CMB}} \sim 0.5 \times 10^{-42}$ and $\chi_m \sim 10^{-56}$, this approximation is well justified.

Hereafter, we consider relativistic photons whose angular frequency $\omega$ is much larger than $m_{\mathrm{pl}}$. In this regime, the equation of motion \eqref{eqeomAm} can be further approximated as
\begin{align}
    -\ddot{A}_m+n_m^{-2}A^{\prime\prime}_m-\frac{m_\mathrm{pl}^2}{\varepsilon_m}A_m
    &\approx(\partial^2-m_\mathrm{pl}^2+2(\chi_\mathrm{CMB}+\chi_m)\omega^2)A_m
    \notag\\
    &\equiv(\partial^2-m^2_{\gamma m})A_m~,
\end{align}
where $\partial^2\equiv-\partial_0^2+\partial_z^2$.
In deriving this expression, we have expanded $n_m^{-2}$ and $\varepsilon_m^{-1}$ around unity, and used the approximation $(\chi_\mathrm{CMB}+\chi_m) A_m'' \approx -(\chi_\mathrm{CMB}+\chi_m) \omega^2 A_m$, which holds for relativistic photons.

\section{Review of axion-photon conversion}
\label{sec:axionreview}

In this section, we derive the equation for the propagation of the axion-photon system.
We then formulate the polarization state of photons undergoing conversion into axions while propagating through background magnetic fields.
Since we focus on the process of photon-to-axion conversion in this work, we assume that the initial condition consists solely of photons, as given in Eq.~\eqref{eq:iniphoton}.

\subsection{Derivation of equation of the axion-photon propagation}
\label{appendix B}
The free Lagrangian for the axion field $a$ with mass $m_a$ is given by
\begin{align}
\mathcal{L}_a&=-\frac{1}{2}(\partial a)^2-\frac{1}{2}m_a^2a^2~.
\end{align}
The axion couples to the photon through the interaction term
\begin{align}
    \mathcal{L}_{\mathrm{int}}&=-\frac{1}{4}g_{a\gamma\gamma}aF_{\mu\nu}\tilde{F}^{\mu\nu}~,
\end{align}
where $g_{a\gamma\gamma}$ denotes the axion-photon coupling constant.
When a background magnetic field as in Eq.~\eqref{eqbgBexp} is present, the above interaction Lagrangian reduces to a mixing term between the axion and the photon,
\begin{align}
    \mathcal{L}_{\mathrm{int}}=-g_{a\gamma\gamma}B_\parallel a\dot{A}_\parallel~,
\end{align}
where $A_\parallel$ is the photon field component defined in Eq.~\eqref{eqphotonAexp}.
The resulting equations of motion for the axion-photon system, in the presence of the background magnetic field together with the CMB and homogeneous plasma, are given by
\begin{align}
    \begin{cases}
    (\partial^2-m_a^2)a(z,t)=-g_{a\gamma\gamma}B_\parallel(z)\dot{A}_\parallel(z,t)~,\\
    (\partial^2-m^2_{\gamma\parallel}) A_\parallel(z,t)=g_{a\gamma\gamma}B_\parallel(z)\dot{a}(z,t)~,\\
        (\partial^2-m^2_{\gamma\perp}) A_\perp(z,t)=0~,\\
    \end{cases}
    \label{eqaxiphoteom1}
\end{align}
where the effective photon mass is defined in Eq.~\eqref{eqAmeffmass} as $ m_{\gamma m}^2=m_{\mathrm{pl}}^2-2(\chi_{\mathrm{CMB}}+\chi_m)\omega^2$.

We expand the photon and axion fields into plane waves with angular frequency $\omega$ as
\begin{align}
     A_m (t,z)= iA_m(z) \, e^{-i\omega t} + \text{c.c.}~,\quad a(t,z)=a(z)e^{-i\omega t}+ \text{c.c.}~.
\end{align}
Under the relativistic approximation, the equations of motion \eqref{eqaxiphoteom1} can be rewritten as 
\begin{align}
    \begin{cases}
    (2\omega(\omega +i\partial_z)-m_a^2)a(z)\approx -\omega g_{a\gamma\gamma}B_\parallel(z)A_\parallel(z)~,\\
    (2\omega(\omega +i\partial_z)-m^2_{\gamma\parallel}) A_\parallel(z)\approx -\omega g_{a\gamma\gamma}B_\parallel(z)a(z)~,\\
    (2\omega(\omega +i\partial_z)-m^2_{\gamma\perp}) A_\perp(z)\approx0~,\\
    \end{cases}
\end{align}
where we have used $\partial^2 = (\omega + i\partial_z) (\omega - i\partial_z)=(\omega + i\partial_z)(\omega+k_z) \approx 2 \omega(\omega + i\partial_z)$.
Then we arrive at a Schr\"{o}dinger-like equation,
\begin{align}
    i\partial_z
    \begin{pmatrix}
        a(z)\\A_\parallel(z)\\A_\perp(z)
    \end{pmatrix}
    =
    \begin{pmatrix}
        -\omega+\frac{m^2_a}{2\omega} & -\frac{1}{2}g_{a\gamma\gamma}B_\parallel(z) & 0\\
         -\frac{1}{2}g_{a\gamma\gamma}B_\parallel(z) &-\omega+\frac{m^2_{\gamma\parallel}}{2\omega} & 0\\
         0 & 0 & -\omega+\frac{m^2_{\gamma\perp}}{2\omega} 
    \end{pmatrix}
    \begin{pmatrix}
         a(z)\\A_\parallel(z)\\A_\perp(z)
    \end{pmatrix}~.
\end{align}
In the following, to express the system in a general coordinate basis within the plane orthogonal to the propagation direction, we rotate the photon components defined in the $(\parallel, \perp)$ basis as
\begin{align}
    \begin{pmatrix}
        a(z)\\A_x(z)\\A_y(z)
    \end{pmatrix}
    =R(\phi)
    \begin{pmatrix}
        a(z)\\A_\parallel(z)\\A_\perp(z)
    \end{pmatrix}~,
\end{align}
where the rotation matrix $R(\phi)$, corresponding to the rotation by the angle $\phi$ in the $x$-$y$ plane [see Eq.~\eqref{eqparaperp}], is given by
\begin{align}
    R(\phi)=\begin{pmatrix}
        1 & 0 & 0\\
         0 & \cos\phi & -\sin\phi\\
         0 & \sin\phi & \cos\phi
    \end{pmatrix}
    ~.
\end{align}
We thus obtain the equation describing axion-photon propagation in the $z$-direction as
\begin{align}
    i\partial_z\Psi(z)=
    \mathcal{M}(z)
    \Psi(z)~.
    \label{eq:equation}
\end{align}
where the ``wave function'' $\Psi$ is a three-component field with two photon degrees of freedom ($A_x, A_y$) and the axion degree of freedom $a$;
\begin{align}
    \Psi(z)\equiv
    \begin{pmatrix}
        a(z)\\A_x(z)\\A_y(z)
    \end{pmatrix}~.
    \label{eqdefPsi}
\end{align}
The mixing matrix $\mathcal{M}$ corresponds to the ``Hamiltonian'' in the Schr\"{o}dinger-like equation~\cite{Galanti:2024lfn}, and is given by
\begin{align}
    \mathcal{M}(z)=R(\phi)
    \begin{pmatrix}
        -\omega+\frac{m^2_a}{2\omega} & -\frac{1}{2}g_{a\gamma\gamma}B_\parallel(z) & 0\\
         -\frac{1}{2}g_{a\gamma\gamma}B_\parallel(z) &-\omega+\frac{m^2_{\gamma\parallel}}{2\omega} & 0\\
         0 & 0 & -\omega+\frac{m^2_{\gamma\perp}}{2\omega} 
    \end{pmatrix}
    R^{-1}(\phi)\equiv \begin{pmatrix}
       \Pi_a & \Pi_{Mx} & \Pi_{My}\\
         \Pi_{Mx} &\Pi_\gamma+\Pi_{xx} & \Pi_{xy}\\
         \Pi_{My} & \Pi_{yx} &\Pi_{\gamma}+\Pi_{yy}
    \end{pmatrix}~.
\end{align}
The diagonal elements of the mixing matrix correspond to the dispersion relations of the axion and photons given by
\begin{align}
\Pi_a&=-\omega+\frac{m^2_a}{2\omega}~,\\
\Pi_{\gamma}&=-(1+\chi_\mathrm{CMB})\,\omega+\frac{m_{\rm pl}^2}{2\omega}~.
\end{align}
We comment on the modification of the dispersion relation of photons in a plasma due to the background magnetic field, known as the Cotton–Mouton effect and the Faraday rotation (see, e.g., Ref.~\cite{Ejlli_2019}).
These effects induce a correction to the effective plasma mass of the photons, $\delta m$, thereby leading to $\Pi_\gamma\approx (m_\mathrm{pl}^2/\omega)(1+\delta m^2/m^2_\mathrm{pl})$.
However, for the photon frequencies $\omega\sim10^{10}~\mathrm{eV}$ and the cosmic magnetic field strength $B \sim 1~\mathrm{\mu G}$ of our interest, these contributions can be neglected.
\footnote{
In fact, the orders of magnitude of these two subleading corrections to the effective plasma mass can be estimated using $\omega_B\equiv \frac{e B}{m_e}$ as
\begin{align*}
\frac{\delta m_\mathrm{CM}^2}{m^2_\mathrm{pl}}  \approx\frac{\omega_B^2}{\omega^2}\sim 10^{-54}  ~\bigg(\frac{\omega}{10^{10}\mathrm{eV}}\bigg)^{-2}~\bigg(\frac{B}{1~\mathrm{\mu G}}\bigg)^{2}~,\quad
\frac{\delta m_\mathrm{F}^2}{m^2_\mathrm{pl}}\approx\frac{\omega_B}{\omega}\sim 10^{-27} ~\bigg(\frac{\omega}{10^{10}\mathrm{eV}}\bigg)^{-1}~\bigg(\frac{B}{1~\mathrm{\mu G}}\bigg)~,
\end{align*}
where $\delta m_\mathrm{CM}$ and $\delta m_\mathrm{F}$ denote the contributions from the Cotton–Mouton effect and the Faraday rotation.
Here, $\omega$ represents a typical photon frequency to be specified later (which corresponds to $\omega_\mathrm{eq}$).
For such frequencies, these corrections are negligible in the following analysis.
}

The elements of the $2 \times 2$ submatrix in the bottom right of the
matrix $\mathcal{M}$ are
\begin{align}
\Pi_{xx}&=-(\chi_\parallel\cos^2\phi+\chi_\perp\sin^2\phi)\,\omega~,
\\
\Pi_{yy}&=-(\chi_\parallel\sin^2\phi+\chi_\perp\cos^2\phi)\,\omega~,
\\
    \Pi_{xy}&=\Pi_{yx}=-(\chi_\parallel-\chi_\perp)\sin\phi\cos\phi\,\omega~,
\end{align}
which correspond to the photon-photon interaction induced by a virtual electron loop in the presence of a background magnetic field.
The other off-diagonal elements are given by
\begin{align}
    \Pi_{Mx}(z)=-\frac{1}{2}g_{a\gamma\gamma}B_\parallel(z)\cos\phi~,\quad \Pi_{My}(z)=-\frac{1}{2}g_{a\gamma\gamma}B_\parallel(z)\sin\phi~,
\end{align}
which represent the axion-photon coupling through a background magnetic field.
This coupling induces conversion between the axion and photons during propagation.

Notice that we do not consider attenuation of photons through pair production of charged particles, in order to isolate the effect of conversion into axions imprinted on the photon polarization; we direct readers interested in this phenomenon in the present system to, e.g., Refs.~\cite{Galanti:2022chk,Li:2024jlc}.

\subsection{Born approximation}

As long as the background magnetic field is sufficiently small and thus the photon-photon or axion-photon interactions are sufficiently weak, Eq.~\eqref{eq:equation} can be solved with a perturbative approach (Born approximation).
We decompose the mixing matrix $\mathcal{M}$ into the zeroth and perturbative parts as

\begin{align}
    \mathcal{M}(z)=
  \begin{pmatrix}
\Pi_{a} & 0 & 0\\
0 & \Pi_{\gamma} & 0\\
0& 0& \Pi_{\gamma} \\
\end{pmatrix}
+\begin{pmatrix}
0 & 0 & 0 \\
0 & \Pi_{xx} & \Pi_{xy} \\
0 & \Pi_{yx} & \Pi_{yy}\\
\end{pmatrix}
+
\begin{pmatrix}
0 & \Pi_{Mx} &\Pi_{My} \\
\Pi_{Mx} & 0 & 0 \\
\Pi_{My} & 0 & 0\\
\end{pmatrix}
\equiv\mathcal{M}_0+\mathcal{M}_{\gamma\gamma}(z)+\mathcal{M}_{a\gamma}(z)~,
\end{align}
where the perturbative Hamiltonian $\mathcal{M}_{\gamma\gamma}$ corresponds to the photon-photon self-interaction induced by the background magnetic field, while $\mathcal{M}_{a\gamma}$ corresponds to the axion-photon interaction that causes conversion between them. The present system resembles the Schr\"{o}dinger equation with a time-dependent perturbation Hamiltonian added to the unperturbed Hamiltonian, which can be solved in terms of the transfer matrix $\mathcal{U}(z,0)$ defined by
\begin{align}
\Psi(z) = \mathcal{U}(z,0) \Psi(0)~.
\end{align}
We expand it in terms of the perturbation order as
\begin{align}
\mathcal{U}(z,0) &= \mathcal{U}_0(z,0) + \mathcal{U}_1(z,0) + \cdots~.
\end{align}
The zeroth order part $\mathcal{U}_0$ can readily be obtained as \begin{align}
\mathcal{U}_0(z,0) &= e^{-i\mathcal{M}_0 z}~.
\end{align}
Using this expression, the transfer matrix up to first order can be written as
\begin{align}
  \mathcal{U}(z,0)&=\mathcal{U}_0(z,0)\Big[I_{3\times3}-i\int_0^zds~\mathcal{U}_0^{\dagger}(s,0) \, (\mathcal{M}_{\gamma\gamma}(s)+\mathcal{M}_{a\gamma}(s)) \, \mathcal{U}_0 (s,0) ]\notag\\
  &\equiv\mathcal{U}_0(z,0)\Big[I_{3\times3}+\mathcal{U}_{\gamma\gamma} (z,0) +\mathcal{U}_{a\gamma} (z,0)
  \Big]~,
  \label{eqUsol}
\end{align}
where $I_{3 \times 3}$ is the $3 \times 3$ identity matrix.
The second term is defined by
\begin{align}
    \mathcal{U}_{\gamma\gamma} (z,0) &=-i\int_0^zds~\mathcal{U}_0^{\dagger}(s,0) \, \mathcal{M}_{\gamma\gamma}(s) \,\mathcal{U}_0 (s,0) 
    \notag
    \\
    &=-i\int^z_0 ds\begin{pmatrix}
      0&0&0\\
      0&\Pi_{xx}(s)&\Pi_{xy}(s)\\
      0&\Pi_{yx}(s)&\Pi_{yy}(s)\\
  \end{pmatrix}~,
  \label{eqUBsol}
\end{align}
while the third term is defined by
\begin{align}
    \mathcal{U}_{a\gamma} (z,0) &=-i\int_0^zds~\mathcal{U}_0^{\dagger}(s,0) \, \mathcal{M}_{a\gamma}(s) \, \mathcal{U}_0(s,0)
    \notag
    \\
  &=-i\int^z_0 ds
  \begin{pmatrix}
      0&e^{i\Pi s}\Pi_{Mx}(s)&e^{i\Pi s}\Pi_{My}(s)\\
      e^{-i\Pi s}\Pi_{Mx}(s)&0&0\\
      e^{-i\Pi s}\Pi_{My}(s)&0&0\\
  \end{pmatrix}~,
  \label{eqUasol}
\end{align}
where $\Pi = \Pi(\omega)$ is a function of the photon angular frequency $\omega$ defined by
\begin{align}
\Pi(\omega) \equiv\Pi_a-\Pi_\gamma=\frac{|m_a^2-m_{\mathrm{pl}}^2|}{2\omega}+\chi_\mathrm{CMB}\,\omega~.
\label{eqdefPi}
\end{align}

\subsection{Density matrix and observables}

To discuss the polarization state of the propagating photons, we introduce the Stokes parameters defined over an ensemble of photons.
This ensemble is characterized by the ``density matrix'' $\rho \equiv \sum_i p_i\Psi_i\otimes\Psi^\dagger_i$ with $p_i$ being the probability to have the ``state'' $\Psi_i$ in Eq.~\eqref{eqdefPsi}.
We assume that only photons are present initially, and in this case the density matrix can be written as
\begin{align}
    \rho_{\mathrm{ini}}\equiv\frac{1}{2}
\begin{pmatrix}
0&0&0\\
0&I_0+Q_0&U_0-iV_0\\
0&U_0+iV_0&I_0-Q_0\\
\end{pmatrix}.\label{eq:iniphoton}
\end{align}
Here, the four real parameters $I_0$, $Q_0$, $U_0$, and $V_0$ are referred to as the Stokes parameters: $I_0$ denotes the total intensity of the initial photons, $Q_0$ and $U_0$ denote their linear polarizations, and $V_0$ denotes their circular polarization.
Since in this paper we are interested in how non-trivial polarizations arise from axion-photon conversion during propagation, we take the initial photons to be unpolarized,
\begin{align}
I_0 &= 1~, \qquad Q_0 = U_0 = V_0 = 0~. 
\label{eqrhoini}
\end{align}

The state of the system after propagating a finite distance through a magnetic field can be calculated using the transfer matrix $\mathcal{U}(z,0)$ as
\begin{align}
\rho(z)
=
\mathcal{U}(z,0) \, \rho_{\mathrm{ini}} \, \mathcal{U}^\dagger(z,0)
=
\frac{1}{2}
\mathcal{U}(z,0) \, 
I_{3\times3} \,
\mathcal{U}^\dagger(z,0)
-
\frac{1}{2}
\mathcal{U} (z,0)
\begin{pmatrix}
1&0&0\\
0&0&0\\
0&0&0\\
\end{pmatrix}
\mathcal{U}^\dagger (z,0)~.
\label{eqrho}
\end{align}
The Stokes parameters after propagation can be read off as
\begin{align}
\left\{
\begin{array}{cl}
    I(\omega)&=\rho_{22}+\rho_{33}=1-\frac{1}{2}(|\mathcal{U}_{21}|^2+|\mathcal{U}_{31}|^2)
    ~,
    \\[0.1cm]
    Q(\omega)&= \rho_{22}-\rho_{33}=-\frac{1}{2}(|\mathcal{U}_{21}|^2-|\mathcal{U}_{31}|^2)~,\\[0.1cm]
    U(\omega)&=\rho_{23}+\rho_{32}=-\mathrm{Re}\left[\mathcal{U}_{21}\mathcal{U}_{31}^*\right]~,\\[0.1cm]
    V(\omega)&=  i (\rho_{23}-\rho_{32})=\mathrm{Im}\left[\mathcal{U}_{21}\mathcal{U}_{31}^*\right]
    ~.
\end{array}
\right.
  \label{eqdefStokes}
\end{align}
Here, $I$ represents the intensity of photons after propagating over a finite distance, while $Q$ and $U$ describe their linear polarization and $V$ their circular polarization.
From Eqs.~\eqref{eqdefStokes}, we find that the Stokes parameters after propagation are independent of the components of the matrix $\mathcal{U}_{\gamma\gamma}$ (see Eqs.~\eqref{eqUasol}).

Substituting the solution for $\mathcal{U}(z,0)$ from Eqs.~\eqref{eqUsol}--\eqref{eqUasol} into Eq.~\eqref{eqdefStokes}, we obtain 
\begin{align}
  \begin{cases}
    I(\omega)= 1-\frac{1}{2}a(\omega)~,\\
    Q(\omega)= -\frac{1}{2}b(\omega)~,\\
    U(\omega)=-\mathrm{Re}\left[c(\omega)\right]~,\\
    V(\omega)=\mathrm{Im}\left[c(\omega)\right]~,\\
  \end{cases}
  \label{eqStokes}
\end{align}
where $a$, $b$, and $c$ are given by
\begin{align}
a(\omega)&=\Big(\frac{g_{a\gamma\gamma}}{2}\Big)^2\int_0^z\int_0^z dsds' e^{-i\Pi(\omega) (s-s')}[B_{x}(s)B_{x}(s')+B_{y}(s)B_{y}(s')]~,
\label{eqdeca}
\\
  b(\omega)&=\Big(\frac{g_{a\gamma\gamma}}{2}\Big)^2\int_0^z\int_0^z dsds' e^{-i\Pi(\omega) (s-s')}[B_{x}(s)B_{x}(s')-B_{y}(s)B_{y}(s')]~,
  \label{eqdecb}
  \\
  c(\omega)&=\Big(\frac{g_{a\gamma\gamma}}{2}\Big)^2\int_0^z\int_0^z dsds'e^{-i\Pi(\omega) (s-s')}B_{x}(s)B_{y}(s')~.
  \label{eqdefc}
\end{align} 
These are second order quantities in the magnetic field, and are expressed as the coordinate integral of the product of the magnetic field $B_i(s)~(i = x, y)$ and the oscillatory factor $e^{- i\Pi(\omega)s}$ with frequency $\Pi(\omega)=|m_a^2-m_{\mathrm{pl}}^2|/(2\omega)+\chi_\mathrm{CMB}\omega$.
Since the initial intensity (total number of photons) has been normalized to $I_0 = 1$, the probability of photon-to-axion conversion is given by $1-I(\omega)=a(\omega)/2$.

This result indicates that, when the initial photons are unpolarized, the effect on photon polarization can be calculated solely from the axion-photon interaction represented by $\mathcal{M}_{a\gamma}$, without using the photon-photon interaction represented by $\mathcal{M}_{\gamma\gamma}$, under the Born approximation.
In other words, for initially unpolarized photons, the interaction term $\mathcal{M}_{\gamma\gamma}$, which arises from the Euler–Heisenberg (EH) term in Eq.~\eqref{eqEHLag} in the presence of background magnetic fields, does not contribute to the evolution of photon polarization within the validity of the Born approximation.
On the other hand, the EH term also modifies the photon dispersion relation through the background CMB radiation field (see Sec.~II C). In particular, this effect contributes to the difference between the two dispersion relations $\Pi(\omega)=\Pi_a-\Pi_\gamma$.
As a result, even though the contribution of the EH term induced by magnetic fields $\mathcal{M}_{\gamma\gamma}$ does not affect the polarization evolution of initially unpolarized photons under the Born approximation, the contribution arising from the CMB modifies the dispersion relation and thereby influences observables through $\Pi(\omega)$.

Before concluding this section, we note the identity
\begin{align}
    (1-I)^2=Q^2+U^2+V^2~,
    \label{con0}
\end{align}
which follows from Eqs.~\eqref{eqStokes}--\eqref{eqdefc} irrespectively of the magnetic field configuration.
This shows that, for any magnetic field configuration, initially unpolarized photons inevitably acquire a nontrivial polarization through axion-photon conversion.

\section{Conversion in stochastic magnetic fields}
\label{sec:random}

In this section, we introduce stochastic magnetic fields characterized by their power spectra and study axion-photon conversion in such backgrounds.
Owing to the stochastic nature of the magnetic fields, the Stokes parameters affected by the conversion also become stochastic.
We compute the expectation values and variances of the Stokes parameters and express them in terms of the magnetic field power spectra.

\subsection{Magnetic field power spectra}
\label{subsec:abcd}
Hereafter, we assume that the background magnetic field is a statistically homogeneous, isotropic, Gaussian random field.
Such a field can be characterized by the two-point correlation function~\cite{Durrer:2013pga,Brandenburg:2018ptt},
\begin{align}
  \left< B_i(\vec{x})B_j(\vec{x}')\right>=\int \frac{d^3k}{(2\pi)^3}e^{i\vec{k}\cdot(\vec{x}-\vec{x}')}\Big[(\delta_{ij}-\hat{k}_i\hat{k}_j )P_B(k)-i\epsilon_{ijm}\hat{k}_mP_{aB}(k)\Big]~,
  \label{2ptfunc}
\end{align}
where $k\equiv\sqrt{k_x^2+k_y^2+k_z^2}$ and $\hat{k}_i\equiv k_i/k$.
The bracket $\left< \cdots \right>$ denotes an ensemble average over infinitely many realizations of the magnetic field.
$P_B(k)$ and $P_{aB}(k)$ are the symmetric and antisymmetric parts of the magnetic field spectrum, respectively.
The factors $(\delta_{ij} - \hat{k}_i \hat{k}_j)$ and $\epsilon_{ijm}\hat{k}_m$ in Eq.~\eqref{2ptfunc} ensure that the magnetic field is divergence-free.
Furthermore, the term proportional to $\epsilon_{ijm}\hat{k}_m$ captures the rotational structure of the magnetic field and represents the parity-odd part of the correlator.
The imaginary unit $i$ in the second term is included so that $P_{aB}(k)$, like $P_B(k)$, remains real.
When $P_{aB}(k) \neq 0$, the magnetic field is referred to as helical.

\subsection{Stochastic properties of Stokes parameters}
\label{subsec:stokes_parameters}

Let us consider photon propagation over a finite distance 
$d$ and investigate how their polarization state is affected by conversion into axions in a background magnetic field.
From Eqs.~\eqref{eqdeca}--\eqref{eqdefc}, this effect is expressed as the integral along the propagation path of length $d$. For brevity, hereafter we use the shorthand notation for integrations along the path, as well as for integrations over wave vector space,
\begin{align}
    \int_s \equiv \int_0^d ds 
    ~, \qquad
    \int_k\equiv\int \frac{d^3k}{(2\pi)^3}
    ~.
\end{align}

When the background magnetic field is stochastic, the resulting Stokes parameters also become stochastic quantities.
We thus calculate the expectation values and variances of the Stokes parameters with respect to the ensemble of magnetic field realizations.
The expectation value $\mathrm{Exp}[O]$ and variance $\mathrm{Var}[O]$ of an observable $O$ with respect to this ensemble are written in terms of the bracket $\left< \cdots \right>$ as
\begin{align}
    \mathrm{Exp}[O]\equiv\left< O \right>~,
    \quad \mathrm{Var}[O]\equiv\left<(O-\left< O \right>)^2\right>=\left< O^2 \right>-\left< O \right>^2
    ~.
\end{align}
Since the Stokes parameters are written as functions of the magnetic field as Eqs.~\eqref{eqStokes}--\eqref{eqdefc}, computing their statistical properties reduces to evaluating the correlation functions of the magnetic field. Specifically,
the expectation values of the Stokes parameters depend on two-point correlation functions of the magnetic field, whereas computing their variances generally requires four-point functions.
Under the Gaussian assumption for the magnetic field, however, the four-point functions factorize into products of two-point functions through Wick contractions, and hence the calculation ultimately reduces to evaluating only the two-point correlations.
As shown in Appendix \ref{app:A}, the expectation values and variances of the Stokes parameters are obtained as
\begin{align}
    \begin{cases}
     \mathrm{Exp}[1-I](\omega)&=\alpha~,\\
    \quad\mathrm{Exp}[Q](\omega)&=0~,\\
    \quad\mathrm{Exp}[U](\omega)&=0~,\\
     \quad\mathrm{Exp}[V](\omega)&=-\beta~,\\
  \end{cases}
\quad
  \begin{cases}
     \mathrm{Var}[1-I](\omega)&=\frac{1}{2}(\alpha^2+\beta^2+\gamma^2-\delta^2)=\frac{1}{2}(\alpha^2+\beta^2+\gamma^2)~,\\
     \quad\mathrm{Var}[Q](\omega)&=\frac{1}{2}(\alpha^2-\beta^2+\gamma^2+\delta^2)=\frac{1}{2}(\alpha^2-\beta^2+\gamma^2)~,\\
     \quad\mathrm{Var}[U](\omega)&=\frac{1}{2}(\alpha^2-\beta^2+\gamma^2+\delta^2)=\frac{1}{2}(\alpha^2-\beta^2+\gamma^2)~,\\
     \quad\mathrm{Var}[V](\omega)&=\frac{1}{2}(\alpha^2+\beta^2-\gamma^2+\delta^2)=\frac{1}{2}(\alpha^2+\beta^2-\gamma^2)~,\\
  \end{cases}
  \label{eq:Stokes_statistical}
\end{align}
where $\alpha$, $\beta$, $\gamma$, and $\delta$ are defined as the following integrals over wave vector space:

\begin{align}
  \begin{cases}
    \alpha(\omega)\equiv\displaystyle\Big(\frac{g_{a\gamma\gamma}}{2}\Big)^2 \int_k\mathcal{I}(k_z;\omega)\frac{1+\hat{k}^2_z}{2}P_{B}(k)~,\\[0.3cm]
    \beta(\omega)\equiv\displaystyle\Big(\frac{g_{a\gamma\gamma}}{2}\Big)^2\int_k\mathcal{I}(k_z;\omega)\hat{k}_z P_{aB}(k)~,\\[0.3cm]
    \gamma(\omega)\equiv\displaystyle\Big(\frac{g_{a\gamma\gamma}}{2}\Big)^2\int_k\mathcal{J}(k_z;\omega)\frac{1+\hat{k}_z^2}{2}P_{B}(k)~,\\[0.3cm]
    \delta(\omega)\equiv\displaystyle\Big(\frac{g_{a\gamma\gamma}}{2}\Big)^2\int_k\mathcal{J}(k_z;\omega)\hat{k}_z P_{aB}(k)=0~.
  \end{cases}
    \label{eq:abcd0}
\end{align}
In these expressions, the integral kernels $\mathcal{I}$ and $\mathcal{J}$ are given by
\begin{align}
  \mathcal{I}(k_z;\omega) &\equiv \int_{s,s'}e^{-i(\Pi(\omega)-k_z)(s-s')}=
  \bigg|\int_{s}e^{-i(\Pi(\omega)-k_z)s} \bigg|^2~,
  \label{eqkernelI}\\
  \mathcal{J}(k_z;\omega)
  &\equiv e^{i\Pi(\omega) d}\int_se^{-i(\Pi(\omega) - k_z)s}\int_{s'}e^{-i(\Pi(\omega) + k_z)s'}~.
  \label{eqkernelJ}
\end{align}
All these integrals depend on the photon angular frequency $\omega$ through $\Pi(\omega)= |m_a^2-m_{\mathrm{pl}}^2| / (2\omega) +\chi_\mathrm{CMB}\omega$.
Note that both kernels $\mathcal{I}$ and $\mathcal{J}$ are real functions.
Furthermore, the kernel $\mathcal{J}$ is an even function with respect to $k_z$, and this is why the integral $\delta$ vanishes identically.
Performing the integration over $s$, these kernels reduce to
\begin{align}
  \mathcal{I}(k_z;\omega)
  &= \left[ \frac{2d \sin (\theta_- (k_z;\omega)/2)}{\theta_- (k_z;\omega)} \right]^2~,\\
  \mathcal{J}(k_z;\omega)
  &= \left[ \frac{2d \sin (\theta_+ (k_z;\omega)/2)}{\theta_+ (k_z;\omega)} \right]\left[ \frac{2d \sin (\theta_- (k_z;\omega)/2)}{\theta_- (k_z;\omega)} \right]~,
\end{align}
where
\begin{align}
    \theta_\pm(k_z;\omega)\equiv(\Pi(\omega)\pm k_z)d~.
\end{align}
The four quantities $\alpha$, $\beta$, $\gamma$, and $\delta$ arise from the computation involving the two-point function of the magnetic field and constitute the basic building blocks of the result.
The origin of the four patterns can be understood based on whether the two magnetic-field components in the two-point function are identical or different, and whether the coordinate integrals yield the kernel $\mathcal{I}$ or $\mathcal{J}$. 
These two choices uniquely determine the pattern. 
More specifically,
\[
(\alpha,\beta,\gamma,\delta)
\;\longleftrightarrow\;
\begin{cases}
\text{magnetic-field components:}~
(\text{same},\,\text{different},\,\text{same},\,\text{different})~,\\[2mm]
\text{kernel type:}~
(\mathcal{I},\,\mathcal{I},\,\mathcal{J},\,\mathcal{J})
~,
\end{cases}
\]
respectively.

The power spectra $P_B(k)$ and $P_{aB}(k)$ depend only on the magnitude of the wave vector due to the assumption of statistical isotropy.
Thus, the four integrands in Eq.~\eqref{eq:abcd0} are rotationally symmetric in the $k_x$-$k_y$ plane and we can reduce the number of integration variables to obtain
\begin{align}
  \begin{cases}
    \alpha= \displaystyle\Big(\frac{g_{a\gamma\gamma}}{4\pi}\Big)^2 \int_0^\infty dk \, P_B(k) \, \mathcal{C}_{\alpha}(\Pi d,kd)~,\\[0.3cm]
    \beta= \displaystyle\Big(\frac{g_{a\gamma\gamma}}{4\pi}\Big)^2 \int_0^\infty dk \, P_{aB}(k) \, \mathcal{C}_{\beta}(\Pi d,kd)~,\\[0.3cm]
    \gamma= \displaystyle\Big(\frac{g_{a\gamma\gamma}}{4\pi}\Big)^2 \int_0^\infty dk \, P_B(k) \, \mathcal{C}_{\gamma}(\Pi d,kd)~,\\[0.3cm]
    \delta= \displaystyle\Big(\frac{g_{a\gamma\gamma}}{4\pi}\Big)^2 \int_0^\infty dk \, P_{aB}(k) \, \mathcal{C}_{\delta}(\Pi d,kd)~.
  \end{cases}
  \label{eq:abcd}
\end{align}
Here, $\mathcal{C}_{i}$'s $(i=\alpha, \beta, \gamma, \delta)$ denote convolution kernels that depend on $\Pi d$ and $kd$, and are given by
\begin{align}
  \begin{cases}
       \mathcal{C}_{\alpha}(\Pi d,kd)\equiv\displaystyle(kd)^2\int_{-1}^1 dz~\left[ \frac{2\sin (\theta_-/2)}{\theta_-} \right]^2 \frac{1+z^2}{2}~,\\[0.3cm]
       \mathcal{C}_{\beta}(\Pi d,kd)\equiv \displaystyle(kd)^2\int_{-1}^1dz~\left[ \frac{2 \sin (\theta_-/2)}{\theta_-} \right]^2~z~,\\[0.3cm]
       \mathcal{C}_{\gamma}(\Pi d,kd)\equiv\displaystyle(kd)^2\int_{-1}^1dz~\left[ \frac{2\sin (\theta_+/2)}{\theta_+} \right]\left[ \frac{2 \sin (\theta_-/2)}{\theta_-} \right]\frac{1+z^2}{2}~,\\[0.3cm]
       \mathcal{C}_{\delta}(\Pi d,kd)\equiv\displaystyle(kd)^2\int_{-1}^1dz~\left[ \frac{2\sin (\theta_+/2)}{\theta_+} \right]\left[ \frac{2 \sin (\theta_-/2)}{\theta_-} \right]~z=0~,
    \end{cases}
    \label{eq:C}
\end{align}
with 
\begin{align}
    \theta_{\pm} = \Pi (\omega) d \pm kd \cdot z~.
\end{align}
The integration over $z$ can be carried out analytically, but the resulting expressions are rather lengthy; for clarity, we compile them in Appendix~\ref{app:C}.
We emphasize that the same integral kernels also appear in graviton-photon conversion (see Ref.~\cite{Chiba:2025odu}), indicating that the kernel structure derived here is not specific to the axion-photon system but is shared by other systems in which light particles convert into one another via a background magnetic field.

The four integrals described thus far completely determine the behavior of the Stokes parameters.
As seen from Eq.~\eqref{eq:Stokes_statistical}, the expectation value of the conversion probability $1-I$ takes nontrivial values and is determined solely by the integral $\alpha$. 
Furthermore, in the case of helical magnetic fields $(P_{aB} \neq 0)$, the expectation value of the circular polarization $V$ also becomes nontrivial and is described solely by the integral $\beta$.
On the other hand, the expectation values of the linear polarization components $Q$ and $U$ vanish due to the statistical isotropy of the magnetic field.
The variances of the Stokes parameters are given by specific combinations of the squares of the four integrals, $\alpha$, $\beta$, $\gamma$, and $\delta$.
The variances of the intensity, linear polarization, and circular polarization involve negative contributions from the terms proportional to $\delta^2$, $\beta^2$, and $\gamma^2$, respectively.

\subsection{Consistency relations}
\label{subsec:Stokes}

In this subsection, we discuss consistency relations among the statistical quantities that follow from Eq.~\eqref{eq:Stokes_statistical} independently of the specific forms of the magnetic field power spectra. 
First, we immediately obtain
\begin{align}
    \mathrm{Var}[Q]=\mathrm{Var}[U]~.
    \label{con1}
\end{align}
Note that, while this equality reflects the statistical rotational symmetry in the $x$-$y$ plane and is rather trivial, the appearance of nonzero variances itself is a non-trivial consequence in the axion-photon system: in the case of the graviton-photon system (see Ref.~\cite{Chiba:2025odu}), the corresponding quantities vanish identically.

We obtain another relation by eliminating $\gamma$ through the combination $\mathrm{Var}[1-I]+\mathrm{Var}[V]$ and expressing $\alpha$ and $\beta$ in terms of the expectation values of $I$ and $V$,
\begin{align}
  \mathrm{Var}[1-I]+\mathrm{Var}[V]=\mathrm{Exp}[1-I]^2+\mathrm{Exp}[V]^2~.
  \label{con2}
\end{align}

We obtain yet another relation as follows.
We eliminate $\alpha$ and $\beta$ and express the integral $\gamma$ in terms of the variances of $I$ and $V$ as $\gamma^2=\mathrm{Var}[1-I]-\mathrm{Var}[V]$.
Alternatively, $\gamma$ can also be written by combining the variances of $Q$ and $U$ with the expectation values of $I$ and $V$ as $\gamma^2=\mathrm{Var}[Q]+\mathrm{Var}[U]-\mathrm{Exp}[1-I]^2+\mathrm{Exp}[V]^2$.
Equating these two expressions for $\gamma^2$ leads to 
\begin{align}
    \mathrm{Var}[1-I]+\mathrm{Exp}[1-I]^2=\mathrm{Var}[Q]+\mathrm{Var}[U]+\mathrm{Var}[V]+\mathrm{Exp}[V]^2~.
     \label{con3}
\end{align}
This relation can also be obtained directly by taking an ensemble average of Eq.~\eqref{con0}, which can be rewritten as
\begin{align}
    \left<(1-I)^2\right>=\langle Q^2+U^2+V^2\rangle~.
\end{align}

The set of the above three relations, Eqs.~\eqref{con1}, \eqref{con2}, and \eqref{con3}, can be equivalently rewritten as 
\begin{align}
    \begin{cases}
        \mathrm{Var}[Q]=\mathrm{Var}[U]~,\\
        \mathrm{Var}[1-I]=\mathrm{Var}[Q]+\mathrm{Exp}[V]^2~,\\
        \mathrm{Exp}[1-I]^2=\mathrm{Var}[Q]+\mathrm{Var}[V]~.\\
    \end{cases}
    \label{eq:consistency}
\end{align}
We call the set of three equations as consistency relations in the axion-photon system.
If the magnetic field follows Gaussian statistics and the Born approximation is valid, statistical analysis of the Stokes parameters for initially unpolarized photons propagating through the magnetic field should reproduce these relations. This can be achieved by repeatedly measuring the photons that survive photon-to-axion conversion.
Under the above assumptions, these relations hold independently of the specific forms of the magnetic field power spectra.

\section{Numerical analysis}
\label{sec:numerical}

In this section, we consider typical forms of the magnetic field power spectra $P_B (k)$ and $P_{aB} (k)$, and numerically evaluate the resulting Stokes parameters of photons affected by conversion into axions.
Specifically, we show how the Stokes parameters discussed in Section~\ref{sec:random} behave as functions of the photon angular frequency $\omega$.
Below, we examine the behavior of the following quantities separately: the integral kernels $\mathcal{C}_\alpha$, $\mathcal{C}_\beta$, and $\mathcal{C}_\gamma$; the integrals $\alpha$, $\beta$, and $\gamma$, obtained by convolving the kernels with the power spectra; and the expectation values and variances of the Stokes parameters.

\subsection{Kernels $\mathcal{C}_\alpha$, $\mathcal{C}_\beta$, and $\mathcal{C}_\gamma$}
\label{subsec:C}
\begin{figure}
\centering
\includegraphics[width=0.6\linewidth]{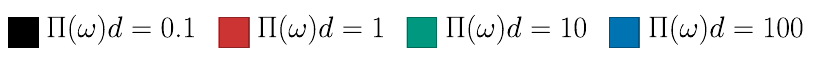}
\\[0.5cm]
\includegraphics[width=0.56\linewidth]{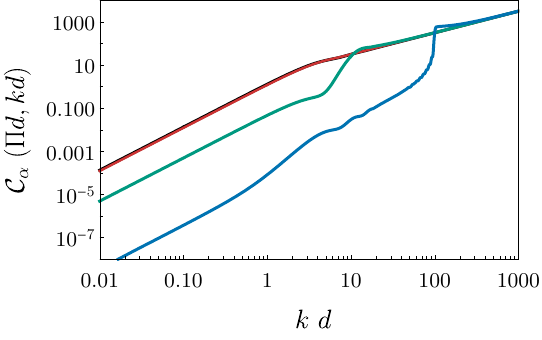}
\\[0.5cm]
\includegraphics[width=0.56\linewidth]{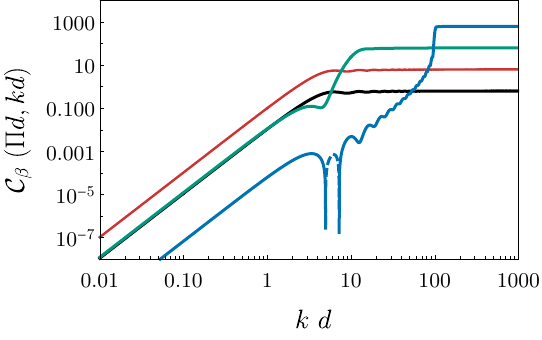}
\\[0.5cm]
\includegraphics[width=0.56\linewidth]{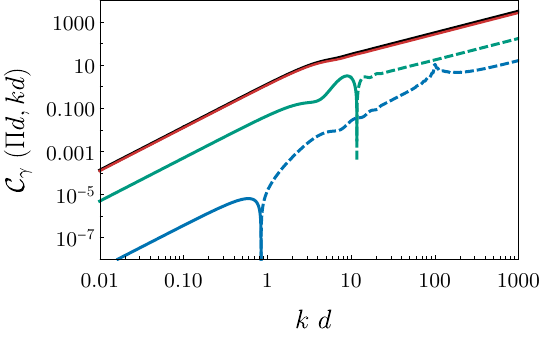}
\caption{
Convolution kernels $\mathcal{C}_\alpha$, $\mathcal{C}_\beta$, and $\mathcal{C}_\gamma$ with various values of $\Pi(\omega) d$ are shown as functions of $kd$. The black, red, green, and blue curves correspond to $\Pi(\omega) d = 0.1$, $1$, $10$, and $100$, respectively.
The dashed curves indicate that the kernel takes negative values.}
\label{fig:C}
\end{figure}

The dimensionless integral kernels $\mathcal{C}_\alpha$, $\mathcal{C}_\beta$, and $\mathcal{C}_\gamma$ are given in Eq.~\eqref{eq:C}.
In Fig.~\ref{fig:C}, we show the behavior of these kernels as functions of $k d$ for several representative values of $\Pi (\omega) d$, while $C_\delta$ is omitted since it vanishes identically.
Note that these kernels are independent of the specific forms of the magnetic field power spectra, and thus provide a universal basis for evaluating the response of the axion-photon system across different magnetic field ensembles.

Let us summarize several key properties of the kernels observed from Fig.~\ref{fig:C}, by considering the two cases $\Pi d \lesssim 1$ and $\Pi d \gg 1$ separately:
\begin{itemize}
\item 
$\Pi d \lesssim 1$: In this case, the kernels $\mathcal{C}_\alpha$ and $\mathcal{C}_\gamma$ converge to the same functional form and become independent of $\Pi d$.
In contrast, the magnitude of the kernel $C_\beta$ increases linearly with $\Pi d$.
For all kernels, a transition in the power-law scaling occurs around $kd = \mathcal{O}(1)$. 
Their behaviors are $\mathcal{C}_\alpha (\approx \mathcal{C}_{\gamma}) \propto (kd)^2$ and $\mathcal{C}_\beta \propto (kd)^3$ in the infrared region ($kd \lesssim 1$), while $\mathcal{C}_\alpha (\approx \mathcal{C}_{\gamma}) \propto kd$ and $\mathcal{C}_\beta = \text{const}.$ in the ultraviolet region ($kd \gtrsim 1$).
\item
$\Pi d \gg 1$: In this case, each kernel changes its behavior around $kd = \mathcal{O}(1)$ and $kd \approx \Pi d$.
For $\mathcal{C}_\alpha$ and $\mathcal{C}_\beta$, their behavior is similar to the case of $\Pi d \lesssim 1$ in the region $kd \gtrsim \Pi d$, whereas in the region $kd \ll \Pi d$ these kernels are strongly suppressed.
On the other hand, $\mathcal{C}_\gamma$ is suppressed over the entire range of $kd$ compared to the case of $\Pi d \lesssim 1$. 
This suppression becomes stronger as $\Pi d$ increases.
In the infrared region ($kd \lesssim 1$), the kernels behave as $\mathcal{C}_\alpha (\approx \mathcal{C}_{\gamma}) \propto (kd)^2$ and $\mathcal{C}_\beta \propto (kd)^3$. 
In the region $1 \lesssim kd \lesssim \Pi d$, nontrivial modulations are observed. 
In particular, in the ultraviolet region ($kd \gtrsim 1$), the kernel $\mathcal{C}_\gamma$ exhibits oscillatory changes in the sign depending on the value of $\Pi d$.
\end{itemize}

\subsection{Integrals $\alpha$, $\beta$, and $\gamma$}
\label{subsec:abc}

Next, we numerically evaluate $\alpha$, $\beta$, and $\gamma$ in Eq.~\eqref{eq:abcd}, obtained by convolving the kernels $\mathcal{C}_\alpha$, $\mathcal{C}_\beta$, and $\mathcal{C}_\gamma$ with the magnetic field power spectra $P_B(k)$ and $P_{aB}(k)$.
For illustration, we adopt the following forms of the power spectra from Ref.~\cite{Durrer:2013pga}:
\begin{align}
    P_B(k) &= P_{B*} \times
    \begin{cases}
    \displaystyle\left(\frac{k}{k_*}\right)^{n_l}~, \quad n_l = 2~, & (k \leq k_*) \\
    \displaystyle\left(\frac{k}{k_*}\right)^{n_h}~, \quad n_h = -\frac{11}{3}~, & (k > k_*) 
    \end{cases}
    \label{eq:PB}
    \\
    P_{aB}(k) &= P_{aB*} \times
    \begin{cases}
    \displaystyle\left(\frac{k}{k_*}\right)^{n_{al}}~, \quad n_{al} = 3~, & (k \leq k_*) \\
    \displaystyle\left(\frac{k}{k_*}\right)^{n_{ah}}~, \quad n_{ah} = - \frac{11}{3}~, & (k > k_*) 
    \end{cases}
    \label{eq:PaB}
\end{align}
where $P_{B*}$ and $P_{aB*}$ denote the peak amplitudes of the symmetric and antisymmetric components of the power spectra, respectively, and $k_*$ is the characteristic wavenumber of the magnetic field.
The indices $n_l$ and $n_h$ ($n_{al}$ and $n_{ah}$) specify the scale dependence of $P_B$ ($P_{aB}$) in the infrared and ultraviolet regimes, respectively.

In the following, we focus on the maximally helical magnetic field with $P_{B*} = P_{aB*}$ as a concrete example.
(For the non-helical case $P_{aB*} = 0$, see Appendix~\ref{app:non-helical}.)
In Fig.~\ref{fig:abc}, we show the profiles of the integrals $\alpha$, $\beta$, and $\gamma$ (from top to bottom), plotted in both log-linear (left column) and log-log (right column) formats as functions of $\Pi (\omega) d$ for several values of $k_* d$. 
On the vertical axes of each panel, we show each integral normalized by two different scales: on the right axis, we use the typical amplitude $g^2_{a\gamma\gamma}P_{B*}k^2_*d/16\pi^2$, while on the left axis we use a combination of physical quantities normalized by their typical values.
For the latter, $B_{*}$ is the characteristic magnetic field strength at the peak scale $k_*$ \cite{Durrer:2013pga},
\begin{align}
    B_* \equiv \sqrt{\frac{k_*^3 P_{B*}}{\pi^2}}~.
\end{align}
In addition, we define the characteristic correlation length of the magnetic field as $\lambda_* \equiv 2\pi / k_*$. 
Regarding the horizontal axes, the dimensionless variable $\Pi(\omega)d$ and the photon angular frequency $\omega$ are shown on the upper and lower axes in each panel, respectively.
Note that $\Pi(\omega)d$ is a function of $\omega$ given by 
\begin{align}
    \Pi(\omega)d=\bigg(\frac{|m_a^2-m_{\mathrm{pl}}^2|}{2\omega}+\chi_\mathrm{CMB}\omega\bigg)d~,
\end{align}
which is parabolic in $\omega$ when plotted on a logarithmic scale, unless $m_a = m_{\mathrm{pl}}$.
Thus, $\Pi(\omega) d$ takes its minimum value, $\Pi(\omega_{\mathrm{eq}})d=\sqrt{2\chi_\mathrm{CMB}|m_a^2-m_{\mathrm{pl}}^2|}~d$, at
\begin{align}
\omega_\mathrm{eq}\equiv\sqrt\frac{|m_a^2-m_{\mathrm{pl}}^2|}{2\chi_{\mathrm{CMB}}}=1\times10^{10}~\mathrm{eV}~\bigg(\frac{|m_a^2-m_{\mathrm{pl}}^2|}{10^{-22}\mathrm{eV^2}}\bigg)^{\frac{1}{2}}~\bigg(\frac{2\chi_\mathrm{CMB}}{10^{-42}}\bigg)^{-\frac{1}{2}}~.
\end{align}
Accordingly, when the integrals $\alpha$, $\beta$, and $\gamma$ are plotted as functions of $\omega$, the curves fold at the point $\omega=\omega_\mathrm{eq}$. In Fig.~\ref{fig:abc}, we use a representative set of parameter values, $|m_a^2-m_{\mathrm{pl}}^2| = 10^{-22}\, \mathrm{eV}^2$, $2\chi_\text{CMB} = 10^{-42}$, and $d = 100\,\mathrm{Mpc}$, for illustration. The point $\omega = \omega_{\mathrm{eq}}$ is indicated by the orange dashed line in each panel.
The angular frequency $\omega$ on the lower horizontal axis is expressed in units of $\omega_{\mathrm{eq}}$, and the axis is folded at $\omega=\omega_{\mathrm{eq}}$.

From Fig.~\ref{fig:abc}, we observe several key properties of the integrals $\alpha$, $\beta$, and $\gamma$:
\begin{itemize}
\item 
In the case of $k_* d \lesssim 1$, the overall amplitude of the integrals $\alpha$, $\beta$, and $\gamma$ monotonically decreases as $k_* d$ decreases.
The behavior of these integrals qualitatively changes around $\Pi d = \mathcal{O} (1)$.
In the region $\Pi d \gg 1$, $\alpha$ and $\beta$ approach the same functional form, while $\gamma$ is smaller.
The frequency dependence is estimated as $\alpha = \beta \propto (\Pi d)^{- 5/3}$ while $\gamma \propto (\Pi d)^{- 2}$ in this region.
In the region $\Pi d \ll 1$, on the other hand, $\alpha$ and $\gamma$ approach the same form, while $\beta$ is suppressed.
The frequency dependence is $\alpha = \gamma = \text{const}.$, while $\beta \propto \Pi d$ in this region.

\item
In the case of $k_* d \gg 1$, $\Pi d \approx k_* d$ provides a characteristic scale for all three integrals, $\alpha$, $\beta$, and $\gamma$. Moreover, $\Pi d = \mathcal{O}(1)$ serves as an additional characteristic scale specific to $\gamma$.
In the first region ($\Pi d \gg k_* d$), $\alpha$ and $\beta$ approach the same form $\alpha = \beta \propto (\Pi d)^{- 5/3}$, while $\gamma$ approaches $\gamma \propto (\Pi d)^{- 2}$.
In the second region ($k_* d \gg \Pi d \gg 1$), $\alpha$ is dominant over the other two integrals.
In the third region ($1 \gg \Pi d$), $\alpha$ and $\gamma$ approach the same functional form while $\beta$ is suppressed.
\end{itemize}

\begin{figure}[htb]
\centering
\includegraphics[width=0.4\linewidth]{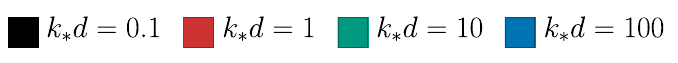}
\\
\includegraphics[width=0.43\linewidth]{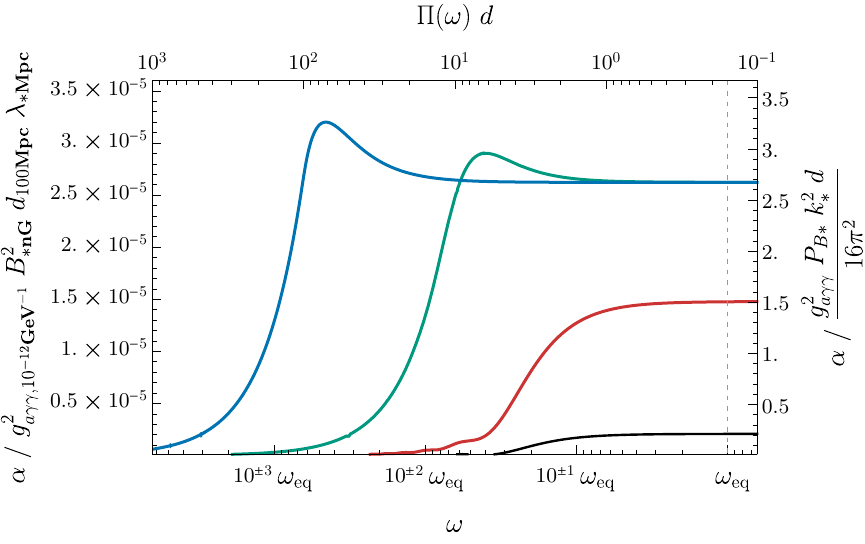}
\hskip 0.4cm
\includegraphics[width=0.4\linewidth]{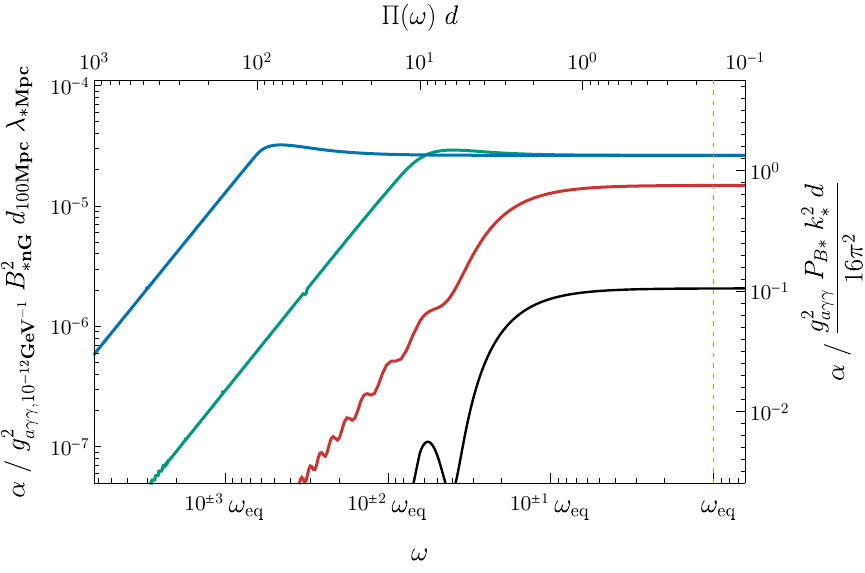}
\\
\includegraphics[width=0.43\linewidth]{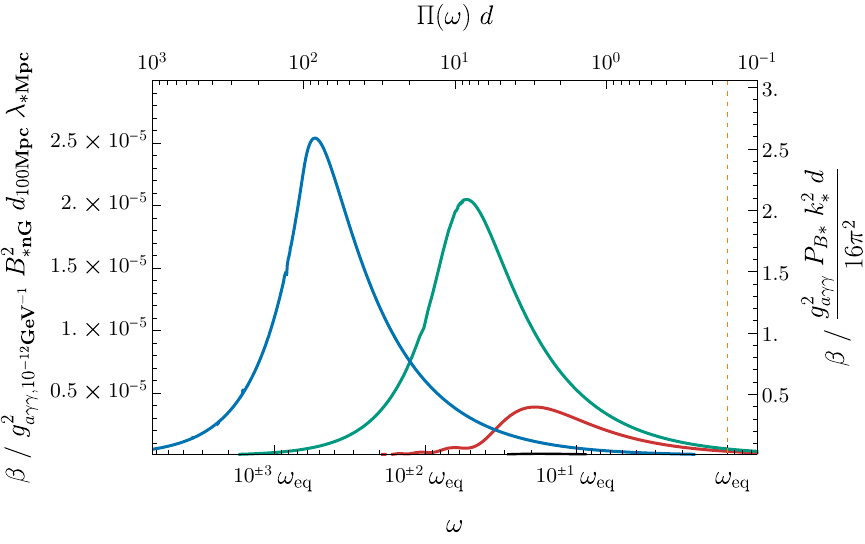}
\hskip 0.4cm
\includegraphics[width=0.4\linewidth]{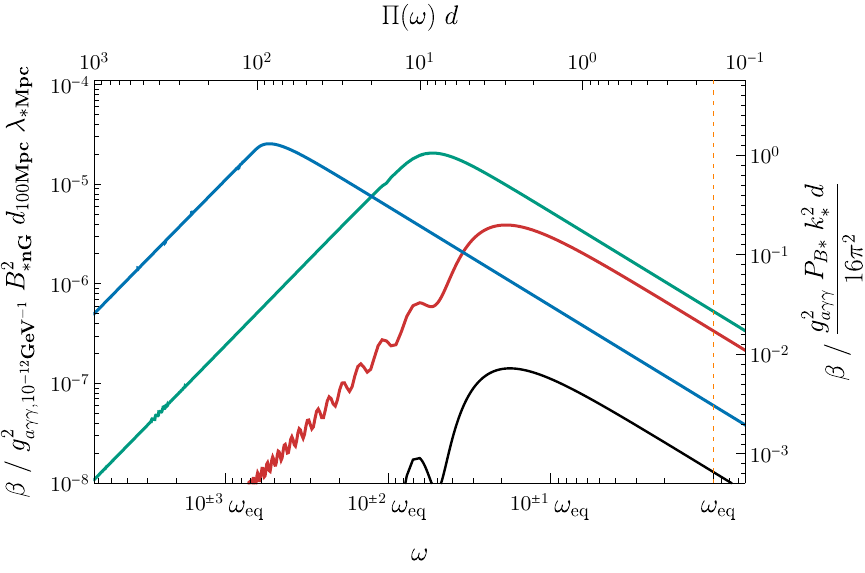}
\\
\includegraphics[width=0.43\linewidth]{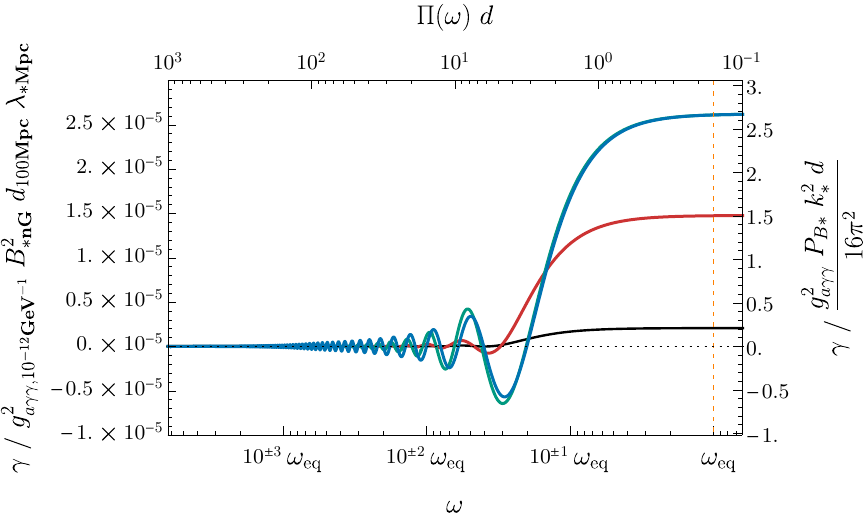}
\hskip 0.4cm
\includegraphics[width=0.4\linewidth]{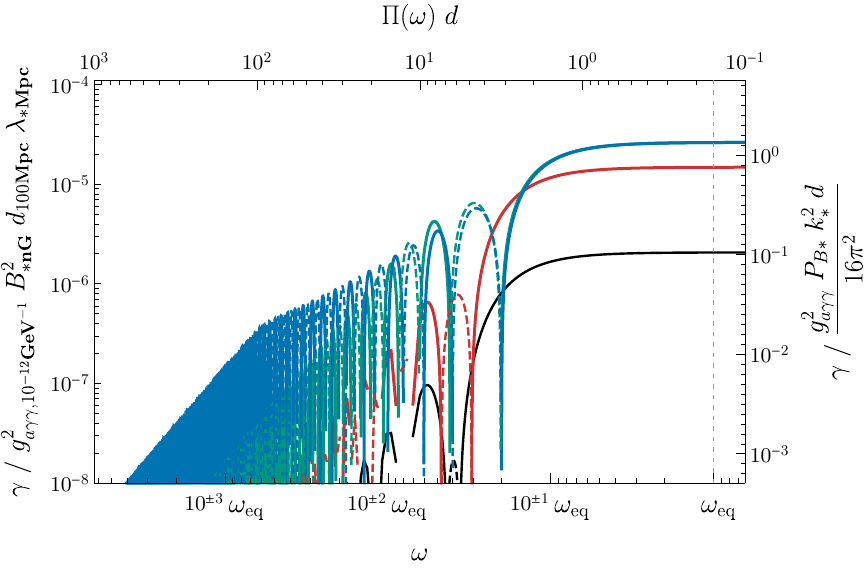}
\caption{
Integrals $\alpha$ (top), $\beta$ (middle), and $\gamma$ (bottom) with the magnetic field power spectra \eqref{eq:PB} and \eqref{eq:PaB} are plotted.
The black, red, green, and blue curves correspond to the case of $k_* d = 0.1$, $1$, $10$, and $100$, respectively.
In each panel, the upper horizontal axis is the dimensionless variable $\Pi(\omega) d$, and the lower horizontal axis is $\omega$ in units of $\omega_\mathrm{eq}\equiv (|m_a^2 - m_\text{pl}^2| / 2\chi_{\mathrm{CMB}})^\frac12$, where $(|m_a^2 - m_\text{pl}^2|)^\frac12=10^{-11}\mathrm{eV}$, $2\chi_{\mathrm{CMB}}=10^{-42}$, and $d = 100\, \mathrm{Mpc}$ are used.
The orange dashed line represents $\omega = \omega_\mathrm{eq}$.
The right vertical axis is the value of $\alpha$, $\beta$, or $\gamma$ normalized by $g_{a\gamma\gamma}^2 P_{B*} k_*^2 d / 16\pi^2$.
For $\beta$, we set $P_{aB*} = P_{B*}$ assuming the maximally helical case.
The left vertical axis is the value of $\alpha$, $\beta$, or $\gamma$ normalized by $g^2_{a\gamma\gamma,10^{-12}\mathrm{GeV}^{-1}}\equiv(g_{a\gamma\gamma}/10^{-12}\mathrm{GeV}^{-1})^2 $, $B_{*\text{nG}}^2 \equiv (B_* / \text{nG})^2$, $d_{100\text{Mpc}} \equiv d / 100\,\text{Mpc}$, and $\lambda_{*\text{Mpc}} \equiv \lambda_{*} / \text{Mpc}$.
In each row, the left and right panels show the same quantity: the vertical axis is linear in the left, but logarithmic in the right.
In the bottom-right panel, the dashed curves indicate that $\gamma$ takes negative values.
}
\label{fig:abc}
\end{figure}

\subsection{Expectation values and variances of Stokes parameters}
\label{subsec:NumCalcStokes}

\begin{figure}
\centering
\includegraphics[width=0.4\linewidth]{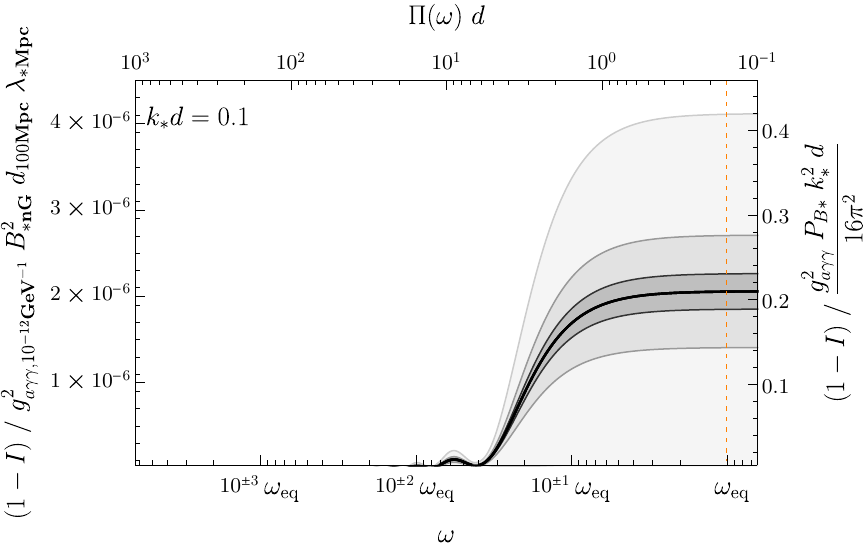}
\hskip 1cm
\includegraphics[width=0.4\linewidth]{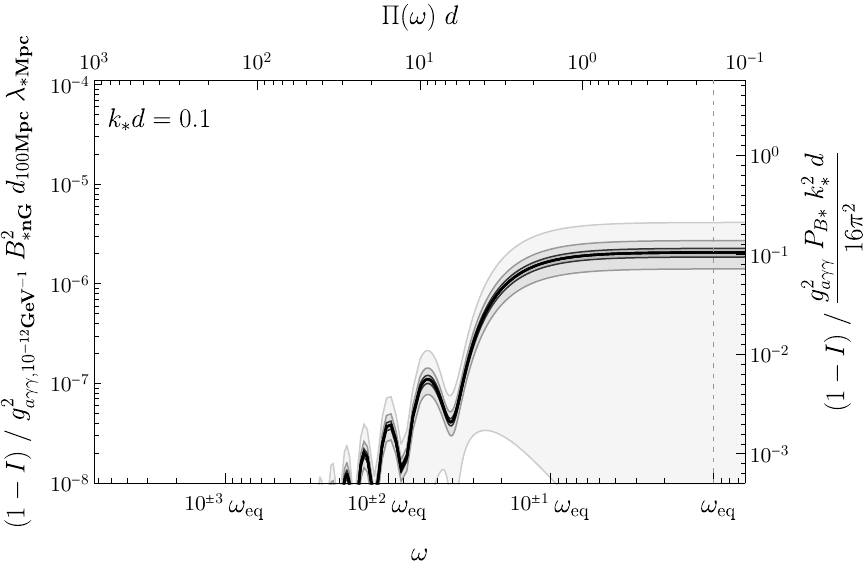}
\\
\includegraphics[width=0.4\linewidth]{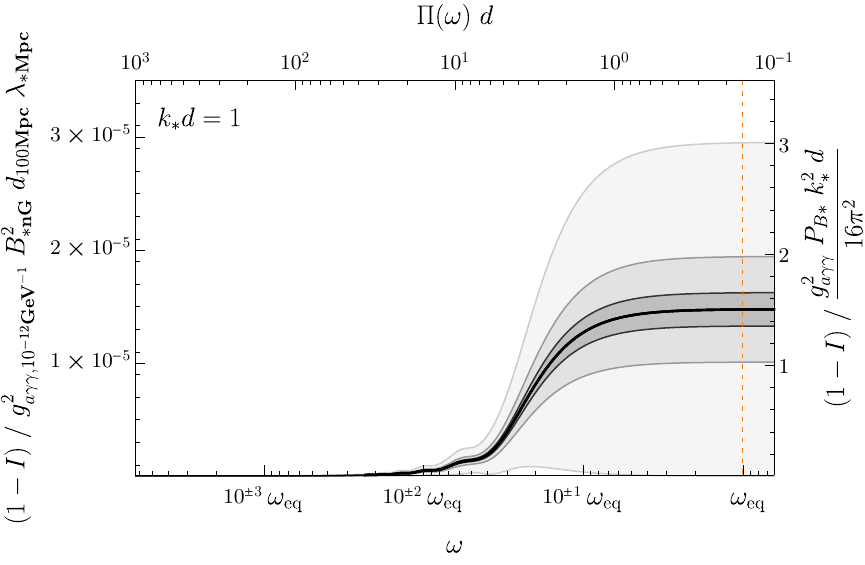}
\hskip 1cm
\includegraphics[width=0.4\linewidth]{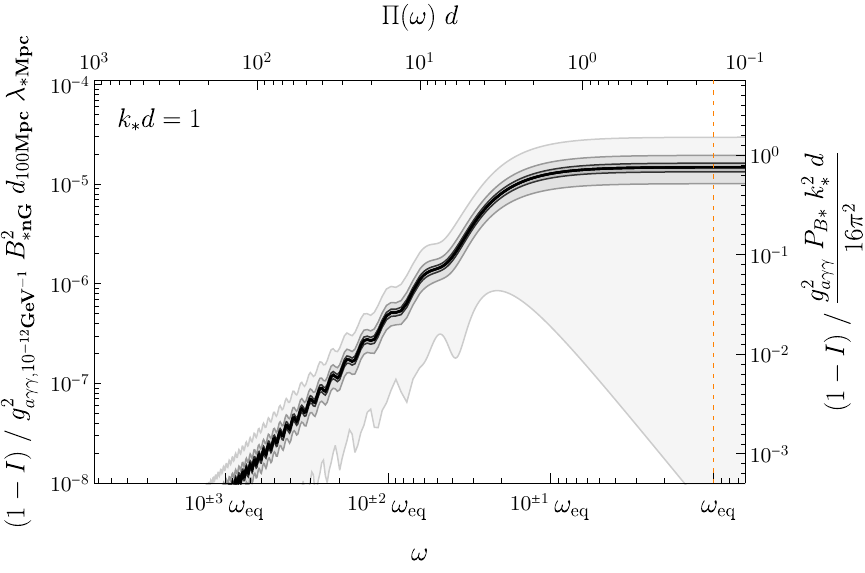}
\\
\includegraphics[width=0.4\linewidth]{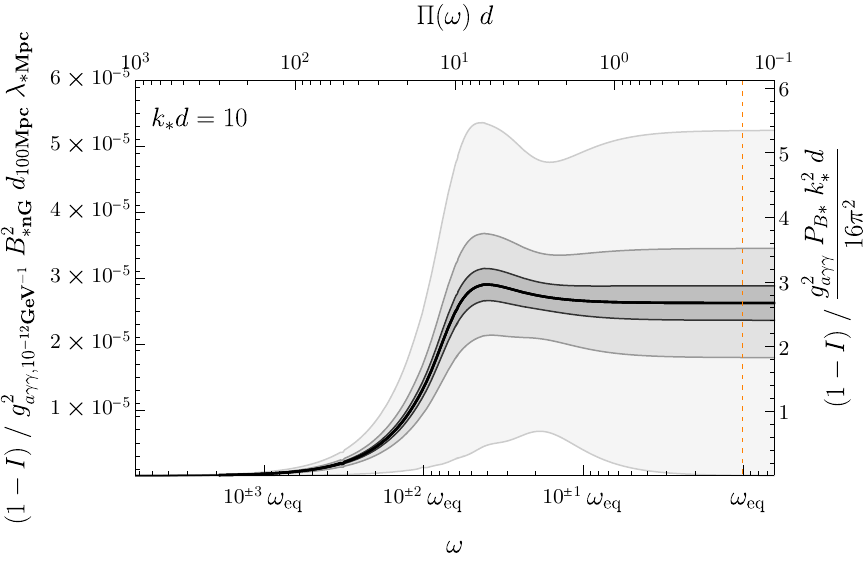}
\hskip 1cm
\includegraphics[width=0.4\linewidth]{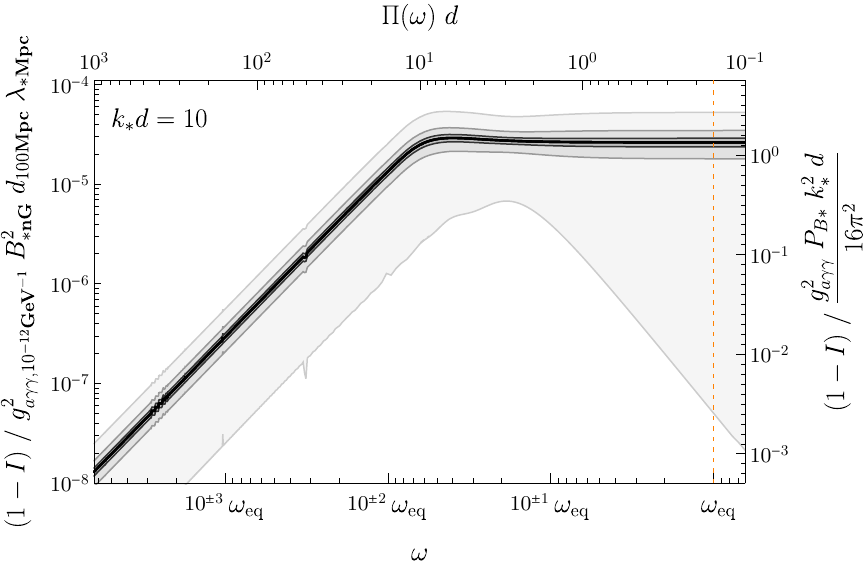}
\\
\includegraphics[width=0.4\linewidth]{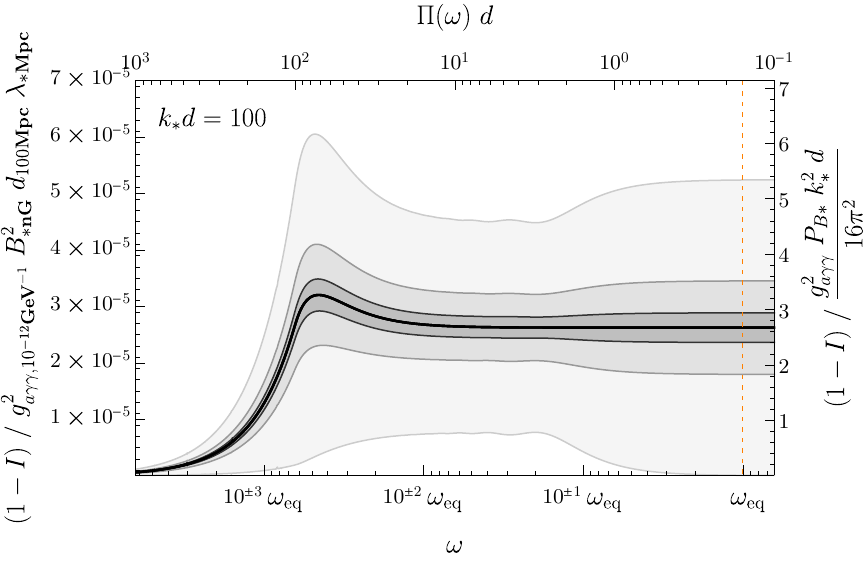}
\hskip 1cm
\includegraphics[width=0.4\linewidth]{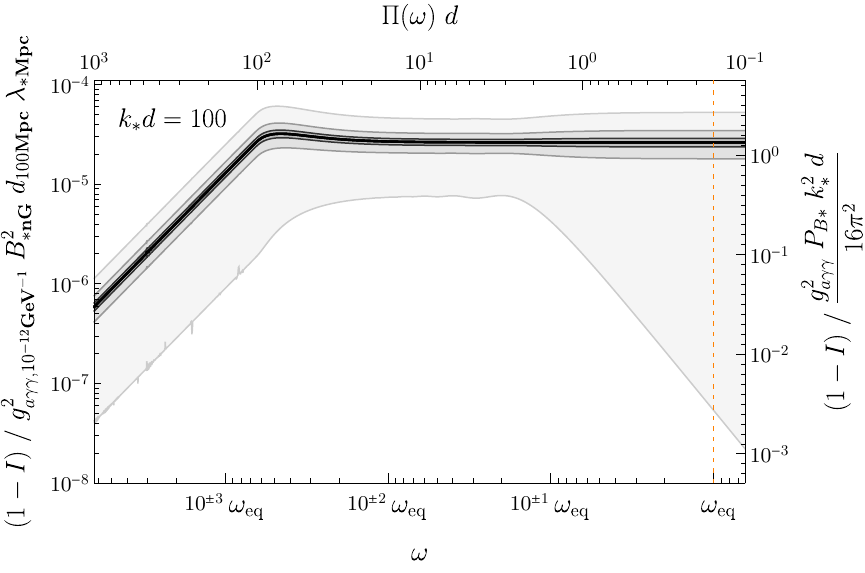}
\caption{
Conversion probability $P_{\gamma \to a} \equiv 1-I$ with the maximally helical ($P_{aB*} = P_{B*}$) magnetic field power spectra \eqref{eq:PB} and \eqref{eq:PaB} is plotted for $k_* d = 0.1$, $1$, $10$, and $100$ from top to bottom.
In each panel, the black line represents the expectation value $\text{Exp}[1-I]$, and the shaded bands represent the standard deviations $\sqrt{\text{Var}[1-I]}$, $\sqrt{\text{Var}[1-I]/10}$, and $\sqrt{\text{Var}[1-I]/100}$.
The upper horizontal axis is the dimensionless variable $\Pi (\omega) d$, and the lower horizontal axis is $\omega$ in units of $\omega_\mathrm{eq}\equiv (|m_a^2 - m_\text{pl}^2| / 2\chi_{\mathrm{CMB}})^{\frac12}$ with $(|m_a^2 - m_\text{pl}^2|)^{\frac12}=10^{-11}\mathrm{eV}$, $2\chi_{\mathrm{CMB}}=10^{-42}$, and $d = 100\, \text{Mpc}$. 
The orange dashed line indicates $\omega = \omega_\mathrm{eq}$.
The right vertical axis is $1-I$ normalized by $g_{a\gamma\gamma}^2 P_{B*} k_*^2 d / 16\pi^2 $. 
The left vertical axis is $1-I$ normalized by $g^2_{a\gamma\gamma,10^{-12}\mathrm{GeV}^{-1}}\equiv(g_{a\gamma\gamma}/10^{-12}\mathrm{GeV}^{-1})^2 $, $B_{*\text{nG}}^2 \equiv (B_* / \text{nG})^2$, $d_{100\text{Mpc}} \equiv d / 100\,\text{Mpc}$, and $\lambda_{*\text{Mpc}} \equiv \lambda_{*} / \text{Mpc}$.
In each row, the left and right panels show the same quantity: the vertical axis is linear in the left, but logarithmic in the right.}
\label{fig:I}
\end{figure}

\begin{figure}
\centering
\includegraphics[width=0.4\linewidth]{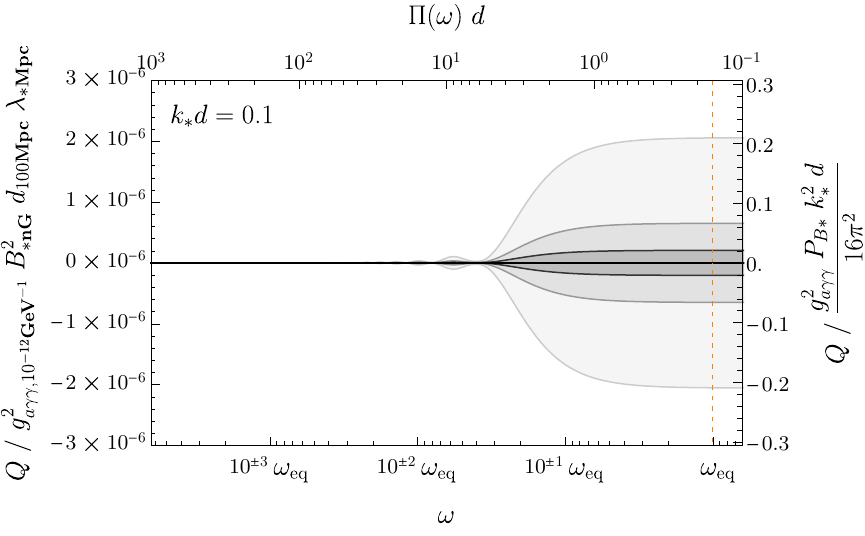}
\hskip 1cm
\includegraphics[width=0.4\linewidth]{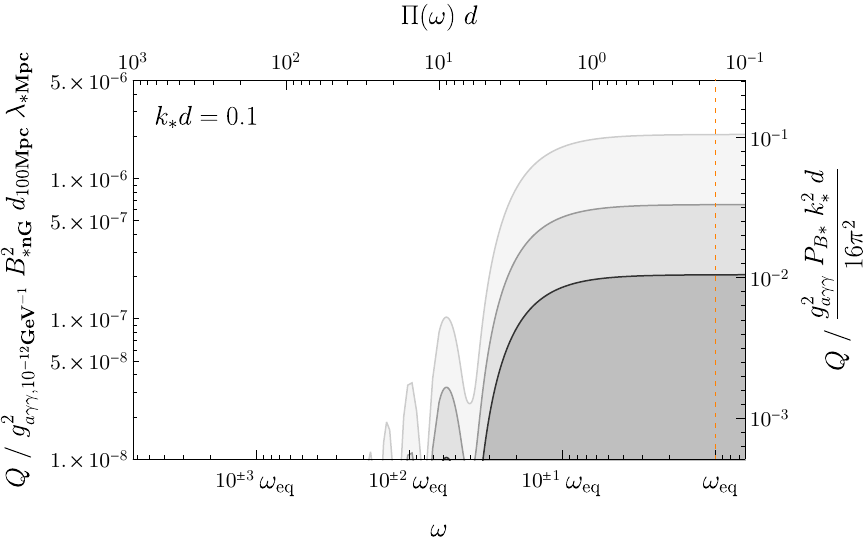}
\\
\includegraphics[width=0.4\linewidth]{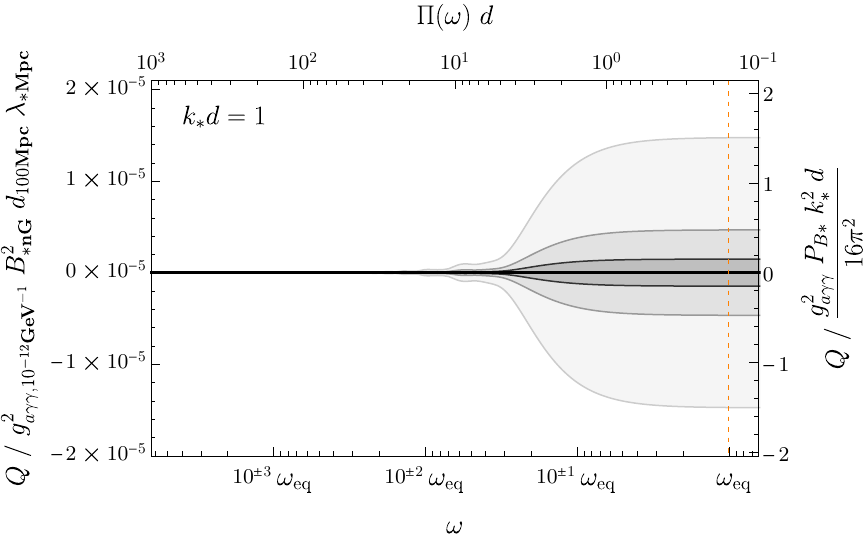}
\hskip 1cm
\includegraphics[width=0.4\linewidth]{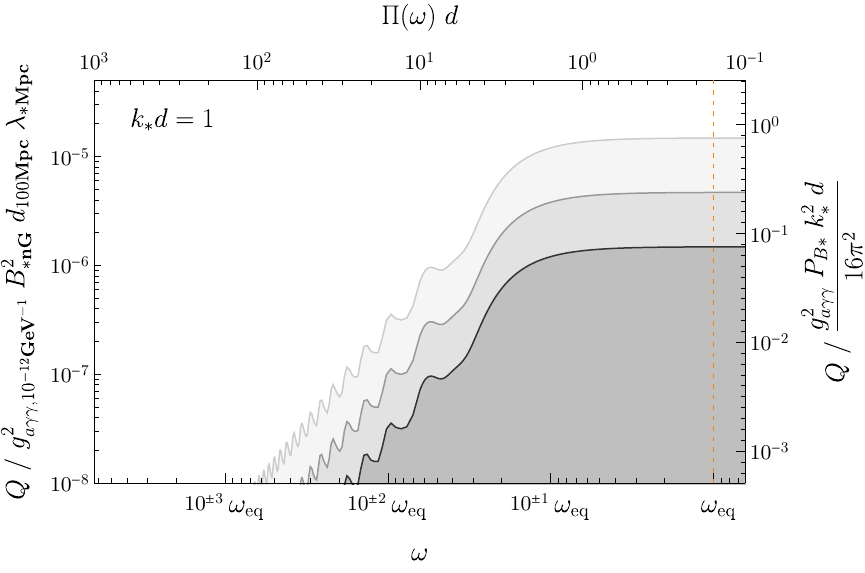}
\\
\includegraphics[width=0.4\linewidth]{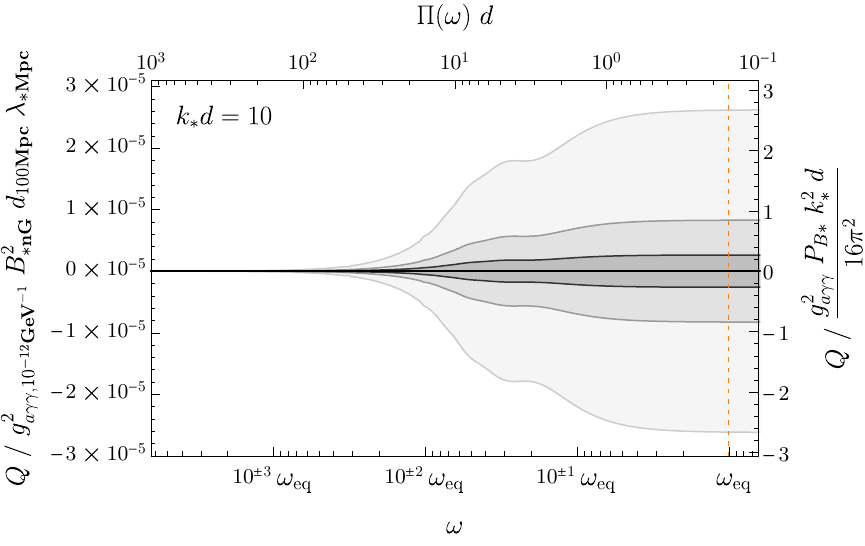}
\hskip 1cm
\includegraphics[width=0.4\linewidth]{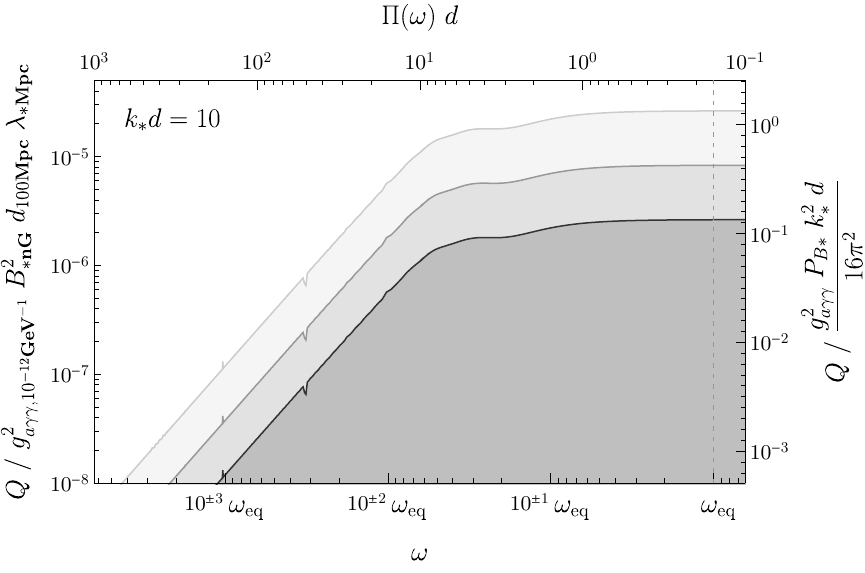}
\\
\includegraphics[width=0.4\linewidth]{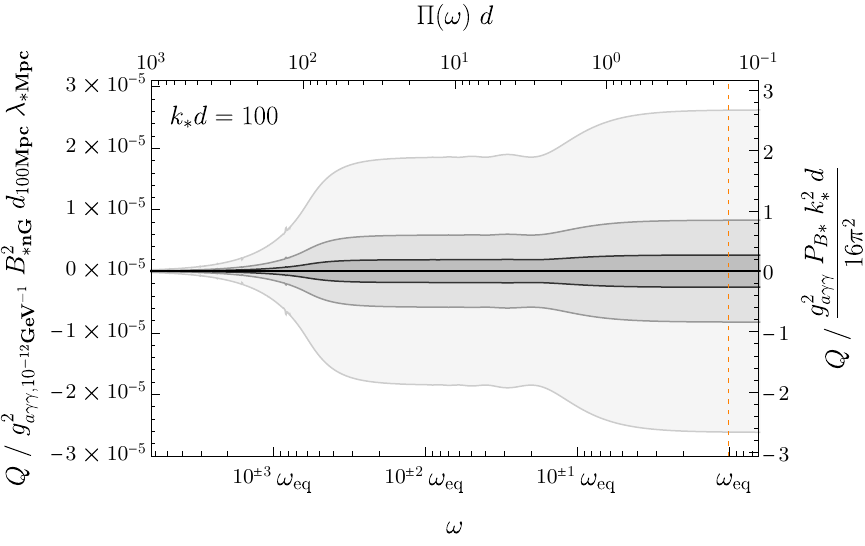}
\hskip 1cm
\includegraphics[width=0.4\linewidth]{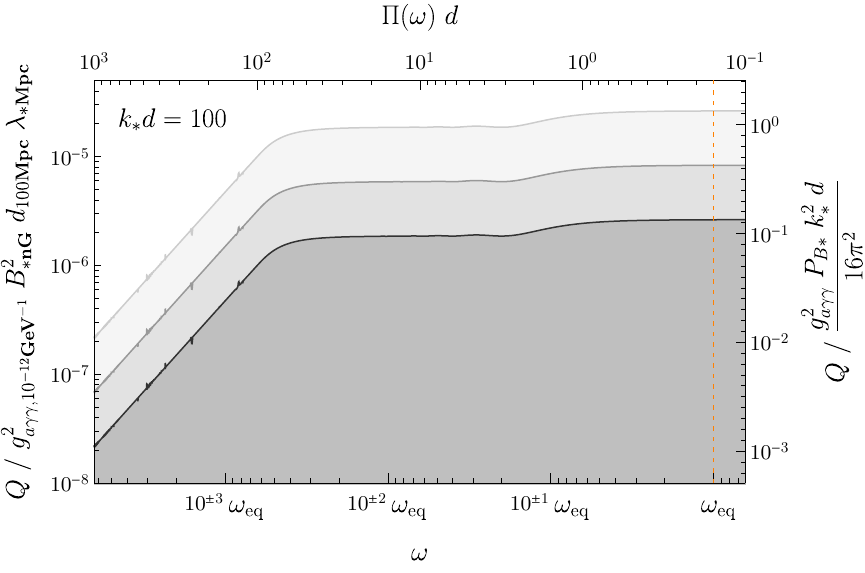}
\caption{
Stokes parameter $Q$ (linear polarization) of photons with the maximally helical ($P_{aB*} = P_{B*}$) magnetic field power spectra \eqref{eq:PB} and \eqref{eq:PaB} is plotted for $k_* d = 0.1$, $1$, $10$, and $100$ from top to bottom.
In each panel, the black line represents the expectation value $\text{Exp}[Q]=0$, and the shaded bands represent the standard deviations $\sqrt{\text{Var}[Q]}$, $\sqrt{\text{Var}[Q]/10}$, and $\sqrt{\text{Var}[Q]/100}$.
The upper horizontal axis is the dimensionless variable $\Pi (\omega) d$, and the lower horizontal axis is $\omega$ in units of $\omega_\mathrm{eq}\equiv (|m_a^2 - m_\text{pl}^2| / 2\chi_{\mathrm{CMB}})^{\frac12}$ with $(|m_a^2 - m_\text{pl}^2|)^{\frac12}=10^{-11}\mathrm{eV}$, $2\chi_{\mathrm{CMB}}=10^{-42}$, and $d = 100\, \text{Mpc}$. 
The orange dashed line indicates $\omega = \omega_\mathrm{eq}$.
The right vertical axis is $Q$ normalized by $g_{a\gamma\gamma}^2 P_{B*} k_*^2 d / 16\pi^2$. 
The left vertical axis is $Q$ normalized by $g^2_{a\gamma\gamma,10^{-12}\mathrm{GeV}^{-1}}\equiv(g_{a\gamma\gamma}/10^{-12}\mathrm{GeV}^{-1})^2 $, $B_{*\text{nG}}^2 \equiv (B_* / \text{nG})^2$, $d_{100\text{Mpc}} \equiv d / 100\,\text{Mpc}$, and $\lambda_{*\text{Mpc}} \equiv \lambda_{*} / \text{Mpc}$.
In each row, the left and right panels show the same quantity: the vertical axis is linear in the left, but logarithmic in the right.
}
\label{fig:Q}
\end{figure}

\begin{figure}
\centering
\includegraphics[width=0.4\linewidth]{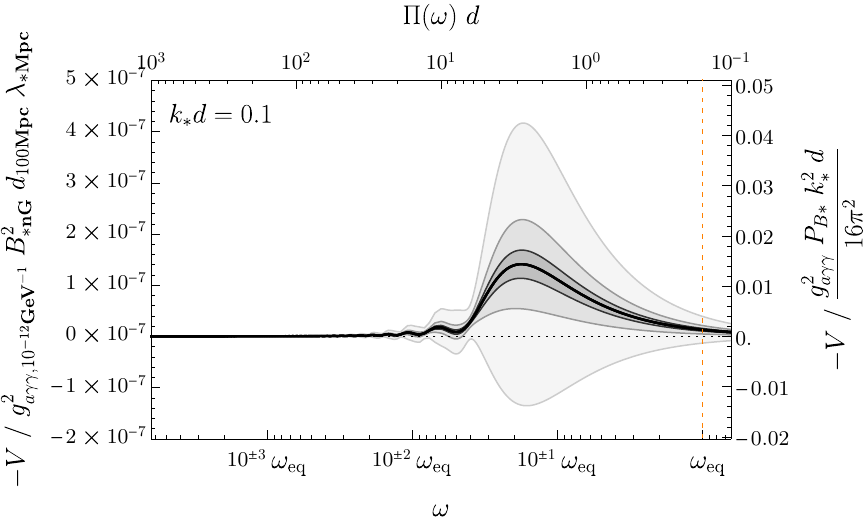}
\hskip 1cm
\includegraphics[width=0.4\linewidth]{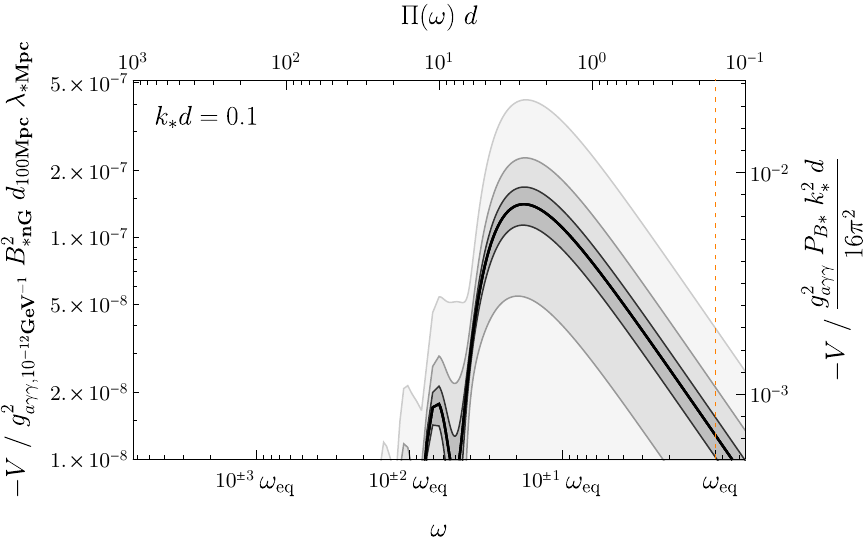}
\\
\includegraphics[width=0.4\linewidth]{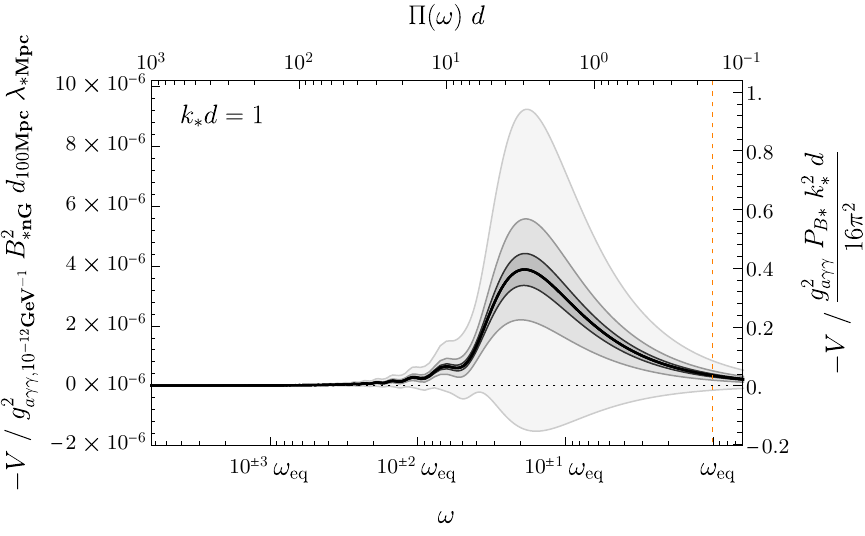}
\hskip 1cm
\includegraphics[width=0.4\linewidth]{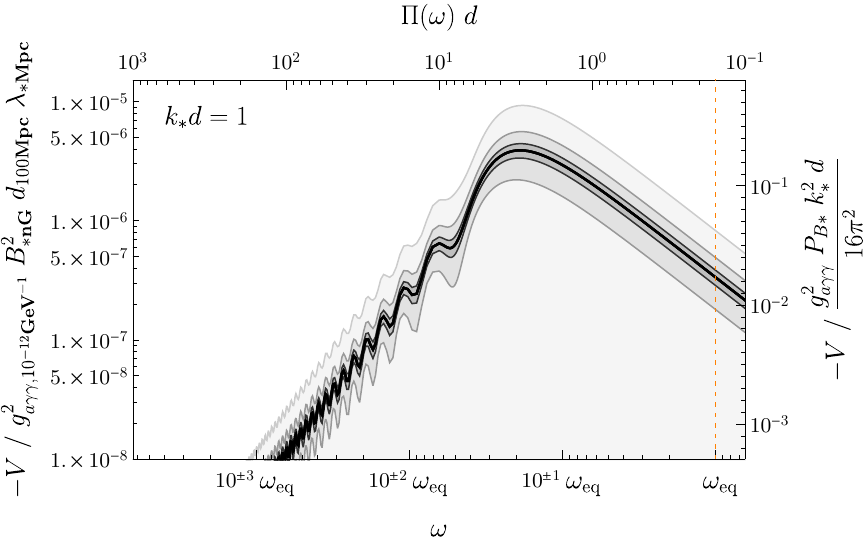}
\\
\includegraphics[width=0.4\linewidth]{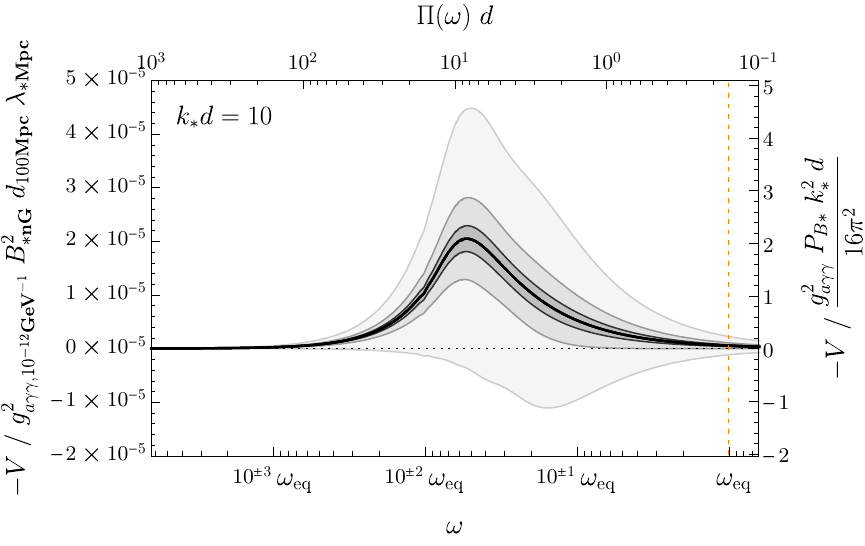}
\hskip 1cm
\includegraphics[width=0.4\linewidth]{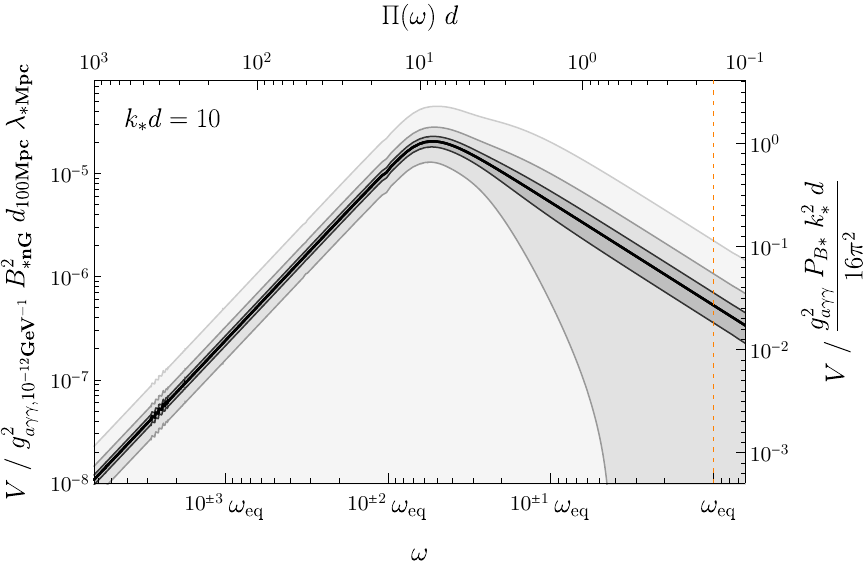}
\\
\includegraphics[width=0.4\linewidth]{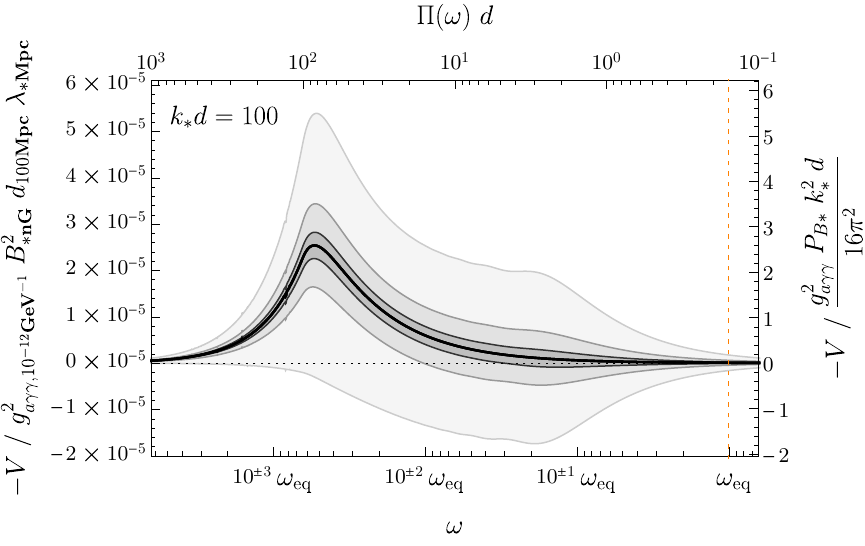}
\hskip 1cm
\includegraphics[width=0.4\linewidth]{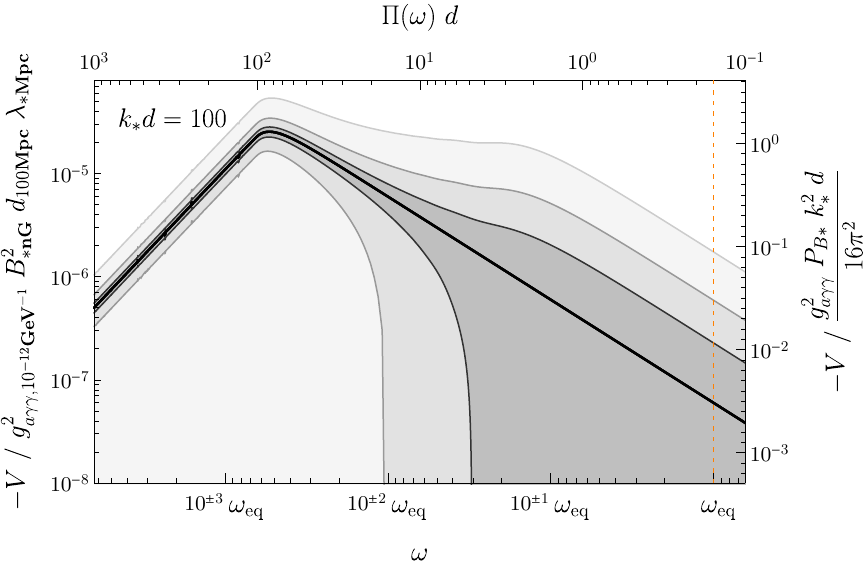}
\caption{
Stokes parameter $-V$ (circular polarization) of photons with the maximally helical ($P_{aB*} = P_{B*}$) magnetic field power spectra \eqref{eq:PB} and \eqref{eq:PaB} is plotted for $k_* d = 0.1$, $1$, $10$, and $100$ from top to bottom.
In each panel, the black line represents the expectation value $-\text{Exp}[V]$, and the shaded bands represent the standard deviations $\sqrt{\text{Var}[V]}$, $\sqrt{\text{Var}[V]/10}$, and $\sqrt{\text{Var}[V]/100}$.
The upper horizontal axis is the dimensionless variable $\Pi (\omega) d$, and the lower horizontal axis is $\omega$ in units of $\omega_\mathrm{eq}\equiv (|m_a^2 - m_\text{pl}^2| / 2\chi_{\mathrm{CMB}})^{\frac12}$ with $(|m_a^2 - m_\text{pl}^2|)^{\frac12}=10^{-11}\mathrm{eV}$, $2\chi_{\mathrm{CMB}}=10^{-42}$, and $d = 100\, \text{Mpc}$. 
The orange dashed line indicates $\omega = \omega_\mathrm{eq}$.
The right vertical axis is $-V$ normalized by $g_{a\gamma\gamma}^2 P_{B*} k_*^2 d / 16\pi^2$. 
The left vertical axis is $-V$ normalized by $g^2_{a\gamma\gamma,10^{-12}\mathrm{GeV}^{-1}}\equiv(g_{a\gamma\gamma}/10^{-12}\mathrm{GeV}^{-1})^2 $, $B_{*\text{nG}}^2 \equiv (B_* / \text{nG})^2$, $d_{100\text{Mpc}} \equiv d / 100\,\text{Mpc}$, and $\lambda_{*\text{Mpc}} \equiv \lambda_{*} / \text{Mpc}$.
In each row, the left and right panels show the same quantity: the vertical axis is linear in the left, but logarithmic in the right.
}
\label{fig:V}
\end{figure}

As shown in Eq.~\eqref{eq:Stokes_statistical}, the expectation values and variances of the Stokes parameters $I$, $Q$, $U$, and $V$ are expressed in terms of the integrals $\alpha$, $\beta$, and $\gamma$.
In Figs.~\ref{fig:I}, \ref{fig:Q}, and \ref{fig:V} we present the behavior of the Stokes parameters as functions of $\Pi(\omega) d$ for several values of $k_* d$.
Note that we selectively present the Stokes parameter $Q$ as a representative of linear polarization, since the parameter $U$ exhibits an identical statistical structure.
In each panel, the thick black curves show the expectation values $\mathrm{Exp}[1-I]$, $\mathrm{Exp}[Q]$, and $\mathrm{Exp}[V]$, while the gray shaded regions represent the standard deviations $\sqrt{\mathrm{Var}[1-I]}$, $\sqrt{\mathrm{Var}[Q]}$, and $\sqrt{\mathrm{Var}[V]}$.
The three bands correspond to the standard deviations scaled by factors of $(1, 1/\sqrt{10}, 1/\sqrt{100})$, illustrating how the statistical uncertainty decreases as the number of independent measurements increases.
These shaded regions model a situation in which photons are observed from multiple, nearly identical sources located in different directions.
In this case, the only differences among the observations arise from different realizations of the magnetic field, and thus the variance decreases inversely with the number of observations.
Accordingly, the three bands can be interpreted as corresponding to approximately $1$, $10$, and $100$ independent observations.

For the expectation values of the Stokes parameters, not only the intensity $I$ but also the circular polarization $V$ can acquire nontrivial values, which are determined solely by the integrals $\alpha$ and $\beta$, respectively (see Eq.~\eqref{eq:Stokes_statistical}).
Accordingly, the behavior of the black curves in Figs.~\ref{fig:I} ($\mathrm{Exp}[1-I]$) and \ref{fig:V} ($\mathrm{Exp}[V]$) directly reflects these integrals shown in Fig.~\ref{fig:abc}.
The nontrivial expectation value of the circular polarization $V$ originates from the helical component of the magnetic field power spectrum, $P_{aB}$.
By contrast, as shown in Eq.~\eqref{eq:Stokes_statistical}, the expectation values of the linear polarizations $Q$ and $U$ vanish identically, as a consequence of the statistical isotropy of the magnetic field.

In contrast, the variances are determined by combinations of the integrals $\alpha$, $\beta$, and $\gamma$ as shown in Eq.~\eqref{eq:Stokes_statistical}, and hence their behavior is somewhat more complicated.
Let us summarize their key properties:
\begin{itemize}
\item 
The variance of $1 - I$ is determined by the combination $\frac{1}{2} (\alpha^2 + \beta^2 + \gamma^2)$.
In the case of $k_* d \lesssim 1$, its behavior changes around $\Pi d = \mathcal{O} (1)$: in the frequency range $\Pi d \gg 1$, $\alpha^2$ and $\beta^2$ contribute to the variance in the same order, but $\gamma^2$ is suppressed, and thus, the variance becomes of the same order as the square of the expectation value, $\frac{1}{2} (\alpha^2 + \beta^2 + \gamma^2) \approx \frac{1}{2} (\alpha^2 + \beta^2 )\sim\alpha^2=\text{Exp}[ 1 - I ]^2$.
While in the range $\Pi d \ll 1$, $\gamma^2$ grows and reaches $\alpha^2$, but $\beta^2$ becomes suppressed, and hence the variance finally reaches the square of the expectation value, $\frac{1}{2} (\alpha^2 + \beta^2 + \gamma^2) \approx \frac{1}{2} (\alpha^2 + \gamma^2 )\rightarrow\alpha^2=\text{Exp}[ 1 - I ]^2$ as $\Pi d \to 0$.

In the case of $k_* d \gg 1$, on the other hand, a nontrivial structure appears in the intermediate range ($k_* d \gg \Pi d \gg 1$).
In this range, the variance is $\frac{1}{2} (\alpha^2 + \beta^2 + \gamma^2) \approx \frac{1}{2}\alpha^2$, and hence the variance and the expectation value satisfy $\text{Var} [1 - I] \approx \frac12 \text{Exp} [1 - I] ^2 $, which means that the variance in this range is smaller by a factor of two than that in the range $\Pi d \ll 1$.

\item 
The variance of $Q$ is determined by the combination $\frac{1}{2} (\alpha^2 - \beta^2 + \gamma^2)$.
In the case of $k_* d \lesssim 1$, only $\Pi d = \mathcal{O} (1)$ appears as a characteristic scale:
in the frequency range $\Pi d \gg 1$, the variance becomes $\frac{1}{2} (\alpha^2 -\beta^2 + \gamma^2) \approx \frac{1}{2} (\alpha^2 - \beta^2 )\sim \alpha^2=\text{Exp}[ 1-I ]^2$, 
while in the range $\Pi d \ll 1$,
the variance approaches $\frac{1}{2} (\alpha^2 - \beta^2 + \gamma^2) \approx \frac{1}{2} (\alpha^2 + \gamma^2 )\rightarrow\alpha^2=\text{Exp}[ 1-I ]^2$ as $\Pi d \to 0$.

In the case of $k_* d \gg 1$, in addition to the scale $\Pi d = \mathcal{O} (1)$, a new characteristic scale $\Pi d \approx k_*d$ appears.
In the intermediate range ($k_* d \gg \Pi d \gg 1$), the variance takes the form $\frac{1}{2} (\alpha^2 - \beta^2 + \gamma^2) \approx \frac{1}{2} \alpha^2 =\frac12 \text{Exp} [1 - I] ^2$, which indicates that the variance in this range is reduced compared to that in the the range $\Pi d \ll 1$, similarly to $1 - I$.

\item 
The variance of $V$ is determined by the combination $\frac{1}{2} (\alpha^2 + \beta^2 - \gamma^2)$.
In the case of $k_* d \lesssim 1$, the variance $\text{Var} [V]$ is comparable to the square of the expectation value $\text{Exp} [V]^2$ in the entire frequency range.

For $k_* d \gg 1$, in the intermediate range ($k_* d \gg \Pi d \gg 1$), the variance is dominated by $\alpha^2$ and thus it is roughly constant with respect to $\Pi d$, while it decreases in the range $\Pi d \ll 1$ due to the negative contribution from $- \gamma^2$.
\end{itemize}

Before concluding this section, we comment on the relative sign between the circular polarization parameter $V$ and the helicity of the background magnetic field $P_{aB*}$.
In our definition, positive signs of $V$ and $P_{aB*}$ correspond to right-handed helicity~\cite{Durrer:2013pga}.
Figure~\ref{fig:V} shows that the photon circular polarization $V$ that appears after propagation is opposite to the helicity of the background magnetic field $P_{aB*}$.
This behavior can be understood in the rest frame of the produced axion.
In this frame, the two participating photons must have opposite spins to produce a spin-zero particle (axion), and hence photons with the same helicity as the background photon tend to get converted into the axion.
Thus, starting from an unpolarized ensemble (equal numbers of right- and left-handed photons), the conversion selectively removes photons with the same helicity as the background magnetic field when the latter contains an antisymmetric (helical) part.

\section{Discussion and conclusions}
\label{sec:discussions}
In this work, we have investigated the conversion of photons into axions in the presence of stochastic magnetic fields, with particular attention to the cosmological magnetic environments expected in the Universe.
When photons propagate in a magnetized region, the axion-photon coupling induces oscillations between these degrees of freedom.
If the magnetic field is stochastically realized, the randomness of the magnetic field structure plays a crucial role in shaping the statistical properties of the resulting photon signals.
Therefore, a statistical framework becomes essential to describe the conversion process, which we have developed in the present work.
Following this framework, we derive the formulae for the expectation values and variances of observables associated with photons affected by the conversion, using ensemble averages over Gaussian magnetic field realizations.
Furthermore, by including the possible helical component of the magnetic field, we provide a more complete description of the axion-photon conversion than has been developed in earlier works~\cite{Li:2022mcf,setabuddin2025axionphotonconversionflrwprimordial}.
This improvement extends the range of situations to which our formalism can be applied, thereby enhancing the prospects for axion searches that exploit photons propagating through cosmic magnetic fields as a probe.

In Sec.~\ref{sec:photonreview} and Sec.~\ref{sec:axionreview}, we review photon propagation in several relevant situations, and derive the propagation equations for the axion-photon system.
We then solve them using the Born approximation.
To describe the photon polarization state, we introduce the Stokes parameters and study how initially unpolarized photons are affected by the conversion into axions.

In Sec.~\ref{sec:random}, we derive expressions for the expectation values and variances of the photon Stokes parameters in terms of the magnetic field power spectra, assuming a Gaussian magnetic field and an initial state given by unpolarized photons.
As shown in Eq.~\eqref{eq:Stokes_statistical}, these quantities are compactly expressed in terms of four basic components, $\alpha$, $\beta$, $\gamma$, and $\delta$, where $\delta$ vanishes identically.
Each component is expressed as a convolution of the magnetic field power spectrum with its own kernel (see Eq.~\eqref{eq:abcd}).
We find that the expectation value of the intensity $I$ becomes a nonzero signal, and that of the circular polarization $V$ also becomes nonzero when the magnetic field has a helical component, whereas the expectation values of the linear polarizations $Q$ and $U$ vanish.
In contrast, we find that the variances of all Stokes parameters, $I$, $Q$, $U$, and $V$ are nonzero, even when the initial photon state is statically unpolarized.
We also obtain a set of three consistency relations \eqref{eq:consistency} among the Stokes parameters.
One difference we find between these results and those in the graviton-photon system investigated in Ref.~\cite{Chiba:2025odu} is that, in the latter system, the linear polarizations $Q$ and $U$ are not generated from unpolarized gravitons for any background magnetic field and thus only one consistency relation is obtained.

In Sec.~\ref{sec:numerical}, using the magnetic field power spectra given by Eqs.~\eqref{eq:PB} and \eqref{eq:PaB} as a specific model, we examine the behavior of the Stokes parameters as functions of the parameters in this system.
The parameters are the effective mass difference between the axion and the photon $\Pi = \Pi_a - \Pi_\gamma = |m_a^2-m_{\rm pl}^2| / (2 \omega)+\chi_{\mathrm{CMB}}~\omega$ (with the axion mass $m_a$, the plasma-induced photon mass $m_{\rm pl}$, the CMB-induced susceptibility $\chi_\mathrm{CMB}$, and the angular frequency of the waves $\omega$), the propagation distance $d$, and the typical wavenumber of the magnetic field $k_*$.
Among these, two dimensionless combinations $\Pi d$ and $k_* d$ can be constructed, which characterize the axion-photon system and the background magnetic field, respectively.
In Sec.~\ref{subsec:C}, we first show the behavior of the convolution kernels. 
In Sec.~\ref{subsec:abc}, we numerically investigate the dependence on $\Pi d$ (or the angular frequency $\omega$) of the integrals $\alpha$, $\beta$, and $\gamma$ for various values of $k_* d$.
Note that, since $\Pi(\omega) d$ is a non-monotonic function of $\omega$, all the integrals generally possess a folding structure with respect to $\omega$.
Using these results, in Sec.~\ref{subsec:NumCalcStokes}, we show the behavior of the expectation values and variances of the Stokes parameters $I$, $Q$, and $V$. 
(The behavior of $U$ is the same as that of $Q$.)
Note that the Stokes parameters inherit the folding structure with respect to $\omega$ from the integrals $\alpha$, $\beta$, and $\gamma$.
We find that the expectation value of the conversion probability $\text{Exp}[1-I]$ changes its behavior from increasing with oscillation (large $\Pi d$) to constant (small $\Pi d$) at $\Pi d = \mathcal{O}(1)$ for $k_* d \ll 1$, whereas from increasing (large $\Pi d$) to constant (small $\Pi d$) at $\Pi d \approx k_* d$ for $k_* d \gg 1$ (see Fig.~\ref{fig:I}).
On the other hand, the variance $\text{Var}[1-I]$ shows a more curious $\omega$-dependence.
Specifically, we find that the particular frequency range $1 \lesssim \Pi d \lesssim k_* d$ has a standard deviation relatively smaller than the expectation value (see the bottom panels of Fig.~\ref{fig:I}).
Thanks to the smaller variance, this frequency range can be favorable for observing signatures of the conversion.
For the linear polarization $Q$, we find that the expectation value $\text{Exp}[Q]$ trivially vanishes for the entire frequency range, while the variance $\text{Var}[Q]$ does not.
For $k_* d \ll 1$, the behavior of the latter changes at $\Pi d = \mathcal{O}(1)$, while for $k_* d \gg 1$, an additional scale $\Pi d \approx k_* d$ appears (see Fig.~\ref{fig:Q}).
For the circular polarization $V$, we find that the expectation value $-\text{Exp}[V]$ appears with a peak structure for any value of $k_* d$.
For $k_* d \ll 1$, it appears around $\Pi d = \mathcal{O}(1)$, while for $k_* d \gg 1$, it shifts to $\Pi d \approx k_* d$.
The variance $\text{Var}[V]$ in the case of $k_* d \gg 1$ again shows a richer structure in the $\omega$-dependence (see Fig.~\ref{fig:V}).

As shown in the fourth row $(k_* d = 100)$ of Figs.~\ref{fig:I}, \ref{fig:Q}, and \ref{fig:V}, for representative values of the system parameters $B_{*} \sim \text{nG}$, $k_* \sim \text{Mpc}^{-1}$, $d \sim 100 \,\text{Mpc}$, $|m_a^2-m_\mathrm{pl}^2| \sim 10^{-22}\text{eV}^{2}$, and $g_{a\gamma\gamma}\sim10^{-12}\mathrm{GeV}^{-1}$, non-trivial signals appear in the Stokes parameters within a specific frequency region, $10^{-3} \omega_{\mathrm{eq}}\lesssim \omega \lesssim 10^{3} \, \omega_{\mathrm{eq}}$, corresponding to $\omega \sim 10\mathrm{MeV}$--$10\mathrm{TeV}$.
Especially, toward both ends of this region, the conversion probability reaches its maximum.
A notable peak structure in the circular polarization also emerges there when the magnetic field is helical.
There are several possible sources of photons belonging to this frequency region.
For example, such high-energy photons may originate from active galactic nuclei jets~\cite{Dermer_2016}, gamma-ray bursts~\cite{luongo2021roadmapgammarayburstsnew}, pulsar wind nebulae~\cite{Kargaltsev_2015}, supernova remnants~\cite{Vink_2011}, star-forming galaxies~\cite{Tacconi_2020}, and so on (for reviews, see Refs.~\cite{Bhattacharjee_2000,Massaro_2015}).
The arrival rate of TeV-scale photons decreases steeply with energy due to the extragalactic background light attenuation \cite{Galanti:2024lfn}.
Consequently, high-energy photons that have successfully avoided this attenuation and propagated over long distances through the cosmological magnetic field may provide a potential probe for axions.

We conclude by mentioning several possible directions for future work.
First, it would be worthwhile to generalize the initial polarization state. 
In the present paper, we assumed unpolarized photons as the initial condition; however, if the initial state possesses a nontrivial polarization, it may affect the resulting observables.
Second, it would be valuable to develop a formulation that applies to multiple magnetic-field domains. 
Although the present analysis assumes a single domain described by a single set of magnetic field power spectra, realistic cosmological magnetic fields are expected to consist of many domains with varying orientations and coherence lengths.
Third, for sources at high redshift, cosmic expansion can leave interesting imprints in the amplitude/polarization structure.
Finally, it would be interesting to extend our framework to the case of axion dark matter.
When both a magnetic field and an axion field act as background fields, the polarization of photons propagating through them may be modified not only by conversion but also by additional axion-induced phenomena such as birefringence.
We leave these studies for future work.

\begin{acknowledgments}

The authors are grateful to Asuka Ito, Kohei Kamada, Akito Kusaka, Fumihiro Naokawa and Jiro Soda for helpful discussions.
The work of W.C. was supported by JST SPRING, Grant Number JPMJSP2148.
R.J. was supported by JSPS KAKENHI Grant Numbers
23K17687 and 24K07013.
K.N. was supported by JSPS KAKENHI Grant Numbers JP24KJ0117 and JP25K17389. 

\end{acknowledgments}

\appendix
\section{Calculations of the expectation values and variances}
\label{app:A}

As shown in Eq.~\eqref{eqStokes}, assuming that the initial photons are unpolarized and adopting the Born approximation, the Stokes parameters of the photons affected by conversion into axions are expressed as
\begin{align}
  \begin{cases}
    I = 1-\frac{1}{2}a~,\\
    Q = -\frac{1}{2}b~,\\
    U = -\mathrm{Re}\left[c\right]~,\\
    V = \mathrm{Im}\left[c\right]~,\\
  \end{cases}
  \label{eqAStokes}
\end{align}
where $a$, $b$, and $c$ are given by
\begin{align}
    \begin{cases}
      a = \displaystyle\Big(\frac{g_{a\gamma\gamma}}{2}\Big)^2\int_{s,s'}e^{-i\Pi (s-s')}[ B_x(s)B_x(s')+B_y(s)B_y(s') ]~,\\
      b = \displaystyle\Big(\frac{g_{a\gamma\gamma}}{2}\Big)^2\int_{s,s'}e^{-i\Pi (s-s')}[ B_x(s)B_x(s')-B_y(s)B_y(s') ]~,\\
      c = \displaystyle\Big(\frac{g_{a\gamma\gamma}}{2}\Big)^2\int_{s,s'}e^{-i\Pi (s-s')}B_x(s)B_y(s')~.\\
  \end{cases}
  \label{eqAabc}
\end{align}
In this Appendix, we compute the expectation values and variances of the Stokes parameters with respect to an ensemble of magnetic fields characterized by the two-point function \eqref{2ptfunc}.
Since photons are assumed to propagate along the $z$-axis and the magnetic fields are statistically isotropic, rotational symmetry in the $x$-$y$ plane holds. Accordingly, in the following calculations, we can use the relation $\hat{k}^2_x=\hat{k}^2_y= (\hat{k}^2_x+\hat{k}^2_y) / 2 = (1-\hat{k}^2_z) / 2$ when performing integrals in the wave vector space.

\subsection{Calculations of the expectation values}

We first compute the expectation value of $a$. Using Eq.~\eqref{2ptfunc}, it is evaluated as
\begin{align}
    \langle a\rangle&=\Big(\frac{g_{a\gamma\gamma}}{2}\Big)^2 \int_{s,s'}e^{-i\Pi(s-s')}\langle B_x(s)B_x(s') + B_y(s)B_y(s') \rangle\notag\\
    &=\Big(\frac{g_{a\gamma\gamma}}{2}\Big)^2 \int_{s,s'}e^{-i\Pi(s-s')}\int_ke^{ik_z(s-s')}(2-\hat{k}^2_x-\hat{k}^2_y)P_B(k)\notag\\
    &=2\Big(\frac{g_{a\gamma\gamma}}{2}\Big)^2~\int_k \Big(\int_{s,s'}e^{-i(\Pi-k_z)(s-s')}\Big)\frac{1+\hat{k}_z^2}{2}P_B(k)\notag\\
    &=2\Big(\frac{g_{a\gamma\gamma}}{2}\Big)^2~\int_k\mathcal{I}(k_z;\omega)\Big(\frac{1+\hat{k}^2_z}{2}\Big)P_{B}(k)
    \equiv 2\alpha~. 
    \label{eqAexpa}
\end{align}
In the last line, we used the kernel $\mathcal{I}(k_z;\omega)$ defined in Eq.~\eqref{eqkernelI}.
This result, together with Eq.~\eqref{eqAStokes}, indicates that
\begin{align}
    \mathrm{Exp}[1-I]=\alpha~.
\end{align}

The expectation value of $b$ is evaluated as 
\begin{align}
    \langle b\rangle&=\Big(\frac{g_{a\gamma\gamma}}{2}\Big)^2\int_{s,s'}e^{-i\Pi(s-s')}\langle B_x(s)B_x(s')-B_y(s)B_y(s')\rangle\notag\\
    &=\Big(\frac{g_{a\gamma\gamma}}{2}\Big)^2 \int_{s,s'}e^{-i\Pi(s-s')}\int_ke^{ik_z(s-s')}(-\hat{k}^2_x+\hat{k}^2_y)P_B(k)\notag\\
    &=0~,
    \label{eqAexpb}
\end{align}
from which we immediately obtain
\begin{align}
    \mathrm{Exp}[Q]=0~.
\end{align}

The expectation value of $c$ is evaluated as 
\begin{align}
    \langle c\rangle&=\Big(\frac{g_{a\gamma\gamma}}{2}\Big)^2\int_{s,s'}e^{-i\Pi(s-s')}\langle B_x(s)B_y(s')\rangle\notag\\
    &=-\Big(\frac{g_{a\gamma\gamma}}{2}\Big)^2
    \bigg[ \int_{s,s'}e^{-i\Pi(s-s')}\int_ke^{ik_z(s-s')}\hat{k}_x\hat{k}_yP_{B}(k)+i \int_{s,s'}e^{-i\Pi(s-s')}\int_ke^{ik_z(s-s')}\hat{k}_zP_{aB}(k)
    \bigg]\notag\\
&=-\Big(\frac{g_{a\gamma\gamma}}{2}\Big)^2
\bigg[ \int_k\mathcal{I}(k_z;\omega)\hat{k}_x\hat{k}_yP_{B}(k)+i \int_k\mathcal{I}(k_z;\omega)\hat{k}_zP_{aB}(k) \bigg]\notag\\
&= -i \Big(\frac{g_{a\gamma\gamma}}{2}\Big)^2 \int_k\mathcal{I}(k_z;\omega)\hat{k}_zP_{aB}(k)
    \equiv -i\beta~,
    \label{eqAexpc}
\end{align}
which shows that $\langle c \rangle$ is purely imaginary. Combining this with Eq.~\eqref{eqAStokes}, we obtain
\begin{align}
    \mathrm{Exp}[U]=0~,\quad\mathrm{Exp}[V]= -\beta~.
\end{align}

\subsection{Calculations of the variances}

Next, we compute the variances of the Stokes parameters. 
The variance of the intensity $I$ is given by $\mathrm{Var}[I] = \langle I^2 \rangle - \langle I \rangle^2 = (\langle a^2\rangle-\langle a\rangle^2) / 4$.
Using the definition of $a$ in Eq.~\eqref{eqAabc}, we can compute $\langle a^2 \rangle$ as 
\begin{align}
   \langle a^2 \rangle &=\Big(\frac{g_{a\gamma\gamma}}{2}\Big)^4
   \bigg< 
   \Big(\int_{s,s'}e^{-i\Pi(s-s')}[B_x(s)B_x(s')+ B_y(s)B_y(s')]\Big)^2
   \bigg>
   \notag\\
   &=\Big(\frac{g_{a\gamma\gamma}}{2}\Big)^4
   \bigg<
   \Big(\int_{s,s'}e^{-i\Pi (s-s')}B_x(s)B_x(s')\Big)~\Big(\int_{s,s'}e^{-i\Pi (s-s')}B_x(s)B_x(s')\Big)
   \bigg>
   \notag\\
   &~+\Big(\frac{g_{a\gamma\gamma}}{2}\Big)^4
   \bigg< 
   \Big(\int_{s,s'}e^{-i\Pi (s-s')}B_x(s)B_x(s')\Big)~\Big(\int_{s,s'}e^{-i\Pi (s-s')}B_y(s)B_y(s')\Big)
   \bigg>
   \notag\\
   &~+\Big(\frac{g_{a\gamma\gamma}}{2}\Big)^4
   \bigg<
   \Big(\int_{s,s'}e^{-i\Pi (s-s')}B_y(s)B_y(s')\Big)~\Big(\int_{s,s'}e^{-i\Pi (s-s')}B_x(s)B_x(s')\Big)
   \bigg>
   \notag\\
   &~+\Big(\frac{g_{a\gamma\gamma}}{2}\Big)^4
   \bigg< 
   \Big(\int_{s,s'}e^{-i\Pi (s-s')}B_y(s)B_
   y(s')\Big)~\Big(\int_{s,s'}e^{-i\Pi (s-s')}B_y(s)B_y(s')\Big)
   \bigg>
   \notag \\
   &\equiv I_1 + I_2 + I_3 + I_4~.
   \label{eqAasquare}
\end{align}
From rotational symmetry in the $x$-$y$ plane, the first and fourth terms are equal, and likewise for the second and third terms.
Thus, the expression simplifies to
\begin{align}
    \langle a^2\rangle = 2(I_1+I_2)~.
    \label{eqAasquare2}
\end{align}
In the following, we set $g_{a\gamma\gamma} / 2 \equiv 1$ for simplicity.
Since the magnetic field is assumed to be Gaussian, both $I_1$ and $I_2$ can be decomposed into products of the two-point function. 
Specifically, $I_1$ can be written as
\begin{align}
    I_1
    &=\int_{s,s',s'',s'''}e^{-i\Pi (s-s'+s''-s''')}\langle B_x(s) B_x(s') B_x(s'') B_x(s''')\rangle\notag\\
    &=\int_{s,s',s'',s'''}e^{-i\Pi (s-s'+s''-s''')}
    \notag \\
    &\hspace{4em}
    \times \Big[\langle B_x(s)B_x(s')\rangle\langle B_x(s'')B_x(s''')\rangle+\langle B_x(s)B_x(s'')\rangle\langle B_x(s')B_x(s''')\rangle +\langle B_x(s)B_x(s''')\rangle\langle B_x(s')B_x(s'')\rangle\Big]\notag\\
    &\equiv A_1+A_2+A_3
    ~.
    \label{eqAI1}
\end{align}

The first term is evaluated as 
\begin{align}
    A_1&=\int_{s,s',s'',s'''}e^{-i\Pi (s-s'+s''-s''')}\langle B_x(s)B_x(s')\rangle\langle B_x(s'')B_x(s''')\rangle\notag\notag\\
    &=\int_{k,k'}\int_{s,s',s'',s'''}e^{-i\Pi (s-s'+s''-s''')}e^{ik_z (s-s')}e^{ik'_z (s''-s''')}(1-\hat{k}_x^2)(1-\hat{k}'^2_x)P_B(k)P_B(k')\notag\\
    &=\int_{k,k'}\mathcal{I}(k_z;\omega) \, \mathcal{I}(k'_z;\omega)\Big(\frac{1+\hat{k}^2_z}{2}\Big)\Big(\frac{1+\hat{k}'^2_z}{2}\Big)P_B(k)P_B(k')\notag\\
    &=\bigg[ \int_k\mathcal{I}(k_z;\omega)\Big(\frac{1+\hat{k}^2_z}{2}\Big)P_B(k)\bigg]^2 = \alpha^2~.
\end{align}
The second term is evaluated as 
\begin{align}
    A_2&=\int_{s,s',s'',s'''}e^{-i\Pi (s-s'+s''-s''')}\langle B_x(s)B_x(s'')\rangle\langle B_x(s')B_x(s''')\rangle\notag\notag\\
    &=\int_{k,k'}\int_{s,s',s'',s'''}e^{-i\Pi (s-s'+s''-s''')}e^{ik_z (s-s'')}e^{ik'_z (s'-s''')}(1-\hat{k}_x^2)(1-\hat{k}'^2_x)P_B(k)P_B(k')\notag\\
    &=\int_{k,k'}\bar{\mathcal{J}}(k_z;\omega) \, \bar{\mathcal{J}}^*(k'_z;\omega)\Big(\frac{1+\hat{k}^2_z}{2}\Big)\Big(\frac{1+\hat{k}'^2_z}{2}\Big)P_B(k)P_B(k')\notag\\
     &=\bigg[ \int_k\mathcal{J}(k_z;\omega)\Big(\frac{1+\hat{k}^2_z}{2}\Big)P_B(k)\bigg]^2 \equiv \gamma^2~.
\end{align}
Here, $\bar{\mathcal{J}}(k_z;\omega)$ is defined by
\begin{align}
    \bar{\mathcal{J}}(k_z;\omega)
    &\equiv \int_{s} e^{- i (\Pi - k_z) s} \int_{s'} e^{-i (\Pi + k_z) s'} 
    = \bar{\mathcal{J}}(-k_z;\omega)~,
\end{align}
and $\mathcal{J}(k_z; \omega)$ denotes its absolute value, given in Eq.~\eqref{eqkernelJ}.
The last term in Eq.~\eqref{eqAI1} becomes
\begin{align}
    A_3&=\int_{s,s',s'',s'''}e^{-i\Pi (s-s'+s''-s''')}\langle B_x(s)B_x(s''')\rangle\langle B_x(s')B_x(s'')\rangle\notag = A_1 =\alpha^2~.
\end{align}
Therefore, $I_1$ is expressed as
\begin{align}
    I_1=2\alpha^2+\gamma^2~.
    \label{eqAI1res}
\end{align}

The integral $I_2$ in Eq.~\eqref{eqAasquare} can be decomposed as 
\begin{align}
    I_2&=\int_{s,s',s'',s'''}e^{-i\Pi (s-s'+s''-s''')}\langle B_x(s)B_x(s')B_y(s'')B_y(s''')\rangle\notag\\
    &=\int_{s,s',s'',s'''}e^{-i\Pi (s-s'+s''-s''')}
    \notag \\
    &\hspace{4em}
    \times \Big[\langle B_x(s)B_x(s')\rangle\langle B_y(s'')B_y(s''')\rangle+\langle B_x(s)B_y(s'')\rangle\langle B_x(s')B_y(s''')\rangle +\langle B_x(s)B_y(s''')\rangle\langle B_x(s')B_y(s'')\rangle\Big]\notag\\
    &\equiv B_1+B_2+B_3
    ~.
\end{align}
Performing calculations analogous to those for $A_1$, $A_2$, and $A_3$, we obtain
\begin{align}
    B_1
    &=\bigg[ \int_k\mathcal{I}(k_z;\omega)\Big(\frac{1+\hat{k}^2_z}{2}\Big)P_B(k)\bigg] ^2=\alpha^2~,
    \\
    B_2
     &=- \bigg[ \int_k\mathcal{J}(k_z;\omega) \,\hat{k}_z \, P_{aB}(k) \bigg]^2 \equiv -\delta^2~,
    \\
    B_3
    &=\bigg[ \int_k\mathcal{I}(k_z;\omega) \, \hat{k}_z \, P_{aB}(k)\bigg]^2 = \beta^2~.
\end{align}
Thus, $I_2$ is expressed as
\begin{align}
    I_2=\alpha^2-\delta^2+\beta^2~.
    \label{eqAI2res}
\end{align}

Substituting Eqs.~\eqref{eqAI1res} and \eqref{eqAI2res} into Eq.~\eqref{eqAasquare2}, we obtain $\langle a^2 \rangle = 2(3\alpha^2 + \beta^2 + \gamma^2 - \delta^2)$. Combining this with $\langle a \rangle = 2\alpha$ from Eq.~\eqref{eqAexpa}, we find 
\begin{align}
    \mathrm{Var}[I]=\frac{1}{4}(\langle a^2\rangle-\langle a\rangle^2)
    &=\frac{1}{2}(\alpha^2+\beta^2+\gamma^2-\delta^2)~.
\end{align}

We proceed to consider the variance of the Stokes parameter $Q$, which is given by $\text{Var}[Q] = \langle b^2 \rangle / 4$ from Eqs.~\eqref{eqAStokes} and \eqref{eqAexpb}.
Using the definition of $b$ in Eq.~\eqref{eqAabc}, we can compute $\langle b^2 \rangle$ as 
\begin{align}
   \langle b^2 \rangle &=\Big(\frac{g_{a\gamma\gamma}}{2}\Big)^4
   \bigg<
   \Big(\int_{s,s'}e^{-i\Pi(s-s')}[B_x(s)B_x(s')- B_y(s)B_y(s')]\Big)^2
   \bigg> \notag\\
   &= I_1-I_2-I_3+I_4
   \notag \\
   &=2(I_1-I_2)~.
\end{align}
Here, we used $I_1$, $I_2$, $I_3$, and $I_4$ defined in Eq.~\eqref{eqAasquare}, together with the relations $I_1 = I_4$ and $I_2 = I_3$, in the same manner as in Eq.~\eqref{eqAasquare2}.
Using Eqs.~\eqref{eqAI1res} and \eqref{eqAI2res}, we obtain 
\begin{align}
    \mathrm{Var}[Q]=\frac{1}{4}\langle b^2\rangle
    =\frac{1}{2}(\alpha^2-\beta^2+\gamma^2+\delta^2)~.
\end{align}

Finally, we consider the variances of the Stokes parameters $U$ and $V$.
To this end, we evaluate the square of $c=-U+iV$ as 
\begin{align}
    c^2 &= (U^2-V^2)-i(2UV)~,
    \\
    |c|^2 &= U^2+V^2~.
\end{align}
From these expressions, it follows that
\begin{align}
    \text{Var}[U] &= 
    \langle U^2\rangle - \langle U \rangle^2 = \frac{1}{2}[\langle |c|^2\rangle+\langle \mathrm{Re}[c^2]\rangle]
    - \langle \mathrm{Re}[c]\rangle^2
    ~,\\
    \text{Var}[V] &= 
    \langle V^2\rangle - \langle V \rangle^2 = \frac{1}{2}[\langle |c|^2\rangle-\langle \mathrm{Re}[c^2]\rangle]
    - \langle \mathrm{Im}[c]\rangle^2
    ~.
\end{align}
Using the definition of $c$ in Eq.~\eqref{eqAabc} and Gaussianity of the magnetic fields, $\langle c^2 \rangle$ and $\langle |c|^2 \rangle$ are expressed as 
\begin{align}
     \langle c^2\rangle&=\int_{s,s',s'',s'''}e^{-i\Pi (s-s'+s''-s''')}\langle B_x(s)B_y(s')B_x(s'')B_y(s''')\rangle\notag\\
    &=\int_{s,s',s'',s'''}e^{-i\Pi (s-s'+s''-s''')}
    \notag \\
    &\hspace{4em} \times
    \Big[\langle B_x(s)B_y(s')\rangle\langle B_x(s'')B_y(s''')\rangle+\langle B_x(s)B_x(s'')\rangle\langle B_y(s')B_y(s''')\rangle
    +\langle B_x(s)B_y(s''')\rangle\langle B_y(s')B_x(s'')\rangle\Big]\notag\\
    &\equiv C_1+C_2+C_3
    ~, \\
    \langle |c|^2\rangle&=\int_{s,s',s'',s'''}e^{-i\Pi (s-s'-s''+s''')}\langle B_x(s)B_y(s')B_x(s'')B_y(s''')\rangle\notag\\
    &=\int_{s,s',s'',s'''}e^{-i\Pi (s-s'-s''+s''')}
    \notag \\
    &\hspace{4em}
    \times \Big[\langle B_x(s)B_y(s')\rangle\langle B_x(s'')B_y(s''')\rangle+\langle B_x(s)B_x(s'')\rangle\langle B_y(s')B_y(s''')\rangle
    +\langle B_x(s)B_y(s''')\rangle\langle B_y(s')B_x(s'')\rangle\Big]\notag\\
    &\equiv D_1+D_2+D_3
    ~.
\end{align}
We again set $g_{a\gamma \gamma} / 2 = 1$ for simplicity.
Each term can be evaluated as
\begin{align}
    C_1 = C_3 = - D_1
    &=- \bigg[ \int_k\mathcal{I}(k_z;\omega) \, \hat{k}_z \, P_{aB}(k) \bigg]^2=-\beta^2~, \\
    C_2
     &=\bigg[\int_k\mathcal{J}(k_z;\omega) \Big( \frac{1+\hat{k}^2_z}{2} \Big) P_B(k) \bigg]^2=\gamma^2~, 
     \\
     D_2
     &= \bigg[ \int_k\mathcal{I}(k_z;\omega)\Big(\frac{1+\hat{k}^2_z}{2}\Big)P_B(k) \bigg]^2 = \alpha^2~,
     \\
     D_3
    &=\bigg[\int_k\mathcal{J}(k_z;\omega) \, \hat{k}_z \, P_{aB}(k)\bigg]^2=\delta^2~.
\end{align}
Accordingly, we obtain $\langle c^2\rangle=- 2\beta^2+\gamma^2$, which is real.
In addition, $\langle |c|^2\rangle=\beta^2+\alpha^2+\delta^2$.
Combining these results with $\langle c \rangle = -i \beta$ from Eq.~\eqref{eqAexpc}, we find the variances of $U$ and $V$ as 
\begin{align}
    \mathrm{Var}[U] &= \frac{1}{2}(\alpha^2-\beta^2+\gamma^2+\delta^2)~,
    \\
    \mathrm{Var}[V] 
    &=\frac{1}{2}(\alpha^2+\beta^2-\gamma^2+\delta^2)~.
\end{align}

\section{Expressions for the convolution kernels}
\label{app:C}

In this appendix, we summarize the analytical expressions for the convolution kernels in Eq.~\eqref{eq:C}, obtained after integration over $z$:
\begin{align}
\mathcal{C}_\alpha (x = \Pi d, y = kd)
&=
- \frac{1}{y (x - y z)}
\left[
2 x (x-y z) (\text{Ci}(| x-y z| ) - \ln (| x-y z| ))+\left(x^2+y^2\right) (x-y z) \text{Si}(x-y z)
\right.
\notag \\
&\qquad\qquad\qquad\quad
\left.
+(x^2+y^2) \cos (x-y z)-(x-y z) \sin (x-y z)-x^2-xy z-y^2+y^2 z^2
\right]
\Big|_{z = -1}^{z = 1}
~,
\\
\mathcal{C}_\beta (x = \Pi d, y = kd)
&=
- \frac{2}{x-y z}
\left[
(x-y z) (\text{Ci}(| x-y z| )-\ln (| x-y z| ))+x (x-y z) \text{Si}(x-y z)+x \cos (x-y z)-x
\right]
\Big|_{z = -1}^{z = 1}
~,
\\
\mathcal{C}_\gamma (x = \Pi d, y = kd)
&=
\frac{1}{2 x y}
\left[
\left(x^2+y^2\right) \cos x (\text{Ci}(| x+y z| )-\ln (| x+y z| )-\text{Ci}(| x-y z| ) + \ln (| x-y z| ))
\right.
\notag \\
&\qquad\qquad
\left.
+\left(x^2+y^2\right) \sin x (\text{Si}(x+y z)-\text{Si}(x-y z))
+2 x (y z \cos x-\sin (y z))
\right]
\Big|_{z = -1}^{z = 1}
~.
\end{align}

\section{Non-helical case}
\label{app:non-helical}

\begin{figure}[H]
\centering
\includegraphics[width=0.4\linewidth]{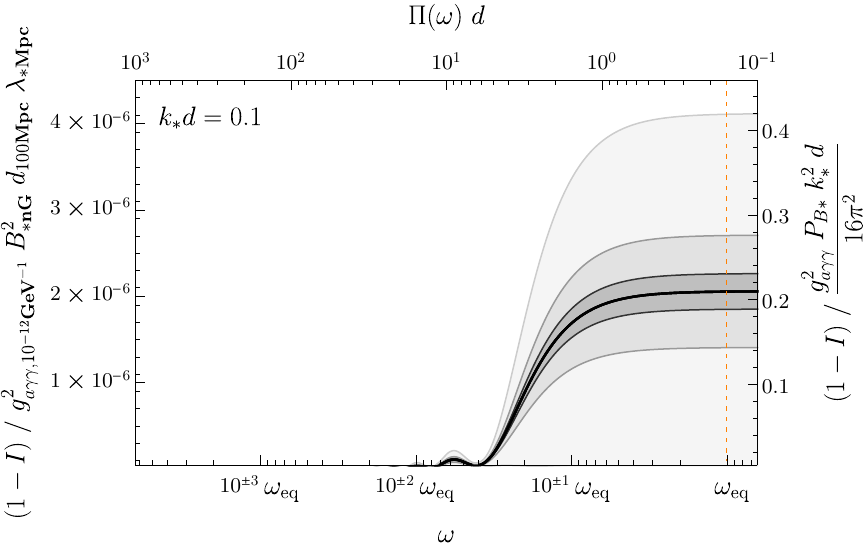}
\hskip 1cm
\includegraphics[width=0.4\linewidth]{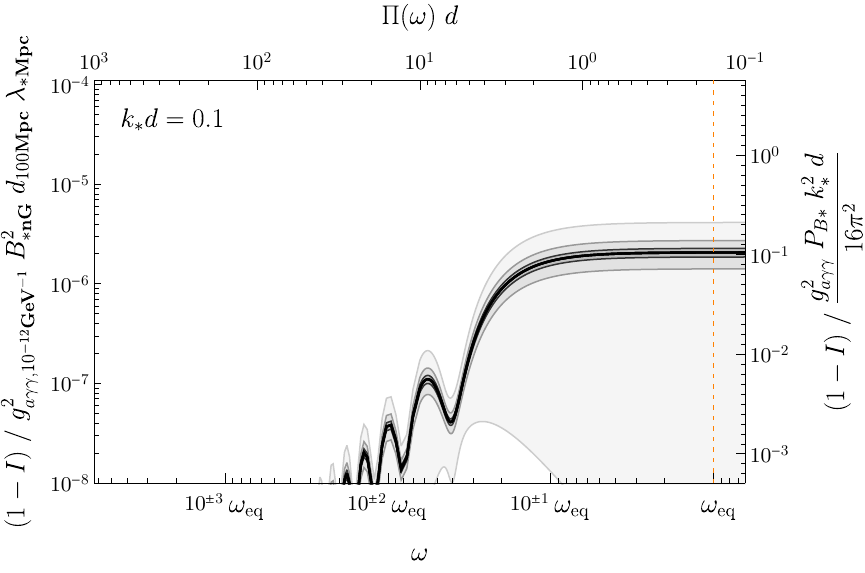}
\\
\includegraphics[width=0.4\linewidth]{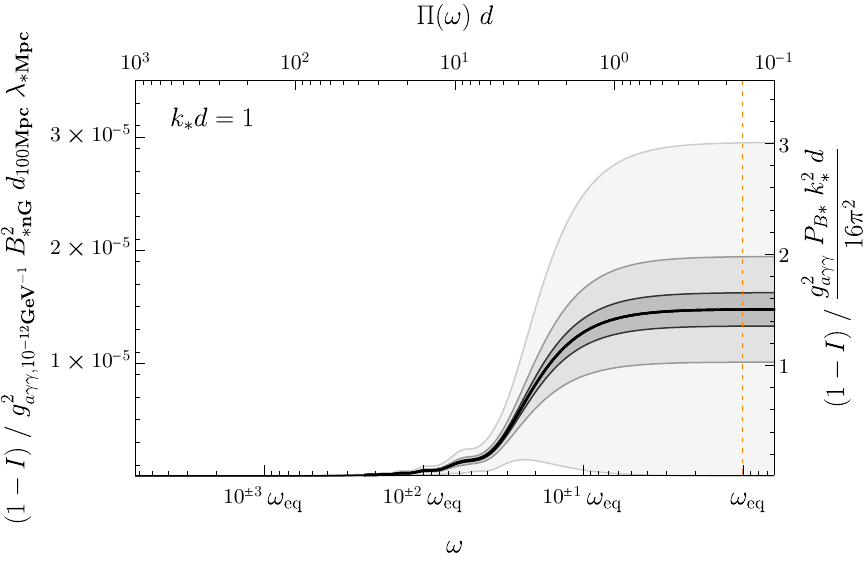}
\hskip 1cm
\includegraphics[width=0.4\linewidth]{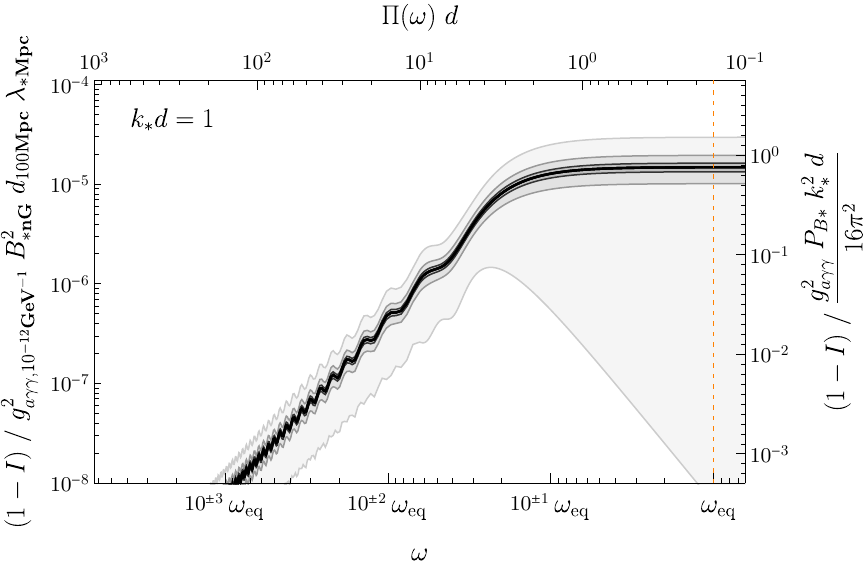}
\\
\includegraphics[width=0.4\linewidth]{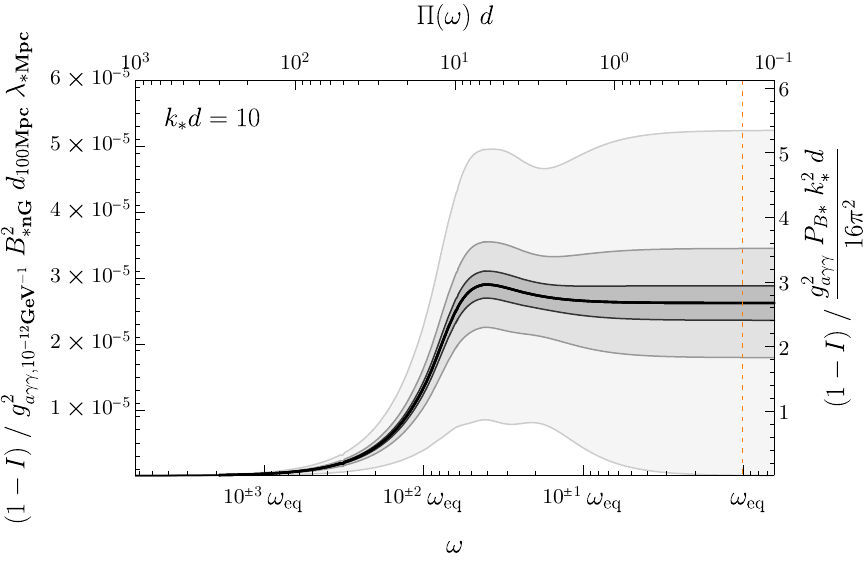}
\hskip 1cm
\includegraphics[width=0.4\linewidth]{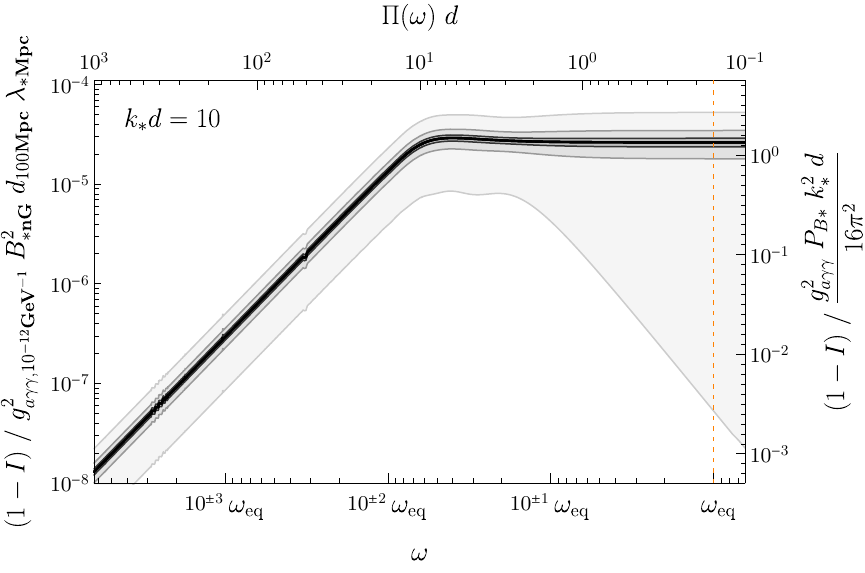}
\\
\includegraphics[width=0.4\linewidth]{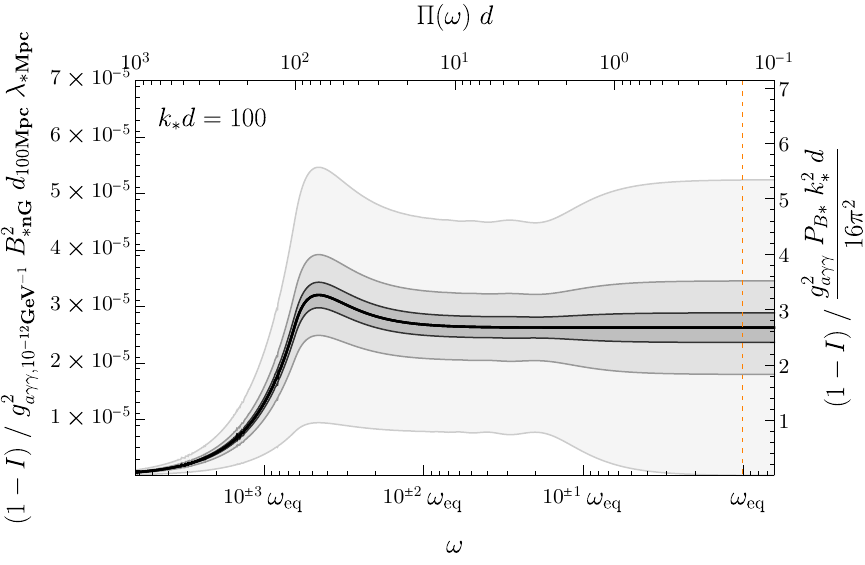}
\hskip 1cm
\includegraphics[width=0.4\linewidth]{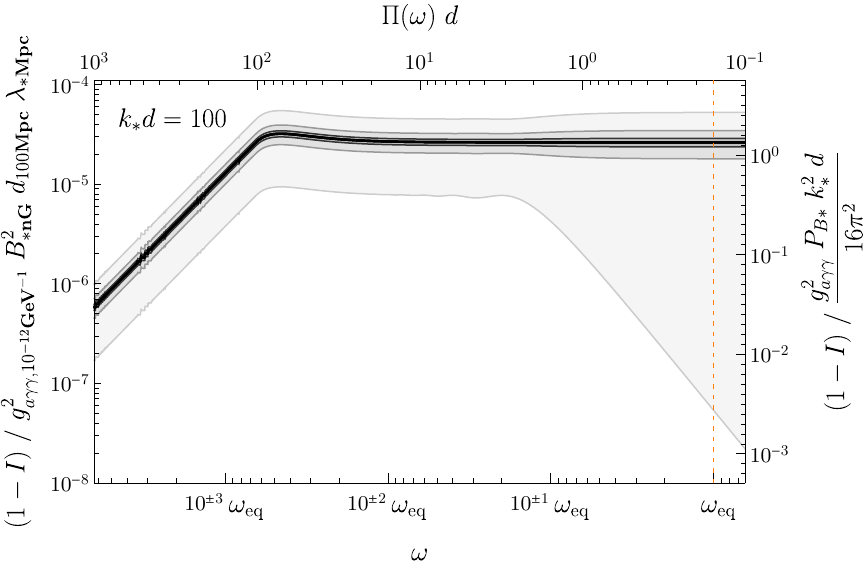}
\caption{
Conversion probability $P_{\gamma \to a} \equiv 1-I$ with the non-helical ($P_{aB*} = 0$) magnetic field power spectrum \eqref{eq:PB} is plotted for $k_* d = 0.1$, $1$, $10$, and $100$ from top to bottom.
In each panel, the black line represents the expectation value $\text{Exp}[1-I]$, and the shaded bands represent the standard deviations $\sqrt{\text{Var}[1-I]}$, $\sqrt{\text{Var}[1-I]/10}$, and $\sqrt{\text{Var}[1-I]/100}$.
The upper horizontal axis is the dimensionless variable $\Pi (\omega) d$, and the lower horizontal axis is $\omega$ in units of $\omega_\mathrm{eq}\equiv (|m_a^2 - m_\text{pl}^2| / 2\chi_{\mathrm{CMB}})^{\frac12}$ with $(|m_a^2 - m_\text{pl}^2|)^{\frac12}=10^{-11}\mathrm{eV}$, $2\chi_{\mathrm{CMB}}=10^{-42}$, and $d = 100\, \text{Mpc}$. 
The orange dashed line indicates $\omega = \omega_\mathrm{eq}$.
The right vertical axis is $1-I$ normalized by $g_{a\gamma\gamma}^2 P_{B*} k_*^2 d / 16\pi^2 $.
The left vertical axis is $1-I$ normalized by $g^2_{a\gamma\gamma,10^{-12}\mathrm{GeV}^{-1}}\equiv(g_{a\gamma\gamma}/10^{-12}\mathrm{GeV}^{-1})^2 $, $B_{*\text{nG}}^2 \equiv (B_* / \text{nG})^2$, $d_{100\text{Mpc}} \equiv d / 100\,\text{Mpc}$, and $\lambda_{*\text{Mpc}} \equiv \lambda_{*} / \text{Mpc}$.
In each row, the left and right panels show the same quantity: the vertical axis is linear in the left, but logarithmic in the right.
}
\label{fig:PlotInon}
\end{figure}

\begin{figure}[H]
\centering
\includegraphics[width=0.4\linewidth]{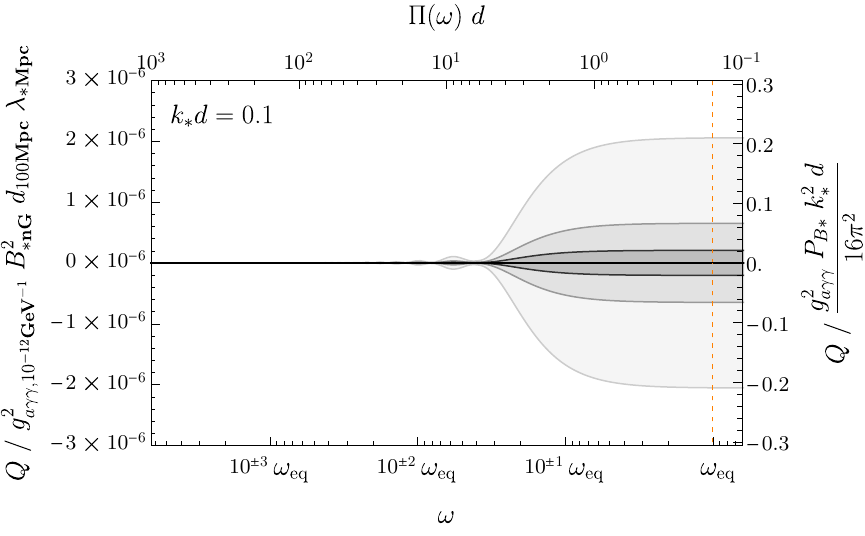}
\hskip 1cm
\includegraphics[width=0.4\linewidth]{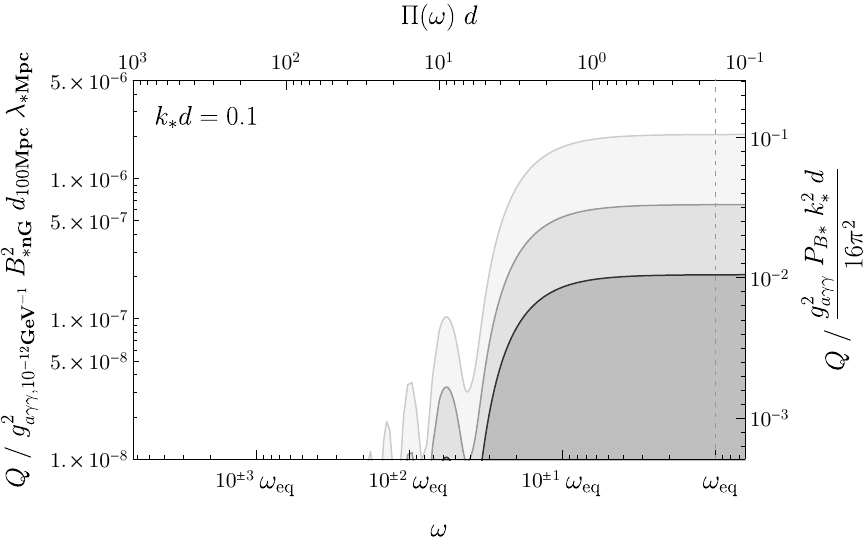}
\\
\includegraphics[width=0.4\linewidth]{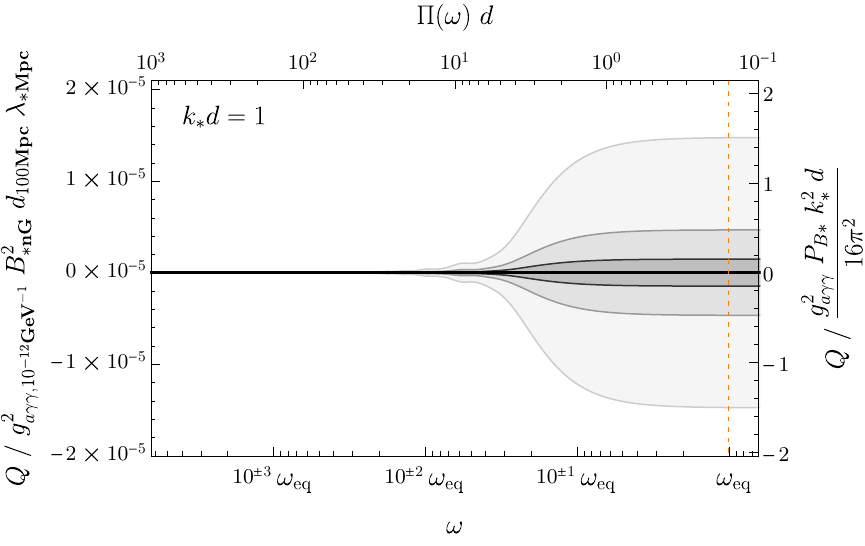}
\hskip 1cm
\includegraphics[width=0.4\linewidth]{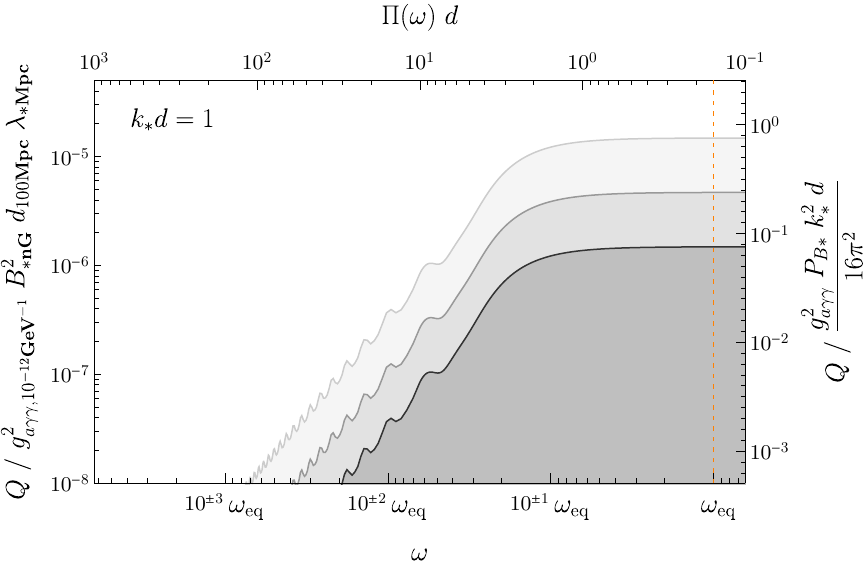}
\\
\includegraphics[width=0.4\linewidth]{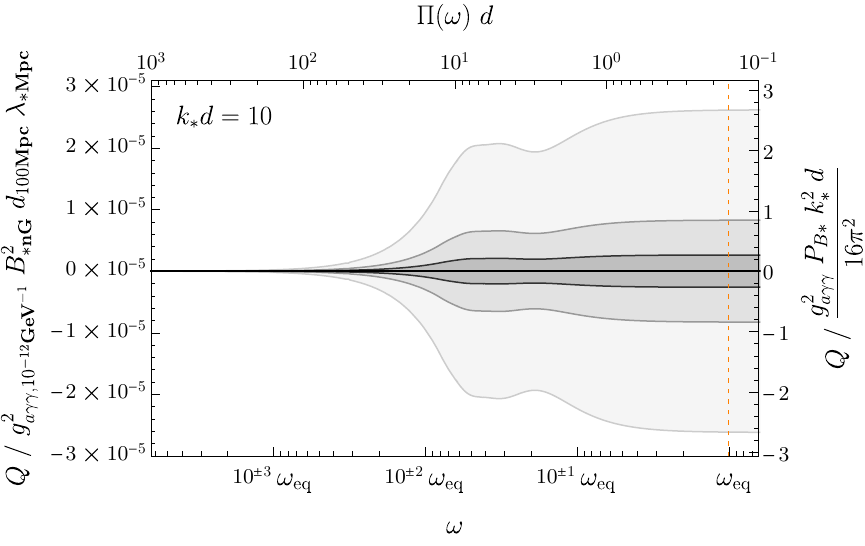}
\hskip 1cm
\includegraphics[width=0.4\linewidth]{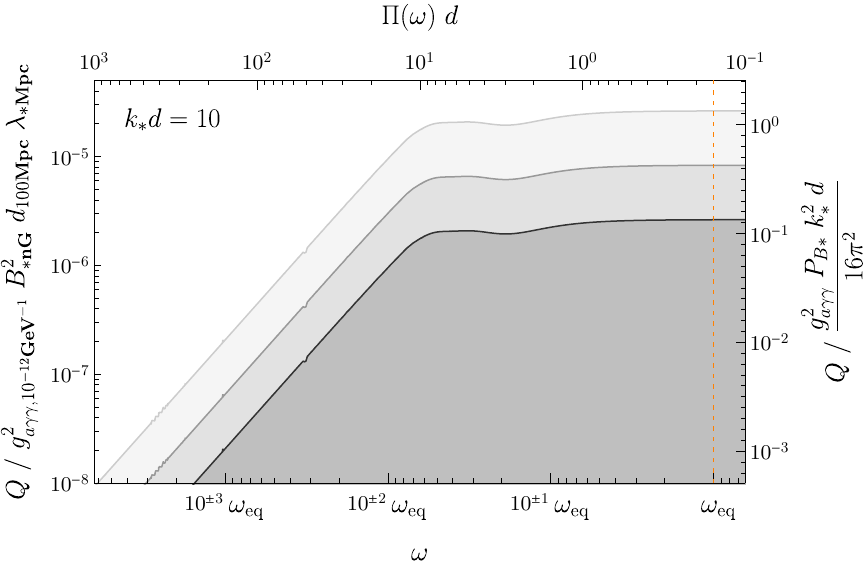}
\\
\includegraphics[width=0.4\linewidth]{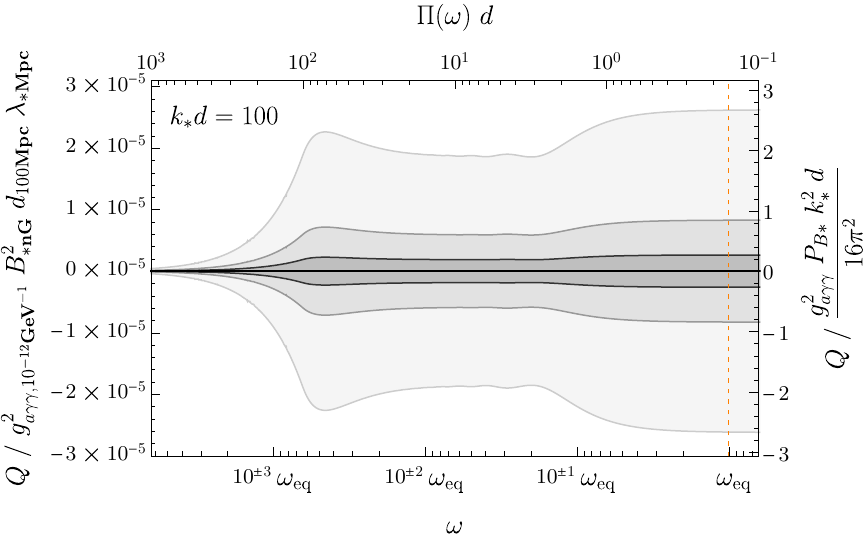}
\hskip 1cm
\includegraphics[width=0.4\linewidth]{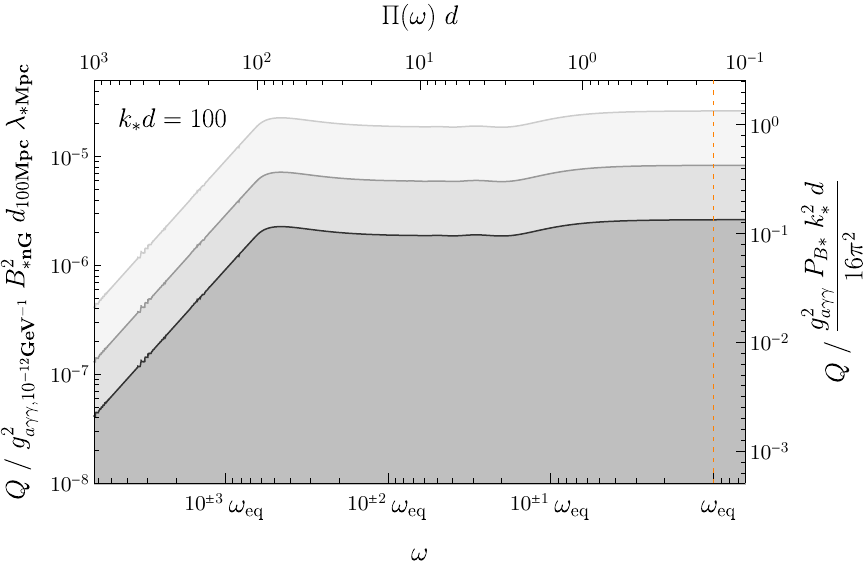}
\caption{
Stokes parameter $Q$ (linear polarization) of photons with the non-helical ($P_{aB*} = 0$) magnetic field power spectrum \eqref{eq:PB} is plotted for $k_* d = 0.1$, $1$, $10$, and $100$ from top to bottom.
In each panel, the black line represents the expectation value $\text{Exp}[Q]=0$, and the shaded bands represent the standard deviations $\sqrt{\text{Var}[Q]}$, $\sqrt{\text{Var}[Q]/10}$, and $\sqrt{\text{Var}[Q]/100}$.
The upper horizontal axis is the dimensionless variable $\Pi (\omega) d$, and the lower horizontal axis is $\omega$ in units of $\omega_\mathrm{eq}\equiv (|m_a^2 - m_\text{pl}^2| / 2\chi_{\mathrm{CMB}})^{\frac12}$ with $(|m_a^2 - m_\text{pl}^2|)^{\frac12}=10^{-11}\mathrm{eV}$, $2\chi_{\mathrm{CMB}}=10^{-42}$, and $d = 100\, \text{Mpc}$. 
The orange dashed line indicates $\omega = \omega_\mathrm{eq}$.
The right vertical axis is $Q$ normalized by $g_{a\gamma\gamma}^2 P_{B*} k_*^2 d / 16\pi^2$. 
The left vertical axis is $Q$ normalized by $g^2_{a\gamma\gamma,10^{-12}\mathrm{GeV}^{-1}}\equiv(g_{a\gamma\gamma}/10^{-12}\mathrm{GeV}^{-1})^2 $, $B_{*\text{nG}}^2 \equiv (B_* / \text{nG})^2$, $d_{100\text{Mpc}} \equiv d / 100\,\text{Mpc}$, and $\lambda_{*\text{Mpc}} \equiv \lambda_{*} / \text{Mpc}$.
In each row, the left and right panels show the same quantity: the vertical axis is linear in the left, but logarithmic in the right.
}
\label{fig:PlotQnon}
\end{figure}

\begin{figure}[H]
\centering
\includegraphics[width=0.4\linewidth]{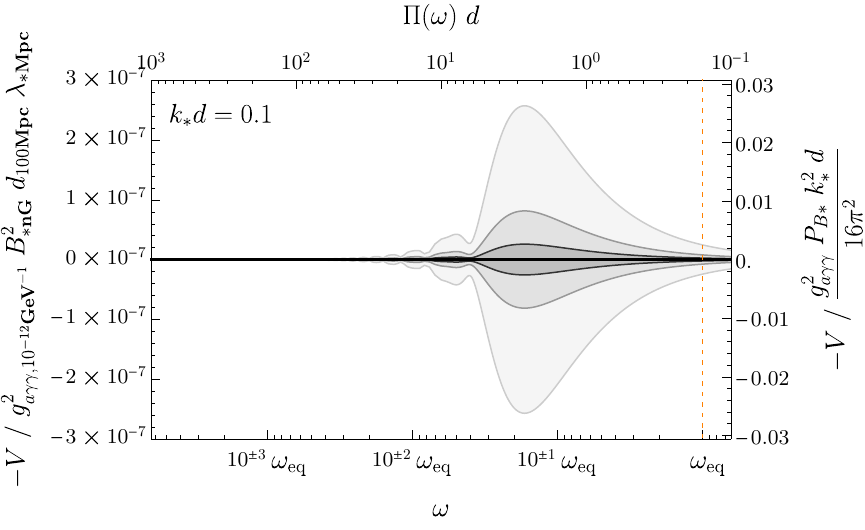}
\hskip 1cm
\includegraphics[width=0.4\linewidth]{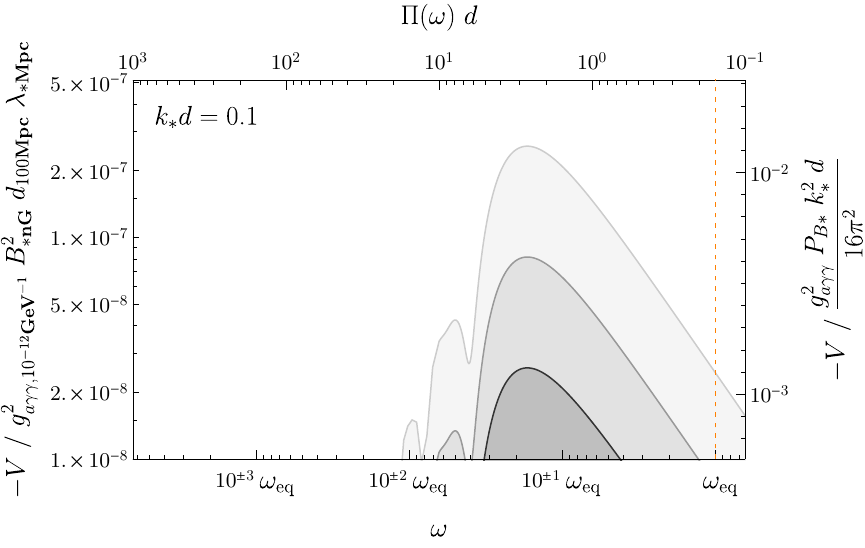}
\\
\includegraphics[width=0.4\linewidth]{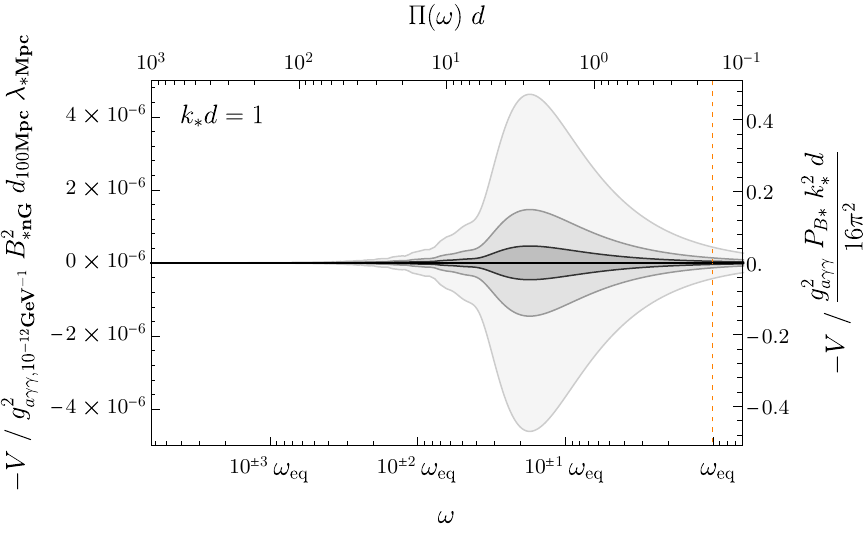}
\hskip 1cm
\includegraphics[width=0.4\linewidth]{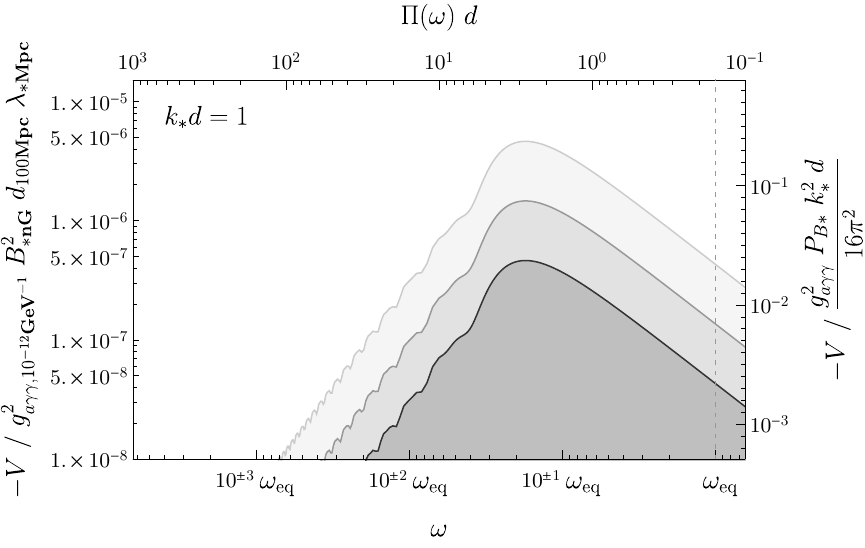}
\\
\includegraphics[width=0.4\linewidth]{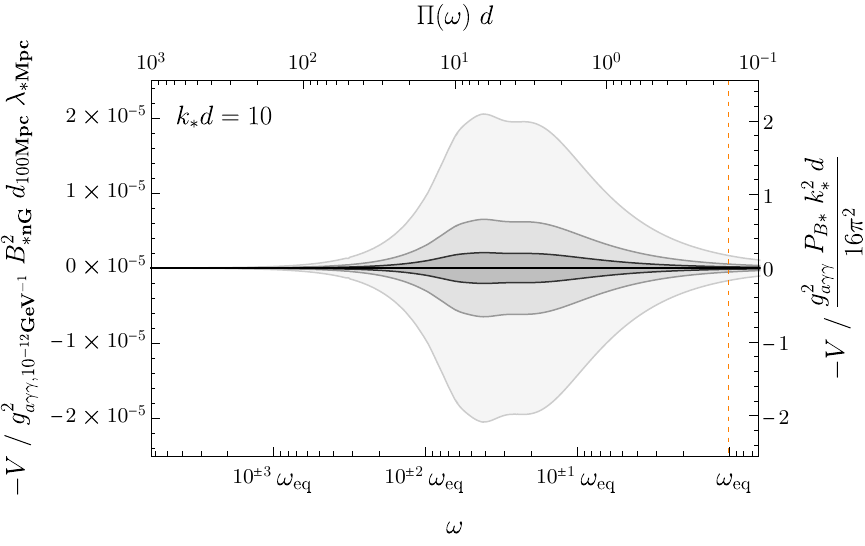}
\hskip 1cm
\includegraphics[width=0.4\linewidth]{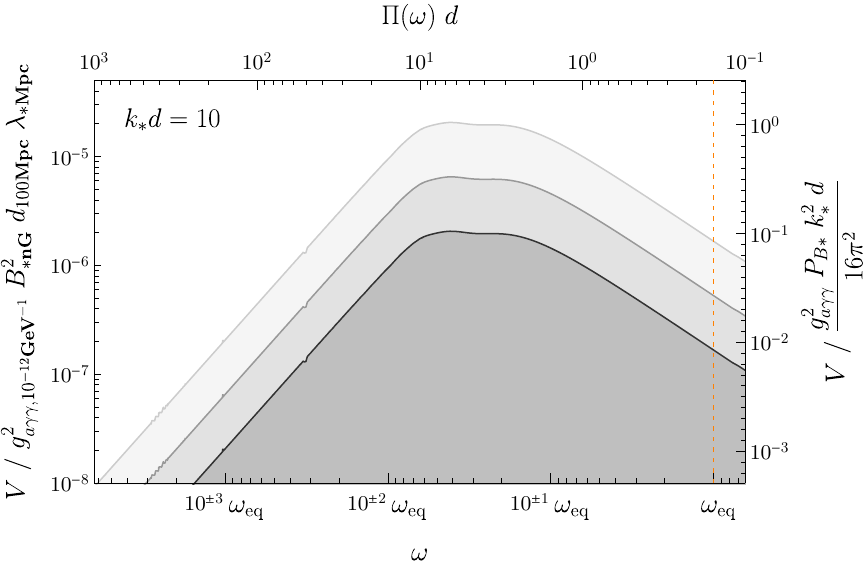}
\\
\includegraphics[width=0.4\linewidth]{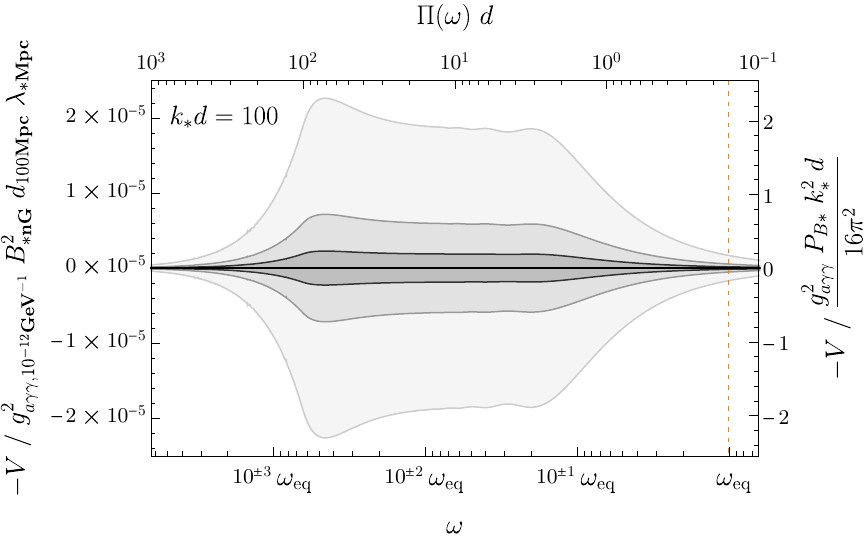}
\hskip 1cm
\includegraphics[width=0.4\linewidth]{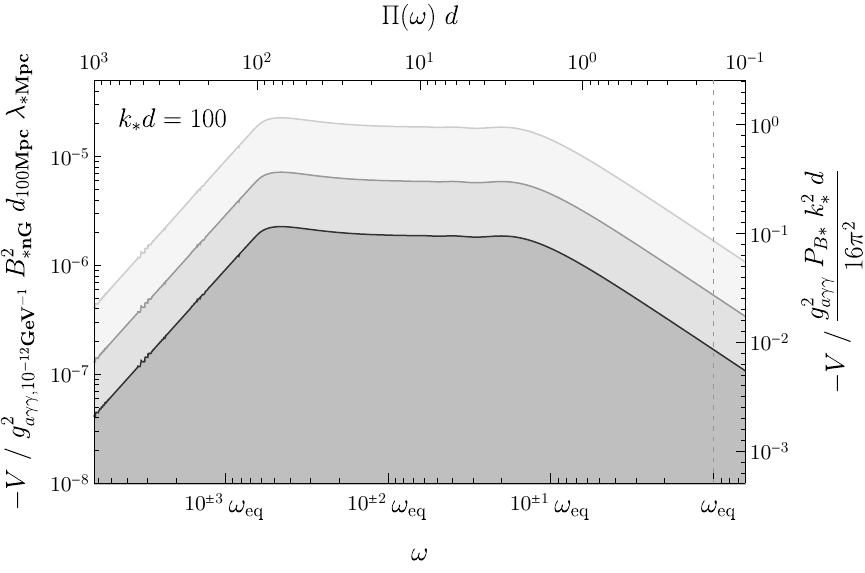}
\caption{
Stokes parameter $-V$ (circular polarization) of photons with the non-helical ($P_{aB*} = 0$) magnetic field power spectrum \eqref{eq:PB} is plotted for $k_* d = 0.1$, $1$, $10$, and $100$ from top to bottom.
In each panel, the black line represents the expectation value $-\text{Exp}[V] = 0$, and the shaded bands represent the standard deviations $\sqrt{\text{Var}[V]}$, $\sqrt{\text{Var}[V]/10}$, and $\sqrt{\text{Var}[V]/100}$.
The upper horizontal axis is the dimensionless variable $\Pi (\omega) d$, and the lower horizontal axis is $\omega$ in units of $\omega_\mathrm{eq}\equiv (|m_a^2 - m_\text{pl}^2| / 2\chi_{\mathrm{CMB}})^{\frac12}$ with $(|m_a^2 - m_\text{pl}^2|)^{\frac12}=10^{-11}\mathrm{eV}$, $2\chi_{\mathrm{CMB}}=10^{-42}$, and $d = 100\, \text{Mpc}$. 
The orange dashed line indicates $\omega = \omega_\mathrm{eq}$.
The right vertical axis is $-V$ normalized by $g_{a\gamma\gamma}^2 P_{B*} k_*^2 d / 16\pi^2$. 
The left vertical axis is $-V$ normalized by $g^2_{a\gamma\gamma,10^{-12}\mathrm{GeV}^{-1}}\equiv(g_{a\gamma\gamma}/10^{-12}\mathrm{GeV}^{-1})^2 $, $B_{*\text{nG}}^2 \equiv (B_* / \text{nG})^2$, $d_{100\text{Mpc}} \equiv d / 100\,\text{Mpc}$, and $\lambda_{*\text{Mpc}} \equiv \lambda_{*} / \text{Mpc}$.
In each row, the left and right panels show the same quantity: the vertical axis is linear in the left, but logarithmic in the right.
}
\label{fig:PlotVnon}
\end{figure}

In this Appendix, we show numerical results for the case of a non-helical magnetic field $P_{aB}=0$.
In Figs.~\ref{fig:PlotInon}, \ref{fig:PlotQnon}, and \ref{fig:PlotVnon}, we present the behavior of the Stokes parameters $I$, $Q$, and $V$ as functions of $\Pi(\omega) d$ for various values of $k_* d$, similarly to Figs.~\ref{fig:I}, \ref{fig:Q}, and \ref{fig:V}.
The black lines represent the expectation values while the gray bands correspond to the standard deviations $\sqrt{\mathrm{Var}[1-I]}$, $\sqrt{\mathrm{Var}[Q]}$, and $\sqrt{\mathrm{Var}[V]}$.
The three different bands are for $(1, 1/\sqrt{10}, 1/\sqrt{100})$ times the original one, respectively.

One observes that Fig.~\ref{fig:PlotInon} remains almost unchanged from Fig.~\ref{fig:I}.
This is because both the expectation value $\text{Exp} [1 - I]$ and the variance $\text{Var} [1 - I]$ are dominated by the integrals $\alpha$ and $\gamma$, which are determined by the power spectrum $P_B$, not by $P_{a B}$.
Fig.~\ref{fig:PlotQnon} also remains qualitatively unchanged from Fig.~\ref{fig:Q}.
In contrast, Fig.~\ref{fig:PlotVnon} behaves quite differently from Fig.~\ref{fig:V}.
The main difference is that the expectation value of $V$ vanishes in Fig.~\ref{fig:PlotVnon}, since it is determined solely by $\beta$.
On the other hand, the variances $\text{Var} [Q]$ and $\text{Var} [V]$ remain nonzero because these are largely determined by $\alpha$ and $\gamma$.

\clearpage

\bibliography{main}

\end{document}